\documentclass[aps,reprint,superscriptaddress,longbibliography,showpacs,superscriptaddress,notitlepage,nofootinbib]{revtex4-1}

\usepackage{mathtools}
\usepackage{dcolumn}
\usepackage{bm}

\usepackage{booktabs}

\usepackage[dvips]{graphicx}
\usepackage{amsmath,amssymb,amsthm,mathrsfs,amsfonts,dsfont,amscd,keyval}
\usepackage{mathtools,mathrsfs}
\usepackage{yfonts} 
\usepackage{ytableau}
\usepackage{subfigure, epsfig}
\usepackage{titletoc} 
\usepackage[subfigure]{tocloft}
\usepackage{titlesec}

\usepackage{braket}
\usepackage{bm}
\usepackage{enumerate}
\usepackage{color}
\usepackage{hyperref}
\usepackage{tabularx}
\usepackage{times}
\hypersetup{breaklinks=true}
\hypersetup{colorlinks=true, linkcolor=blue, citecolor=magenta, urlcolor=black, linktocpage=true}
\usepackage[hyphenbreaks]{breakurl}
\usepackage{multirow}
\usepackage{float}
\usepackage{pdfpages}
\usepackage{pgffor}
\usepackage{changepage}
\usepackage{booktabs}
\usepackage{bbm}

\usepackage{ytableau}
\ytableausetup{boxsize = 7pt}
\usepackage{tikz-cd}

\DeclareMathOperator{\U}{U}
\DeclareMathOperator{\SU}{SU}

\newcommand{\comments}[1]{} 

\theoremstyle{definition}
\newtheorem{definition}{Definition}[section]
\newtheorem*{example}{Example}
\newtheorem{proposition}[definition]{Proposition}

\theoremstyle{plain}
\newtheorem{theorem}[definition]{Theorem}
\newtheorem{lemma}[definition]{Lemma}
\newtheorem{corollary}[definition]{Corollary}
\newtheorem*{conjecture}{Conjecture}
\newtheorem*{fact}{Fact}
\newtheorem*{observation}{Observation}

\theoremstyle{remark}
\newtheorem*{remark}{Remark}

\makeatletter
\AtBeginDocument{\let\LS@rot\@undefined}
\makeatother

\begin{document}

\title{Designs from Local Random Quantum Circuits with $\text{SU}(d)$ Symmetry}

\author{Zimu Li $^{**}$}
\email{lizm@mail.sustech.edu.cn}
\affiliation{Yau Mathematical Sciences Center, Tsinghua University, Beijing 100084, China}

\author{Han Zheng $^{**}$}
\email{hanz98@uchicago.edu}
\affiliation{Pritzker School of Molecular Engineering, The University of Chicago, Chicago, IL 60637, USA}
\affiliation{Department of Computer Science, The University of Chicago, Chicago, IL 60637, USA}

\author{Junyu Liu}
\email{junyuliu@uchicago.edu}
\affiliation{Pritzker School of Molecular Engineering, The University of Chicago, Chicago, IL 60637, USA}
\affiliation{Department of Computer Science, The University of Chicago, Chicago, IL 60637, USA}
\affiliation{Kadanoff Center for Theoretical Physics, The University of Chicago, Chicago, IL 60637, USA}

\author{Liang Jiang}
\email{liangjiang@uchicago.edu}
\affiliation{Pritzker School of Molecular Engineering, The University of Chicago, Chicago, IL 60637, USA}

\author{Zi-Wen Liu}
\email{zwliu0@tsinghua.edu.cn}
\affiliation{Yau Mathematical Sciences Center, Tsinghua University, Beijing 100084, China}
\affiliation{Perimeter Institute for Theoretical Physics, Waterloo, Ontario N2L 2Y5, Canada}

\begin{abstract}
    The generation of $k$-designs (pseudorandom distributions that emulate the Haar measure up to $k$ moments) with local quantum circuit ensembles is a problem of fundamental importance in quantum information and physics. Despite the extensive understanding of this problem for ordinary random circuits, the crucial situations where symmetries or conservation laws are in play are known to pose fundamental challenges and remain little understood. We construct, for the first time, explicit local unitary ensembles that can achieve high-order unitary $k$-designs under transversal continuous symmetry, in the particularly important $\text{SU}(d)$ case. Specifically, we define the Convolutional Quantum Alternating group (CQA) generated by 4-local $\text{SU}(d)$-symmetric Hamiltonians as well as associated 4-local $\text{SU}(d)$-symmetric random unitary circuit ensembles, and prove that they form and converge to $\text{SU}(d)$-symmetric $k$-designs, respectively, for all $k < n(n-3)/2$ with $n$ being the number of qudits. A key technique that we employ to obtain the results is the Okounkov--Vershik approach to $S_n$ representation theory. To study the convergence time of the CQA ensemble, we develop a numerical method using the Young orthogonal form and $S_n$ branching rule. We provide strong evidence for a subconstant spectral gap and certain convergence time scales of various important circuit architectures, which contrast with the symmetry-free case. We also provide comprehensive explanations of the difficulties and limitations in rigorously analyzing the convergence time using methods that have been effective for cases without symmetries, including Knabe's local gap threshold and Nachtergaele's martingale method. This suggests that a novel approach is likely necessary for understanding the convergence time of $\text{SU}(d)$-symmetric local random circuits.
\end{abstract}
\maketitle
\newpage

\def\thefootnote{**}\footnotetext{These authors contributed equally to this work}\def\thefootnote{\arabic{footnote}}


\section{Introduction} \label{sec: intro}

As a ubiquitous principle in nature, symmetry has played a fundamental role in the development of physics. As dictated by the celebrated Noether's theorem, symmetries are linked with conservation laws such as energy and charge conservation. The presence of symmetries and conserved quantities, especially continuous and non-Abelian ones, can drastically alter the physics of quantum information. For instance, transversal continuous symmetries in quantum error-correcting codes can enforce the loss of logical information, leading to the Eastin--Knill theorem \cite{eastin2009restrictions} and its approximate versions \cite{faist20,Woods2020continuousgroupsof,Kubica21,Zhou2021newperspectives,Yang22,Tajima2022,liu2021approximate,liu2023quantum} that have drawn great interest in quantum computing as well as fundamental physics. In addition, continuous symmetries induce inherent constraints on quantum dynamics and scrambling effects, which are of extensive interest at the forefront of many-body physics and quantum gravity (see, e.g.~Refs.~\cite{nahum2018operator,Nahum2,Hayden-Preskill2007,SekinoSusskind2008,khemani2018operator, rakovszky2018diffusive, Huang2019OTOC}). Notably, the effects of non-Abelian symmetries in the aforementioned areas of quantum codes and dynamics have been under active study in recent years \cite{Hayden_2021,faist20,Woods2020continuousgroupsof,Zhou2021newperspectives,kong2022near,Yang22,Kubica21,liu2021approximate,liu2023quantum,MarvianSU2,MarvianSUd,majidy2023non,majidy2023critical,Liu2024Mpemba}.
Among the wide variety of different symmetries, $\text{SU}(d)$ holds exceptional importance in quantum theory, manifesting the group of transformations on a $d$-dimensional quantum system. Furthermore, it represents non-Abelian symmetries (associated with noncommuting charges) that exhibit more complex structures and richer physics. 

The locality of interactions is also a fundamental property of physics. In mathematical characterizations of the legitimate dynamics of physical systems, this is captured by the locality of terms in the Hamiltonian or the locality of elementary gates in circuit models. A particularly desired feature of such models driven by local interactions is the ability to generate intrinsically global dynamics, which is essential in physics as the foundation for the emergence of nontrivial global phenomena, as well as for practical reasons such as the universality of quantum circuits composed of small elementary gates which underpins the viability of quantum computing. Specifically, we call an operator \emph{$r$-local} if it acts on at most $r$ sites of a system, which represents the notion of all-to-all locality, while an operator is said to be \emph{geometrically $r$-local} if these sites are adjacent on the specific graph corresponding to the circuit architecture or geometry.

A fundamental form of this emergence of globalness is the generation of (approximate) \emph{unitary $k$-designs}, i.e.,~ensembles of unitaries that match the Haar (invariant) measure up to the $k$-th homogeneous polynomial moment \cite{Dankert2026PRA,Gross2006} (one is usually interested in $k \geq 2$). (In the rest of this paper, we may refer to ``unitary designs'' simply as ``designs'' without causing confusion.) 
Designs have become a standard concept in quantum information, many-body physics, and quantum gravity, due to their close connection to the notions of decoupling, many-body chaos, thermalization, entanglement generation, circuit complexity, etc., which play key roles in these fields \cite{Hayden-Preskill2007,Yoshida-Kitaev2017,RobertsChaos2017,junyu2017chaos,Liu_2018,Huang:2020tih,PhysRevLett.120.130502,junyu2020chargescrambler,brandao2021models,Haferkamp2024linear}.
Furthermore, there are many more important applications of designs and their formation in separate contexts -- e.g., providing general frameworks for understanding the phenomena of anticoncentration  \cite{Hangleiter2018anticoncentration,harrow2023approximate} and barren plateaus \cite{McClean_2018}, which are of great recent interest in quantum computing. For a diverse range of other applications, see, e.g.~Refs.~\cite{harrow2016local,zhu2016clifford,webb2015clifford,zhu2017multiqubit} and references therein.

Given the importance of both symmetry and locality principles, it is imperative to study the properties of local models such as quantum circuits that respect symmetries, 
the appeal of which extends from physical scenarios \cite{khemani2018operator, rakovszky2018diffusive, Huang2019OTOC,yoshida:soft,Nakata2023, chang2024deep} to areas with practical motivations including covariant quantum error correction \cite{Hayden_2021,faist20,Woods2020continuousgroupsof,Zhou2021newperspectives,kong2022near,Yang22,Kubica21,liu2021approximate,liu2023quantum} and quantum machine learning \cite{Seki_2020,Zheng2021SpeedingUL,ragone2022representation,LSKP23}.
Besides the physical and practical motivations evident from the above discussion, the problem of $k$-design generation under continuous symmetry constraints is highly intriguing from the mathematical perspective. Concretely, two central questions regarding design generation are: (i) whether it is possible to generate designs of a certain order with local ensembles, and (ii) if so, how efficiently this can happen. Here, note that the notion of designs may be either exact or approximate; in particular, as long as an ensemble can converge to a $k$-design, i.e., produce an approximate $k$-design to arbitrary precision under convolution, we consider it capable of generating a $k$-design. Without symmetry, the situation has been extensively studied and well understood. First, as naturally expected, universal 2-local gate ensembles can converge to the Haar measure, i.e., generate $k$-designs of any order $k$ \cite{Dankert2026PRA,Gross2006,Arnold2009,emerson05convergence}. Furthermore, it has been rigorously proven in Refs.~\cite{Znidaric2008,Harrow2design2009,Brown_2010,harrow2016local, harrow2023approximate,Metger2024,Haferkamp2024linear} that $k$-designs of any $k$ can be generated by local random circuits with various circuit architectures in time polynomial in both the system size and $k$, which provides a foundation for the practical appeal of designs: while sufficiently powerful for applications, they can be produced efficiently (in polynomial time) in contrast to the true Haar measure that requires exponential time. 

Interestingly, when continuous symmetries are imposed, the situation becomes fundamentally different. 
A remarkable result of Marvian \cite{MarvianNature} reveals a crucial insight that with continuous symmetries in play, global unitaries that can be generated by local circuits are severely restricted, in stark contrast to the scenario without symmetry, casting serious doubts on the capability of local circuits to generate designs. Non-Abelian symmetries make the problem more intricate. When $d \geq 3$, 2-local $\SU(d)$-symmetric unitaries are unable to form $k$-designs even if $k = 2$ \cite{MarvianSUd,marvian2023nonuniversality}. Besides, it has been observed that various existing results and techniques for designs fail to carry over to the case with symmetry, indicating fundamental difficulties in understanding design generation in the presence of symmetries \cite{kong2022near}. So far, no results for local constructions capable of generating nontrivial designs in the presence of $\text{SU}(d)$ symmetry are known, and even the basic question of whether they are possible at all is largely up in the air.

In this work, we solve this open problem of design generation under symmetry raised in Ref.~\cite{kong2022near} in the case of $\text{SU}(d)$ symmetry, where operators are demanded to commute with $\hat{U}^{\otimes n }$ acting transversally on qudits of \emph{local dimension} $d$ from an $n$-qudit system. The $\text{SU}(d)$ symmetry has garnered significant interest in various related areas of quantum information processing. In covariant quantum error correction, it remains an open question whether efficient constructions of random $\text{SU}(d)$-symmetric  error-correcting codes exist \cite{kong2022near}. In quantum thermodynamics, random quantum circuits with transversal $\text{SU}(d)$ symmetry can be seen as a canonical dynamical model with non-Abelian symmetry, leading to many intriguing and often counterintuitive physical phenomena \cite{agarwal2023charge,majidy2023critical, majidy2023non,Liu2024Mpemba}. All these questions hinge on the efficient construction of local unitary circuits capable of converging to $\text{SU}(d)$-symmetric unitary designs. A key conclusion of our work is that it is possible to construct 4-local ensembles with $\text{SU}(d)$ symmetry that generate $k$-designs for $k$ up to at least $O(n^2)$. We first provide a systematic characterization of unitary $k$-designs using commutants from invariant theory \cite{Goodman2009}, bridging various widely used characterizations of random unitaries including the tensor product expander \cite{HarrowTEP09,harrow2016local,Haferkamp2021} and the frame potential \cite{RobertsChaos2017,junyu2017chaos,hunter2019unitary,junyu2020chargescrambler,brian2022linear}. Together with the utilization of group-representation-theoretic techniques, in particular, the Okounkov--Vershik approach \cite{Okounkov1996}, we are able to circumvent the no-go theorems in the presence of $\text{SU}(d)$ symmetry \cite{MarvianNature,MarvianSU2,MarvianSUd} that forbid the generation of arbitrary global $\text{SU}(d)$-symmetric unitaries via 2-local gates. 

Specifically, we prove that the $S_n$-\emph{Convolutional Quantum Alternating (CQA) group} introduced in Ref.~\cite{Zheng2021SpeedingUL}, which is generated by at most 4-local $\text{SU}(d)$-symmetric Hamiltonians, is able to form exact $k$-designs of the group $\mathcal{U}_\times$ consisting of all $\text{SU}(d)$-symmetric unitaries for all $k<n(n-3)/2$ whenever $d < n$. 
For such $k$, we show that various ensembles constructed based on the group CQA, given by $\text{SU}(d)$-symmetric 4-local Hamiltonians or unitary gates, can converge to $\text{SU}(d)$-symmetric $k$-designs. Moreover, we rigorously prove that for generators with bounded locality, it is impossible to achieve  $\text{SU}(d)$-symmetric $k$-designs for arbitrarily large $k$.
Altogether, our results show that local circuit models with continuous symmetry that generate $k$-designs for fairly large $k$ (sufficient for most applications) do exist, which, to our best knowledge, has not been reported before.  

To understand another key question of the time scales of convergence to $k$-designs, we examine the spectral gaps of the generating ensembles. In particular, in the basic qubit ($d = 2$) case, our numerical analysis suggests that the CQA local random circuits with e.g.~the standard one-dimensional (1D) architecture form $\epsilon$-approximate $2$-designs in $\Theta(n^4 + n^3\log(1/\epsilon))$ time (circuit depth), which is longer than local random circuits without symmetries \cite{Dankert2026PRA,Oliveira2design2007a,Znidaric07:2,Oliveira2design2007b,Znidaric2008,Brown_2010,harrow2016local,harrow2023approximate,hunter2019unitary} by a factor that scales as $n^2$.
We then carefully illustrate the mathematical difficulties in analytically determining the convergence time of CQA ensembles using the previously considered techniques, including the frame potential \cite{RobertsChaos2017,junyu2017chaos,hunter2019unitary,junyu2020chargescrambler,brian2022linear}, the local gap threshold \cite{Knabe1988,Gosset2016,Haferkamp2021}, and the martingale method \cite{Nachtergaele1996,Cirac2006,harrow2016local}, supported by numerical analysis. These understandings may serve as a guide for future research on rigorously determining the convergence times to $k$-designs under continuous symmetry. From an application perspective, our results open up the possibility of constructing random ensembles with $\text{SU}(d)$ symmetry, which holds wide significance in quantum information and physics, as extensively discussed in a companion paper \cite{SUd-k-Design2023Application}.

This paper is organized as follows. In Section \ref{sec: main_results}, we provide formal definitions of various key concepts -- in particular unitary $k$-designs and their symmetric variants -- and carefully discuss the connection and difference between the {frame potential} \cite{RobertsChaos2017,junyu2017chaos,hunter2019unitary,junyu2020chargescrambler,brian2022linear} and {spectral gap} \cite{harrow2016local,Haferkamp2021} characterizations of the convergence to unitary $k$-designs. We also overview the main results of this work as well as related existing works. In Section \ref{sec:ExactDesign}, we elucidate that the {CQA group} forms exact  $\text{SU}(d)$-symmetric unitary $k$-designs for $k < n(n-3)/2$, and demonstrate that unbounded locality is necessary to form $k$-designs with arbitrarily large $k$. In Section \ref{sec: cqa-construction}, we discuss the convergence of dynamical models based on CQA to   $\text{SU}(d)$-symmetric $k$-designs. Technical details of the proofs discussed in the main text and additional mathematical background can be found in the Appendix. 


\section{Preliminaries and summary of results} \label{sec: main_results}

In this section, we will formally lay out the key definitions and provide an overview of the main technical results. For readers' convenience, we summarize several key notations and symbols used in this paper in Table \ref{table:Notations}.
Further details will be explained later within specific contexts.

\begin{table}[ht] 
	\caption{Summary of notations.} 
	\begin{tabular}{c l } 
		\hline\hline 
		Notation & Definition  \\ [0.5ex] 
		\hline 
		$[n]$ & The set of all integers $1,2,...,n$ \\
		$\lfloor x \rfloor$ & The largest integer less than or equal to a real number $x$ \\
		$\lambda \vdash n$ & A partition of $n$ \\
		$S^\lambda$ & An $S_n$ irrep as a subspace of the Hilbert space $\mathcal{H}$  \\ 
		$p(n,d)$ & The number of inequivalent $S_n$ irreps of $n$ qudits \\ 
		$X_l$ & A Young--Jucys--Murphy (YJM) element  \\
		$\tau$ & A Generic transposition or SWAP \\
		$\tau_j$ & A nearest-neighbor transposition or SWAP $(j,j+1)$ \\
		$\mathcal{U}_\times$ & The group of $\text{SU}(d)$-symmetric unitaries \\
		$S\mathcal{U}_\times$ & The subgroup of $\mathcal{U}_\times$ with trivial relative phases \\
        $\mathcal{V}_4$ & The group generated by 4-local $\text{SU}(d)$-symmetric unitaries \\
		CQA & The group generated by 2nd order YJM elements and $\tau_j$ \\
		$\mathrm{CQA}^{(k)}$ & The group generated by $k$-th order YJM elements and $\tau_j$ \\
		$\mathcal{E}_{\mathrm{CQA}}$ & The CQA random walk ensemble \\
        $\mathcal{E}_{\mathcal{V}_4}$ & The CQA 4-local random unitary ensemble \\
		\hline\hline
	\end{tabular}
	\label{table:Notations}
\end{table}

\subsection{Unitary $k$-designs and commutants} \label{sec: k-designs-intro}

Let $\mathcal{E}$ be an ensemble (distribution) consisting of unitaries acting on the Hilbert space $\mathcal{H}$. For any operator $M \in \operatorname{End}(\mathcal{H}^{\otimes k})$, the \emph{$k$-fold (twirling) channel} with respect to $\mathcal{E}$ acting on $M$ is defined by the following integral over $\mathcal{E}$:
\begin{align}\label{eq:tpe}
	T_k^{\mathcal{E}}(M) = \int_{\mathcal{E}} dU U^{\otimes k} M U^{\dagger \otimes k}.
\end{align}
Alternatively, it can be characterized by
\begin{align}
	T_k^{\mathcal{E}} = \int_\mathcal{E} dU U^{\otimes k } \otimes \overline{U}^{ \otimes k},
\end{align}
namely, the \emph{$k$-th moment (super-)operator} acting on $\operatorname{End}(\mathcal{H}^{\otimes k})$. Besides, given any compact group $G$, we use $T_k^G$ to denote the $k$-th moment operator associated with the Haar measure over $G$. An ensemble is called an \emph{(exact) unitary $k$-design} of the group $G$ if $T^{\mathcal{E}}_k = T_k^G$. 

More generally, we call $\mathcal{E}$ an \emph{$\epsilon$-approximate $k$-design} if the strong notion of $\epsilon$-approximation in terms of complete positivity \cite{harrow2016local,harrow2023approximate} holds, namely,
\begin{align}\label{def:approx-design}
	(1 - \epsilon) T_k^G \leq_{\mathrm{cp}} T_k^{\mathcal{E}} \leq_{\mathrm{cp}} (1 + \epsilon) T_k^G. 
\end{align}
where $A \leq_{\mathrm{cp}} B$ means that $B - A$ is completely positive (for various other definitions and their applications, see also Refs.~\cite{Dankert2026PRA,Harrow2design2009,Liu_2018,hunter2019unitary,Haferkamp2021,Gao2022} ). We also denote by $c_{\mathrm{cp}}(\mathcal{E}, k)$ the smallest constant $\epsilon$ achieving the above bound.

A fundamental case is when $G$ is the unitary group $\text{U}(\mathcal{H})$, which has been extensively studied in the quantum information literature. Specifically, the moment operator of interest is $T^{\operatorname{U}(\mathcal{H})}_k$, where we integrate over  $\text{U}(\mathcal{H}) \equiv \operatorname{U}(d^n) \equiv \operatorname{U}(N)$, with $N \equiv \dim \mathcal{H} = d^n$ being the dimension of the system. Important knowledge includes, e.g., that 2-local Haar-random ensembles approximate $k$-designs \cite{Dankert2026PRA,Gross2006,Znidaric2008,HarrowTEP08,Brown_2010,harrow2016local,hunter2019unitary,brandao2021models,harrow2023approximate,Haferkamp2024linear}, and, for qubits, that the Clifford group forms an exact 3-design of $\text{U}(\mathcal{H})$  \cite{webb2015clifford,zhu2016clifford,zhu2017multiqubit}.

Here, we are interested in the more involved situation in which tranversal continuous symmetries are imposed. In this paper, we consider transversal $\mathrm{SU}(d)$ symmetry on qudits as a canonical example.
\begin{definition}
	An operator $O$, including unitaries or Hermitian matrices, on the Hilbert space of $n$ qudits, is \emph{$\text{SU}(d)$-symmetric} if $O \hat{U}^{\otimes n} = \hat{U}^{\otimes n} O$ for any transversal action of $\hat{U} \in \operatorname{SU}(d)$ on the qudits.
\end{definition} 
We say that an ensemble is an \emph{exact $\text{SU}(d)$-symmetric unitary $k$-design} if its $k$-th moment operator matches that of the Haar measure over the symmetry-restricted group $\mathcal{U}_\times$, i.e.,~ $T_k^{\mathcal{E}} = T_k^{\mathcal{U}_\times}$. And again, we consider the strong notion of approximation of $k$-designs based on complete positivity, with its relation to various other natural approximation conditions carefully discussed in Appendix~\ref{sec:Definitions}.
The formal definitions go as follows.

\begin{definition}\label{def:approx-SUd-design}
	An ensemble $\mathcal{E}$ is said to be an \emph{ $\text{SU}(d)$-symmetric unitary $k$-design} if $T_k^{\mathcal{E}} = T_k^{\mathcal{U}_\times}$, where $\mathcal{U}_\times$ is the group consisting of all $\text{SU}(d)$-symmetric unitaries acting on the system. Furthermore, we say that $\mathcal{E}$ is an \emph{$\epsilon$-approximate $\text{SU}(d)$-symmetric unitary $k$-design} if
	\begin{align}
		(1 - \epsilon) T_k^{\mathcal{U}_\times} \leq_{\mathrm{cp}} T_k^{\mathcal{E}} \leq_{\mathrm{cp}} (1 + \epsilon) T_k^{\mathcal{U}_\times},
	\end{align}
	where $A \leq_{\mathrm{cp}} B$ means that $B - A$ is completely positive.
\end{definition}

By the left and right invariance, or simply the bi-invariance property, of Haar measure defined on any compact subgroup $G$, it can be straightforwardly checked that the operator $T_k^G$ is an orthogonal projector, meaning that $T_k^G$ is Hermitian and $(T_k^G)^2 = T_k^G$. Consequently, $T_k^G$ only has zero and unit eigenvalues. The eigenspace of unit eigenvalues, called the \emph{commutant} of $G$ under the representation, is of central importance. It can be verified by definition that
\begin{align}\label{eq:commutant}
	\operatorname{Comm}_k(G) = \{M \in \operatorname{End}(\mathcal{H}^{\otimes k}); U^{\otimes k} M = M  U^{\otimes k} \}.
\end{align}

\begin{fact}
	Given a unitary ensemble $\mathcal{E}$ that is also a compact subgroup of the concerned group $G$, it forms an exact unitary $k$-design if $\operatorname{Comm}_k(\mathcal{E}) = \operatorname{Comm}_k(G)$. Otherwise, it never converges to a unitary $k$-design even approximately.
\end{fact} 

As a concrete example of the commutant, in the case of unitary $k$-designs without concern for symmetry, we consider $\operatorname{U}(d^n)$. Then, by Schur--Weyl duality and the double commutant theorem \cite{Goodman2009,Tolli2009}, $\operatorname{Comm}_k(\operatorname{U}(d^n) )$ is spanned by permutations of symmetric group $S_k$ acting on $\mathcal{H}^{\otimes k}$. A typical element from the spanning set can be expressed as $\ket{\psi_{\sigma,d}}^{\otimes n}$, where $\ket{\psi_{\sigma,d}} = d^{-k/2} (I \otimes \pi_{S_k}(\sigma) ) \sum_{i = 1}^{d^k} \ket{i,i}$ \cite{harrow2016local,Haferkamp2021}
\begin{align}\label{eq:permutation}
	\pi_{S_k}(\sigma) \ket{i_1,...,i_k} = \ket{i_{\sigma^{-1}(1) },...,i_{\sigma^{-1}(k) } }, \sigma \in S_k.
\end{align}
More straightforwardly, the representation $\pi_{S_k}(\sigma)$ of these permutation operators can be derived using either the Pauli basis or matrix units. For example, consider the following transposition on $\mathcal{H}^{\otimes 2}$, \cite{Collins2003,RobertsChaos2017}:
\begin{align}\label{eq:transposition}
	W_{12} = \frac{1}{d}\sum_P P \otimes P^\dag = \sum_{i,j} E_{ij} \otimes E_{ji},
\end{align}
where $(E_{ij})_{kl} = \delta_{ik} \delta_{jl}$. Together with the identity operator, they span $\operatorname{Comm}_2(\operatorname{U}(d^n) )$.
 
For a general ensemble $\mathcal{E}$, it is straightforward to check that $T^{\operatorname{U}(\mathcal{H})}_k$ commutes with $T^{\mathcal{E}}_k$. If $T^{\mathcal{E}}_k$ is furthermore Hermitian, they are simultaneously diagonalizable. To determine whether $\mathcal{E}$ forms unitary $k$-designs, it suffices to show that its unit eigenvalue subspace (it should not be called a commutant here if $\mathcal{E}$ is not a group) is equal to the commutant of the Haar unitaries. As such, the commutant plays a vital role in understanding the design properties and can be connected to other mathematical tools such as the \emph{frame potential}:
\begin{align}\label{eq:FramePotential}
	    F^{(k)}_{\mathcal{E}} = \int_{\mathcal{E}} dU dV | \operatorname{Tr}(UV^\dagger) |^{2k}
    = \operatorname{Tr}(T^{\mathcal{E}}_k T^{\mathcal{E} \dagger}_k).
\end{align}
Since we always have $\operatorname{Comm}(\mathcal{E}) \supset \operatorname{Comm}(\operatorname{U}(d^n))$, it is clear that
\begin{align}\label{eq:FramePotentialInequality}
	F^{(k)}_{\mathcal{E}} \geq F^{(k)}_{\operatorname{Haar}} = \dim \operatorname{Comm}_k(\operatorname{U}(d^n)).
\end{align}
In the case without symmetry, note that $\dim \operatorname{Comm}_k(\operatorname{U}(\mathcal{H}))$ in \eqref{eq:FramePotentialInequality} is equal to $k!$ for $k \leq d^n$ and can even be evaluated through the so-called \emph{increasing subsequence problem} from combinatorics for larger $k$ \cite{Rains1998}.


\subsection{Structures under transversal $\text{SU}(d)$ symmetry}\label{sec:PreliminariesSymmetricDesign}

To deal with the $\text{SU}(d)$ symmetry, we need to employ a powerful mathematical tool---\emph{Schur--Weyl duality} \cite{Goodman2009,Tolli2009} from representation theory---not just to study commutants as before, but as the theoretical foundation to understand $\text{SU}(d)$-symmetric operators. To be precise, Schur--Weyl duality indicates that the Hilbert space $\mathcal{H}$ of an $n$-qudit system is decomposed according to the irreducible representation (irrep) of the symmetric group $S_n$ as
\begin{align}\label{eq:SchurWeyl}
	\mathcal{H} = \bigoplus_\lambda \mathbbm{1}_{ m_{\lambda}} \otimes S^\lambda,
\end{align}
where $S^\lambda$ stands for an irrep  with $\lambda \vdash n$ recording the irrep as a {partition of $n$} into at most $d$ parts \cite{Fulton1997,Sagan01}. The number $m_\lambda$ denotes the multiplicity of $S^\lambda$ and $\dim S_\lambda \equiv d_\lambda$ is its dimension. We also denote by $p(n,d)$ the total number of inequivalent irreps. A key observation from Schur--Weyl duality and the double commutant theorem \cite{Goodman2009,Tolli2009} is that any $\text{SU}(d)$-symmetric unitary happens to take the form 
\begin{align}
	U = \bigoplus_\lambda I_{m_\lambda} \otimes U_\lambda
\end{align}
with $U_\lambda \in \operatorname{U}(S^\lambda)$, the unitary group acting on the irrep $S^\lambda$, and $I_{m_\lambda}$ being the identity matrix acting on the multiplicity space. Accordingly, we also have the compact group $\mathcal{U}_{\times}$ of $\text{SU}(d)$-symmetric unitaries with a well-defined Haar measure. 

On the other hand, we denote by $S\mathcal{U}_\times$ the group consisting of all such $U \in \mathcal{U}_\times$ modulo the \emph{relative phase factors} (i.e.\ $\det(\Pi_\lambda U) = 1$ with $\Pi_\lambda$ being the projection operator onto the specific irrep $S^\lambda$ as well as its multiple copies). Intuitively, this is analogous to the relationship between the unitary and special unitary groups. However, there is a fundamental difference: the unitary group $\U(d^n)$ can be generated by 2-local unitaries \cite{Brylinski2001,Vlasov2001,Sawicki2016}, but it is impossible to generate $\mathcal{U}_{\times}$ by local unitaries under symmetries. Interestingly, we show that $S\mathcal{U}_\times$ can be generated locally (see \eqref{eq:RestrictedUniversal}). A comprehensive description is provided in Section \ref{sec:main}.

We also sketch in the following the typical group elements $\bigoplus U_j = \bigoplus e^{i\phi_j}V_j, \bigoplus V_j$ from $\mathcal{U}_\times$ and $S\mathcal{U}_\times$, respectively. Note that $U_2$ and $V_2$ are written twice as an indication of possible multiplicities:
\begin{figure}[!h]
\centering
\begin{tikzpicture}[x=0.75pt,y=0.75pt,yscale=-0.3,xscale=0.3,font=\fontsize{0.22}{0.4em}\selectfont]
	
	\draw   (2,2) -- (105.33,2) -- (105.33,105.33) -- (2,105.33) -- cycle ;
	\draw   (277.33,277.33) -- (382.33,277.33) -- (382.33,382.33) -- (277.33,382.33) -- cycle ;
	\draw   (105.33,105.33) -- (191.33,105.33) -- (191.33,191.33) -- (105.33,191.33) -- cycle ;
	\draw   (191.33,191.33) -- (277.33,191.33) -- (277.33,277.33) -- (191.33,277.33) -- cycle ;
	\draw  [dash pattern={on 4.5pt off 4.5pt}] (105.33,105.33) -- (277.33,105.33) -- (277.33,277.33) -- (105.33,277.33) -- cycle ;
	\draw   (388,2) -- (491.33,2) -- (491.33,105.33) -- (388,105.33) -- cycle ;
	\draw   (663.33,277.33) -- (768.33,277.33) -- (768.33,382.33) -- (663.33,382.33) -- cycle ;
	\draw   (491.33,105.33) -- (577.33,105.33) -- (577.33,191.33) -- (491.33,191.33) -- cycle ;
	\draw   (577.33,191.33) -- (663.33,191.33) -- (663.33,277.33) -- (577.33,277.33) -- cycle ;
	\draw  [dash pattern={on 4.5pt off 4.5pt}] (491.33,105.33) -- (663.33,105.33) -- (663.33,277.33) -- (491.33,277.33) -- cycle ;
	
	\draw (280,311.43) node [anchor=north west][inner sep=0.75pt]   [align=left] {$\displaystyle e^{i\phi _{3}} V_{3}$};
	\draw (5,33.43) node [anchor=north west][inner sep=0.75pt]   [align=left] {$\displaystyle e^{i\phi _{1}} V_{1}$};
	\draw (98,128.43) node [anchor=north west][inner sep=0.75pt]   [align=left] {$\displaystyle e^{i\phi _{2}} V_{2}$};
	\draw (183,214.43) node [anchor=north west][inner sep=0.75pt]   [align=left] {$\displaystyle e^{i\phi _{2}} V_{2}$};
	\draw (695,321.43) node [anchor=north west][inner sep=0.75pt]   [align=left] {$\displaystyle V_{3}$};
	\draw (420,43.43) node [anchor=north west][inner sep=0.75pt]   [align=left] {$\displaystyle V_{1}$};
	\draw (512,138.43) node [anchor=north west][inner sep=0.75pt]   [align=left] {$\displaystyle V_{2}$};
	\draw (600,224.43) node [anchor=north west][inner sep=0.75pt]  [align=left] {$\displaystyle V_{2}$};
\end{tikzpicture}
\end{figure}

This subspace decomposition with respect to the symmetry hinders one from approaching the problem by commonly used methods developed for $k$-designs without symmetry. Indeed, as shown in Appendix \ref{sec:SnCommutant}, the commutant corresponding to Haar randomness under $\text{SU}(d)$ symmetry has been foliated with respect to tensor products of inequivalent $S_n$ irreps  and multiplicities. Consider the case of 2-design as an enlightening example. Formally, the matrix representing $T^{\mathcal{U}_\times}_2$ is expanded by integrating 
\begin{align}\label{eq:Exapnsion1}
	& U_{\lambda_1, m_1} \otimes U_{\lambda_2, m_2} \otimes \overline{U}_{\mu_1, m_1'} \otimes \overline{U}_{\mu_2, m_2'} \\
	=\,& e^{-i(\phi_1 + \phi_2 - \psi_1 - \psi_2 )} V_{\lambda_1, m_1} \otimes V_{\lambda_2, m_2} \otimes \overline{V}_{\mu_1, m_1'} \otimes \overline{V}_{\mu_2, m_2'} \notag
\end{align}
with various choices of $\lambda_i,\mu_i,m_i$, and $m_i'$ according to Schur--Weyl duality.
For inequivalent irreps , we are free to assign different phase factors such that $U_{\lambda, m_{\lambda}} = e^{-i \phi_\lambda } V_{\lambda, m_{\lambda}}$, which implies that the integral is nonvanishing if and only if $\lambda_1 = \mu_1; \lambda_2 = \mu_2$ or $\lambda_1 = \mu_2; \lambda_2 = \mu_1$. We refer to these pairings of irrep labels $\lambda_i$ and $\mu_i$ as \emph{Wick contractions}, which can be also generalized for arbitrary $k$ (see, e.g., Ref.~\cite{roberts2016chaos}). More importantly, this shows that the commutant with the presence of $\text{SU}(d)$ symmetry is no longer 2-dimensional (cf.\ the discussion at the end of Section \ref{sec: k-designs-intro}). Instead, it is spanned by more distinct elements characterized by projecting $W_{12}$ from Eq.~\eqref{eq:transposition} into each $S_n$ irrep , like 
\begin{align}\label{eq:GneeralTransposition}
	\begin{aligned}
		& \sum_{\alpha^\lambda,\alpha^\mu} & E_{(\alpha^\lambda, m_1), (\alpha^\lambda, m_1')} \otimes E_{(\alpha^\mu, m_2),(\alpha^\mu, m_2')}, \\
		& \sum_{\alpha^\lambda,\alpha^\mu} & E_{(\alpha^\lambda, m_1), (\alpha^\mu, m_2)} \otimes E_{(\alpha^\mu, m_2'),(\alpha^\lambda, m_1')},
	\end{aligned}
\end{align}
where $E_{(\alpha^\lambda, m_1), (\alpha^\lambda, m_1')}$ is still a matrix unit as in Eq.~\eqref{eq:transposition} but labeled by basis vector indices $\alpha^\lambda$ and $\alpha^\mu$ as well as multiplicities $m_i$ for irrep $S^\lambda$ and $S^\mu$, respectively.

Because of the necessity of counting the number of irreps  denoted by $p(n,d)$, the multiplicities, and a certain symmetric factor to obtain a trivial phase from Eq.~\eqref{eq:Exapnsion1}, analytically evaluating $\dim \operatorname{Comm}_k(\mathcal{U}_\times)$ becomes infeasible with the presence of $\text{SU}(d)$ symmetry. When $d = 2$ (qubits), $p(n,2) = \lfloor n/2 \rfloor + 1$ and we have (for more details, see Appendix \ref{sec:FramePotential})
\begin{align}
	\begin{aligned}
		F^{(2)}_{\mathcal{U}_\times} = &~ (n+1)^4  + 2\sum_{1 \leq r \leq \lfloor n/2 \rfloor} (n-2r + 1)^4 \\
		& + 2\sum_{0 \leq r \neq s \leq \lfloor n/2 \rfloor} (n-2r + 1)^2 (n-2s + 1)^2.
	\end{aligned}
\end{align}
The computation for larger $k$ is conceivably involved. When $d$ is arbitrary, there is no closed-form formula for $p(n,d)$ in general and only some asymptotic approximations are known \cite{Ramanujan1918,Uspensky1920,Maroti2003, Wladimir2009}. 

As a result, we opt to analyze the commutant directly and we also assume that $T^{\mathcal{E}}_k$ is both Hermitian and positive semidefinite. This is the case for the CQA ensemble that will be discussed later, as well as various other previously studied cases \cite{HarrowTEP08,HarrowTEP09,harrow2016local, harrow2023approximate,Gross2021,Haferkamp2021}. Suppose that we successfully verify that the unit eigenspace of $T^{\mathcal{E}}_k$ and $\operatorname{Comm}_k(\mathcal{U}_\times)$ are identical. To evaluate the convergence cost of the unitary ensemble, we need to bound its second largest eigenvalue of $T^{\mathcal{E}}_k$, which we call $g(\mathcal{E},k)$. It satisfies the inequality (see Appendix \ref{sec:Definitions} as well as Ref.~\cite{harrow2016local} for the version without symmetry)
\begin{align} \label{eq:cp-infty-norm}
	c_{\mathrm{cp}}(\mathcal{E}, k) \leq N^{2k} g(\mathcal{E},k) = N^{2k} \Vert T^{\mathcal{E}}_k - T^{\mathcal{U}_\times}_k \Vert_{\infty}
\end{align} 
The spectral gap characterizes the rate at which the ensemble $\mathcal{E}$ converges to $k$-designs and, consequently, the number of times needed to sample from the ensemble to $\epsilon$-approximate a $k$-design. (By converging to $k$-designs, we mean that the ensemble can generate an $\epsilon$-approximate $k$-design for arbitrarily small $\epsilon$.) To be more precise, suppose that $g(\mathcal{E},k) \leq 1 - \delta$ with $\delta$ being any lower bound on the spectral gap. Consider a circuit consisting of $p$ steps of random walks where in each step we sample a unitary from the ensemble $\mathcal{E}$. With a careful comparison of several super-operator norms \cite{vanDam2002,Low2010,harrow2016local}, it can be shown by the inequality in \eqref{eq:cp-infty-norm} that when $p \geq \frac{1}{\delta} (2kn\log d + \log 1/\epsilon)$, this random circuit forms an $\epsilon$-approximate $k$-design.


When the quantum system obeys other symmetries or conservation laws, the Hilbert space $\mathcal{H}$ is decomposed according to various inequivalent charge sectors associated with a charge number $\lambda$ (in a general sense) and a multiplicity $m_\lambda$, namely, 
\begin{align}\label{eq:ChargeDecomposition}
	\mathcal{H} \cong \bigoplus_{\lambda } \mathbbm{1}_{m_\lambda} \otimes S^\lambda.
\end{align}
A basic example is the transversal $\text{U}(1)$ symmetry, where the Hilbert space decomposes into a direct sum {of invariant subspaces or charge sectors indexed by the Hamming weights} $Z_{\operatorname{tot}} = \sum^n_{i=1} (I + Z_i)/2$, and each inequivalent charge sector only appears once in the decomposition. Studies on $\text{U}(1)$ or $\text{SU}(2)$-symmetric designs as well as other groups such as $\mathbb{Z}_p$ can be found in Refs.~\cite{U(1)Design2023,PRXQuantum.4.040331,MarvianDesign,mitsuhashi2024Designs,mitsuhashi2024Designs2}. 


\subsection{Main results}\label{sec:main}

We now summarize the main results of this paper. In later sections, we will delineate the proof strategies with examples and numerical computations. All proof details can be found in the Appendix. The main goal of this work is to formally understand the possibility and rate of generating (exact and approximate) unitary designs with $O(1)$-local circuits in the presence of $\text{SU}(d)$ symmetry.

As a warm-up, for the simplest case of qubits, i.e., $d = 2$, we obtain the following result.
\begin{proposition} (Informal) \label{prop:DesignForQubits}
	On an $n$-qubit system, quantum circuits generated by 2-local SU($2$)-symmetric unitaries converge to unitary $k$-designs for all $k < n(n-3)/2$ when $n \geq 9$.
\end{proposition}

Just like the case without symmetry \cite{Brylinski2001,Vlasov2001,Sawicki2016} or with $\text{U}(1)$ symmetry \cite{MarvianNature,U(1)Design2023}, 2-local $\text{SU}(2)$-symmetric unitaries are sufficient for the generation of higher order designs. This conclusion is based on the extensive prior study on $\mathrm{SU}(2)$-symmetric universality in Refs.~\cite{Marin1,MarvianSU2,MarvianSUd} in conjunction with our methods from representation theory described in Section \ref{sec:ExactDesign}.

For general qudits with $d \geq 3$, 2-local unitaries cannot achieve either universality or high-order designs \cite{MarvianNature,MarvianSU2,MarvianSUd} (for the mathematical accounts of special properties of $\SU(d)$ with $d \geq 3$, see also \cite{Biedenharn1,Biedenharn2,Marin1,Marin2}). We explicitly construct a class of 4-local unitary ensembles that exactly form or converge to $\text{SU}(d)$-symmetric $k$-designs for $k$ up to at least $O(n^2)$. To be specific, we consider the $S_n$-Convolutional Quantum Alternating group (CQA) proposed in Ref.~\cite{Zheng2021SpeedingUL}. As the name indicates, the CQA is generated by alternating products of unitary time evolutions generated by (exponentials of) the following Hamiltonians:
\begin{align}\label{eq:CQAGenerators}
	H_S = \sum_{j = 1}^{n-1} (j,j+1), \quad H_{\mathrm{YJM}} = \sum_{k,l} \beta_{kl} X_k X_l,
\end{align}
where $(j,j+1)$ are transposition or SWAP operators on qudits, $\beta_{kl}$ are real-valued parameters and 
\begin{align}\label{eq:YJM}
	X_j = (1,j) + (2,j) + \cdots + (j-1, j)
\end{align}
are the so-called \emph{Young--Jucys--Murphy elements}, or \emph{YJM elements} for short \cite{Young1977,Jucys1974,Murphy1981},
a concept that is central to the Okounkov--Vershik approach \cite{Okounkov1996} to $S_n$ representation theory and underpins most results in this work. Defined with actions of the symmetric group $S_n$, they obey $\text{SU}(d)$ symmetry according to Schur--Weyl duality. Note that we will explicitly write out a transposition as $(i,j)$ when we need to emphasize the sites on which it acts. For abstract computation such as in Eq.~\eqref{eq:Tktau}, we denote a generic transposition by the symbol $\tau$. 

\begin{definition}[CQA group]\label{def:CQAGroup}
	The group CQA, henceforth denoted simply as the CQA, is a compact Lie group whose Lie algebra is generated by Hamiltonians from~\eqref{eq:CQAGenerators}.
\end{definition}

Mathematically, simply taking the unitary time evolutions of components from the Hamiltonians in \eqref{eq:CQAGenerators}, which are at most {4-local} permutations on qudits, also generates the group (for more details, see proofs in Ref.~\cite{Zheng2021SpeedingUL} as well as Appendices \ref{sec:CQA2} and \ref{sec:CQAEnsemble}). More importantly, CQA is universal on arbitrary $n$-qudit systems when ignoring relative phase factors in inequivalent $S_n$ irreps as demonstrated in Ref.~\cite{Zheng2021SpeedingUL}. Letting $\mathcal{V}_4$ be the unitary group generated by  $\text{SU}(d)$-symmetric 4-local unitaries, we have the following inclusion relationship among these groups:
\begin{align}\label{eq:RestrictedUniversal}
	S\mathcal{U}_\times \subsetneqq \mathrm{CQA} \subsetneqq \mathcal{V}_4 \subsetneqq \mathcal{U}_\times.
\end{align}
Note that the generation of $S\mathcal{U}_\times$ relies on the interplay of $H_S$ and $H_{\text{YJM}}$ in CQA.  
Along with \eqref{eq:RestrictedUniversal}, we demonstrate that CQA already enables the generation of high-order $\text{SU}(d)$-symmetric $k$-designs.
Note that the generation of $S\mathcal{U}_\times$ relies on the interplay of $H_S$ and $H_{\text{YJM}}$ while $H_S$ itself is not sufficient: it is already known from Refs.~\cite{MarvianNature,Marin1} that 2-local gates cannot generate global ones even with trivial relative phases when $d \geq 3$. Since YJM elements also provide nontrivial phases, CQA can form high order designs.

\begin{definition}[CQA random walk ensemble]\label{def:CQAEnsemble}
	The CQA ensemble $\mathcal{E}_{\mathrm{CQA}}$ is defined by a random walk, at each step of which we uniformly sample an index $j \in [n]$ and parameters $t, \beta_{kl}, \beta_{kl}' \in [0, 2\pi]$ and implement $\exp(-i \sum_{k, l} \beta_{kl} X_k X_l) \exp(-it (j,j+1) ) \exp(-i \sum_{k, l} \beta_{kl}' X_k X_l)$ on the qudits. 
\end{definition}
We will see in Section \ref{sec: cqa-construction} that this definition ensures that the induced $k$-fold moment operator $T_k^{\mathcal{E}}$ for each step of the random walk is Hermitian and in fact positive semidefinite, which facilitates the approach of comparing its unit eigenspace with the commutant of the group of  $\text{SU}(d)$-symmetric unitaries, as well as evaluating the spectral gap, as discussed in Section \ref{sec:PreliminariesSymmetricDesign}.  

We also define the following explicit 4-local  $\text{SU}(d)$-symmetric random unitary circuit model.  

\begin{definition}[CQA local random unitary circuit]\label{def:V4CQAEnsemble}
	In each step of the random walk, we uniformly sample a 4-local $\operatorname{SU}(d)$-symmetric unitary operator acting on four random locations $i_1,i_2,i_3,i_4\in[n]$. This is a random circuit model consisting of 4-local unitary gates, which we denote as \emph{$\mathcal{E}_{\mathcal{V}_4}$}. 
\end{definition}

Note that the locality can be defined with respect to different geometries (e.g.,~1D or all-to-all adjacency graphs) and boundary conditions (e.g.,~open or periodic boundary conditions). We will make these clear when needed.

Using the properties of YJM elements in $S_n$ representation theory, we first prove that $\operatorname{Comm}_k(\mathcal{E}_{\mathrm{CQA}}) = \operatorname{Comm}_k(\mathcal{U}_\times)$, under certain conditions specified in Theorem \ref{Thm:main} below, which indicates that the CQA ensemble converges to $\text{SU}(d)$-symmetric unitary $k$-designs from the perspective of commutants as discussed earlier. Then, it is clear by definition that
\begin{align}\label{eq:V4Commutant}
	\begin{aligned}
		\operatorname{Comm}_k(\mathcal{U}_\times) \subset \operatorname{Comm}_k(\mathcal{E}_{\mathcal{V}_4}) 
		& \subset \operatorname{Comm}_k(\mathcal{E}_{\mathrm{CQA}}) \\ 
		& = \operatorname{Comm}_k(\mathcal{U}_\times),
	\end{aligned}
\end{align}
which indicates that the commutant or, more precisely, the unit eigenspace of the ensemble $\mathcal{E}_{\mathcal{V}_4}$, is equal to that of $\mathcal{U}_\times$, thus ensuring its convergence to $\text{SU}(d)$-symmetric $k$-designs.

Our main results are summarized as follows. 

\begin{theorem} \label{Thm:main}
	The following statements for unitary $k$-designs with  $\text{SU}(d)$ symmetry hold:
	\begin{enumerate}
		\item For an $n$-qudit system with $n \geq 9$ and $d < n$, the group CQA, as well as $\mathcal{V}_4$, forms exact $\text{SU}(d)$-symmetric unitary $k$-designs for all $k < n(n-3)/2$. When $d \geq n$, the largest possible $k$ is precisely $2n-5$.
        
        \item Analogously, the CQA ensemble $\mathcal{E}_{\mathrm{CQA}}$ given by 4-local Hamiltonian evolutions and the 4-local unitary circuit variant $\mathcal{E}_{\mathcal{V}_4}$ converge to  $\text{SU}(d)$-symmetric unitary $k$-designs with the same bounds on $k$.
		
		\item It is impossible to find an ensemble  of (finitely many or infinite) $\text{SU}(d)$-symmetric unitaries with bounded (finite) locality that converges to an  $\text{SU}(d)$-symmetric unitary $k$-design for an arbitrarily large $k$. {Hence, any ensemble of bounded locality cannot converge to the $\mathrm{SU}(d)$-symmetric Haar measure. }
	\end{enumerate}
\end{theorem}

The conditions on $n$ and $d$ are due to the dimension of certain $S_n$ irreps arising from the direct sum in \eqref{eq:ChargeDecomposition} for $\text{SU}(d)$ symmetry. Intuitively, the larger the local dimension $d$ is, the more the inequivalent $S_n$ irreps there are and it becomes harder to achieve higher order $k$-designs. The bound $n(n-3)/2$ takes the worst case into account and hence works for all $d < n$. We also refer interested readers to Appendices \ref{sec:LR} and \ref{sec:CQA2} for more details. In Table~\ref{table:results}, we showcase the key results for the orders of $k$-designs that can be achieved with certain locality under general $\text{SU}(d)$ symmetry and make a comparison with the symmetry-free case where designs of arbitrary order can already be attained by 2-local gates due to their universality, highlighting their fundamental difference. For cases of other groups such as $\text{U}(1)$ or $\text{SU}(2)$, we refer readers to recent works \cite{U(1)Design2023,MarvianDesign,mitsuhashi2024Designs,mitsuhashi2024Designs2}.

\begin{table}[ht]
	\centering
	\begin{ruledtabular}
		\begin{tabular}{c|c|c|c}
			& No symmetry & \multicolumn{2}{c}{With $\text{SU}(d)$ Symmetry} \\
			\colrule
			Locality & arbitrary $d$ & $3 \leq d < n$ & $n \leq d$ \\
   \colrule
			2-locality & {$\infty$} &  $< 2$  &  $< 2$ \\
			4-locality & {$\infty$} & $\geq n(n-3)/2-1$ & $\leq 2n-5$ \\
			Any bounded locality & {$\infty$} & $< \infty$ & $< \infty$ \\
		\end{tabular}
	\end{ruledtabular}
	
	\caption{A comparison of the orders of unitary $k$-designs that can be achieved under different locality conditions and local dimensions in cases without symmetry and with $\SU(d)$ symmetry.}
 \label{table:results}
\end{table}

A key importance of such local circuit ensembles is that they can be used to model physical dynamics that have an associated time scale. In this context, a central problem is to understand the rate (or time) at which the models converge to certain designs (which indicate pseudorandomness, scrambling effects, etc.). In recent physics literature \cite{RobertsChaos2017,junyu2017chaos,hunter2019unitary,junyu2020chargescrambler,brian2022linear}, frame potentials (Eq.~\eqref{eq:FramePotential}) have been commonly used to establish upper bounds on the convergence rate of an ensemble to normal $k$-designs. However, due to the special decomposition of the Hilbert space under symmetry \eqref{eq:ChargeDecomposition}, computing frame potentials even for the Haar measure of $\mathcal{U}_\times$ becomes highly infeasible, as discussed in Section \ref{sec:PreliminariesSymmetricDesign}. Therefore, we resort to using the infinity norm in \eqref{eq:cp-infty-norm}, which translates the problem of bounding the convergence rate into estimating the spectral gap \cite{Oliveira2design2007a,Oliveira2design2007b,Znidaric07:2,Znidaric2008,Harrow2design2009,Brown_2010,harrow2016local,harrow2023approximate,Haferkamp2021} as also discussed above. Interestingly, our numerical computations provide evidence that for $k=2$ on qubits (convergence to $\text{SU}(2)$-symmetric 2-designs), the spectral gap of the 1D CQA ensemble scales as $\Theta(1/n^2)$, and for the all-to-all CQA ensemble it is $O(1)$. Here, 1D and all-to-all means that the SWAPs act on nearest neighbors in 1D and any two sites, respectively. These results immediately rule out the possibility of using the local gap threshold \cite{Knabe1988,Gosset2016} and martingale methods \cite{Nachtergaele1996} and their adaptations for unitary $k$-designs \cite{harrow2016local,Haferkamp2021} to bound the spectral gaps of interest here, because they only work for systems with a constant gap. A more comprehensive study on the convergence time of CQA and other symmetric local circuit ensembles is left for future work.


\section{Exact $\text{SU}(d)$-symmetric $k$-designs from local ensembles} \label{sec:ExactDesign}

Here, we explain the $\text{SU}(d)$-symmetric design formation properties of the group CQA. To this end, we intuitively decompose (the Lie algebra of) CQA into two parts, one arising from $S\mathcal{U}_\times$ (see \eqref{eq:RestrictedUniversal}), and the other based on diagonal phase matrices consisting of scalar sub-matrices on each $S^\lambda$ with a basis denoted by $\{c_j\}_{j=1}^4$ (for the reason why there are four basis elements, see Appendix \ref{sec:CQA2} ): 
\begin{align}\label{eq:TkCQA}
	\begin{aligned}
		T_k^{\mathrm{CQA}} = & \int_{S\mathcal{U}_\times} dV V^{\otimes k} \otimes \bar{V}^{\otimes k}  \\
		& \cdot \int_{\gamma_j} d\gamma (e^{-i \sum_j \gamma_j c_j})^{\otimes k} \otimes (e^{i \sum_j \gamma_j c_j} )^{\otimes k} ,
	\end{aligned}
\end{align}
where $\gamma \in [0,2\pi]^4$ is integrated over uniform distribution. In the following subsections, we first analyze the first integral inside Eq.~\eqref{eq:TkCQA}, which yields $T_k^{S\mathcal{U}_\times} = T_k^{\mathcal{U}_\times}$ for $k < n-1$. Then we investigate the second part of Eq.~\eqref{eq:TkCQA}, which integrates phases and helps to further raise $k$. These two procedures finally lead to  Theorem \ref{Thm:main}. 

\subsection{Integral of $S\mathcal{U}_\times$ and Littlewood-Richardson rule}

First, we study the moment operator $T_k^{S\mathcal{U}_\times}$. To motivate the problem, when there is no symmetry, the special unitary group $\text{SU}(N)$ trivially forms a unitary $k$-design with respect to $\text{U}(N)$ because
\begin{align}\label{eq:CommonDesign}
	\begin{aligned}
		\int_{\operatorname{SU}(N)} d\nu V^{\otimes k} \otimes \bar{V}^{\otimes k} = \int_{\operatorname{U}(N)} d\mu U^{\otimes k} \otimes \bar{U}^{\otimes k},
	\end{aligned}
\end{align}
where $\mu$ is the Haar measure on $\text{U}(N)$ and $\nu$ is the restriction to $\text{SU}(N)$. However, the relative phases, as illustrated in Eq.~\eqref{eq:Exapnsion1}, can cause problems when generalizing the aforementioned identity for $\mathcal{U}_\times$ and $S\mathcal{U}_\times$. It is only under the mild assumption $k < n-1$ that the relative phases do not matter.

\begin{theorem}\label{thm:littlewood-richardson}
	For $n \geq 5$, $d < n$, and $k < n-1$, we have that $T^{S\mathcal{U}_\times}_k = T^{\mathcal{U}_\times}_k$, i.e.,~$S\mathcal{U}_\times$ is an exact  $\text{SU}(d)$-symmetric $k$-design.
\end{theorem}

The proof utilizes classic tools from representation theory, in particular, the Littlewood--Richardson rule \cite{Fulton1997,Goodman2009}, which has found important physical applications in e.g.~particle physics \cite{Coleman:2011xi, Coleman:1985rnk,Zeilinger2017,Zeilinger2019} and, more recently, quantum information \cite{faist20,kong2022near}. To illustrate the proof idea, let us consider the case of $k = 2$. Comparing with the expansion in Eq.~\eqref{eq:Exapnsion1} of $T^{\mathcal{U}_\times}_{k=2}$ under $\text{SU}(d)$ symmetry, the integrands of $T^{S\mathcal{U}_\times}_{k=2}$ are given by
\begin{align}\label{eq:Exapnsion2}
	V_{\lambda_1, m_1} \otimes V_{\lambda_2, m_2} \otimes \overline{V}_{\mu_1, m_1'} \otimes \overline{V}_{\mu_2, m_2'}
\end{align}
and are not subject to further constraints, as there are no nontrivial phase factors. Therefore, one cannot conclude that the integrals are identical in general. Let us set aside the multiplicities for a moment and examine the following integral for different choices of irreps :
\begin{align}\label{integral}
      \int_{S\mathcal{U}_\times} dV V_{\lambda_1, m_1} \otimes V_{\lambda_2, m_2} \otimes \overline{V}_{\mu_1, m_1'} \otimes \overline{V}_{\mu_2, m_2'}.
\end{align}
When $\lambda_1 \neq \lambda_2 \neq \mu_1 \neq \mu_2$, the above four unitaries are integrated independently in their own irreps .  Namely, \eqref{integral} is given by 
\begin{align}\label{eq:Exapnsion3}
	\int dV_{\lambda_1} V_{\lambda_1} \int dV_{\lambda_2} V_{\lambda_2}  \int dV_{\mu_1} \overline{V}_{\mu_1}  \int dV_{\mu_2} \overline{V}_{\mu_2}. 
\end{align}
If $\lambda_1 = \lambda_2$ and $\mu_1 = \mu_2$, then $ V_{\lambda_1} \otimes V_{\lambda_2}$ and $\overline{V}_{\mu_1} \otimes \overline{V}_{\mu_2}$ are integrated separately:
\begin{align}\label{eq:Exapnsion4}
	\int dV_{\lambda_1} V_{\lambda_1} \otimes V_{\lambda_1} \int dV_{\mu_1} \overline{V}_{\mu_1} \otimes \overline{V}_{\mu_1}. 
\end{align}
Moreover, we have
\begin{align}\label{eq:Exapnsion5}
	\int dV_{\lambda_1} V_{\lambda_1} \otimes \overline{V}_{\lambda_1} \int dV_{\lambda_2} V_{\lambda_2} \otimes \overline{V}_{\lambda_2}
\end{align}
and there are still other ways to choose $\lambda_i$ and $\mu_i$. Recall that $\lambda_1 = \mu_1, \lambda_2 = \mu_2$ or $\lambda_1 = \mu_2, \lambda_2 = \mu_1$ are the only cases that give nonvanishing integrals in Eq.~\eqref{eq:Exapnsion1}. This is also the case for \eqref{integral} even integrating with no relative phase factors, as claimed in the previous theorem for $k = 2$. By \emph{Schur orthogonality} from group representation theory, integrals expanded from \eqref{eq:Exapnsion2} always vanish unless the integrand can be further decomposed into trivial representations. For example, the expansion in \eqref{eq:Exapnsion3} vanishes because at least one of the four irreps  is nontrivial. 
There are other cases like expansions in \eqref{eq:Exapnsion4} and \eqref{eq:Exapnsion5}, and more complicated combinations arise for general large $k$. To identify trivial representations among these cases, we apply the Littlewood--Richardson rule. Let $d_\lambda$ denote the dimension of an irrep . By basic representation theory of the special unitary group $\text{SU}(S^\lambda) \equiv \operatorname{SU}(d_\lambda)$, its irreps can be represented by Young tableaux whose rows are given by the fundamental weights (a total of $d_\lambda-1$ rows), such that a single box represents the fundamental representation  and a diagram with $d_\lambda -1$ boxes in one column represents the conjugate representation. The desired trivial representation is given by $d_\lambda$ boxes in one column. By counting the total number of boxes along with other sophisticated treatments, the Littlewood--Richardson rule unveils the types of irreps that would be obtained from the decomposition of tensor products \cite{Fulton1997,Goodman2009}. We illustrate some basic decomposition rules as follows: 
\begin{align*}
	& \ytableausetup{boxsize=1.25em,aligntableaux=center}
	\ydiagram{1} \quad \otimes \quad  \ydiagram{1} \quad = \quad \ydiagram{1,1} \quad 
	\oplus \quad  \ydiagram{2} \\
	& \ydiagram{1} \quad  \otimes \quad \begin{ytableau} \phantom{1} \\ \none[\vdots] \\ \none[\vdots] \\  \phantom{1} \end{ytableau} \quad = \quad \begin{ytableau} \phantom{1} \\ \none[\vdots] \\ \none[\vdots] \\  \phantom{1} \\ \phantom{1} \end{ytableau} \quad \oplus \quad \begin{ytableau} \phantom{1} & \phantom{1}  \\ \none[\vdots] \\ \none[\vdots] \\  \phantom{1} \end{ytableau}
\end{align*}
The expansion in \eqref{eq:Exapnsion4} vanishes in general for large $n$ because the 2-fold tensor product of either fundamental or conjugate representations cannot be trivial, as they cannot be 2-dimensional. Expansion \eqref{eq:Exapnsion5} does not vanish because trivial representations can always be found from the decomposition of the tensor product of fundamental and conjugate representations, which is consistent with Eq.~\eqref{eq:Exapnsion1} for the group $\mathcal{U}_\times$ of  $\text{SU}(d)$-symmetric unitaries. Full details can be found in Appendix \ref{sec:LR}. 

It is worth noting that the above method works for any symmetry that results in a block diagonalization of the Hilbert space as in \eqref{eq:ChargeDecomposition}. In fact, the upper bound of $n - 1$ equals the second smallest dimension of all inequivalent irreps when the local dimension $d < n$, which is a well-known fact in $S_n$ representation theory \cite{Rasala1977,Sagan01,Sellke2020}. We can replace the condition $k < n-1$ with $k < d(n)$, where $d(n)$ is the second smallest dimension of the sectors for general block decompositions. For instance, under U$(1)$ symmetry, we have polarized up-down states corresponding to two 1-dimensional inequivalent charge sectors. As a result, $d(n) = 1$ and the group of U$(1)$-symmetric unitaries with trivial relative phase can never form any U$(1)$-symmetric $k$-design. Merely applying 2-local U$(1)$-symmetric operators with nontrivial phase factors helps to alleviate the problem and raise $k$ to be at least $n$, which is verified by results in Ref.~\cite{U(1)Design2023} for high dimensional lattices.


\subsection{Structure of relative phases and YJM elements}

In this work, the use of YJM elements is of central importance in constructing $\text{SU}(d)$-symmetric $k$-designs in this work.  Roughly speaking, YJM elements are diagonal under $S_n$ irreps. It has been proved by Okounkov and Vershik \cite{Onishchik1990} that linear combinations of products of YJM elements are able to generate arbitrary diagonal matrix including phase factors. To retain the locality of unitaries acting on qudits, we take up to second-order products of YJM elements, which are at most 4-local. We conduct a more intricate treatment of the integral of phase factors in Eq.~\eqref{eq:TkCQA}, using $S_n$ character theory \cite{Ingram1950,Rasala1977,Roichman1996,Lassalle2008,Giambruno2015,Pak2020} in Appendix \ref{sec:CQA2}. In conclusion, $T^{S\mathcal{U}_\times}_k = T^{\mathcal{U}_\times}_k$ for $k < n-1$ and these additional YJM elements enable CQA, embracing $S\mathcal{U}_\times$, to form an exact $k$-design for all $k < n(n-3)/2$, which has been stated as the first main result in Theorem \ref{Thm:main}. Moreover, defining general order products of YJM elements as $P_l \equiv (\sum^n_{i=1} X_i)^l$, one can explicitly find a basis spanning the relative phases on qubits.

\begin{theorem}(Informal) \label{Thm:basis}
	The set $\{P_l\}$ with $l = 0, ..., \lfloor n/2 \rfloor$ constitutes a basis that spans the space of all $\text{SU}(2)$-symmetric relative phase matrices on an $n$-qubit system.
\end{theorem}

As a result, $\{ c_j\}$ in Eq.~\eqref{eq:TkCQA} can be expanded by $\{P_l \}$ in the case of qubits. We note that there are other bases obtained in Ref.~\cite{MarvianSU2} using products of disjoint transpositions which lead to a full characterization of $\text{SU}$(2)-symmetric Hamiltonians realizable with $r$-local $\text{SU}$(2)-symmetric unitaries:
\begin{align}
	& \hspace{-2mm} B_\ell = \frac{1}{(2/\ell)!} \hspace{-2mm} \sum_{i_1 \neq \cdots \neq i_\ell} (i_1,i_2) \cdots (i_{\ell-1},i_\ell), \\
	& \hspace{-2mm} C_\ell = \frac{1}{(2/\ell)!} \hspace{-2mm} \sum_{i_1 \neq \cdots \neq i_\ell} \hspace{-3mm} \Big[ (i_1,i_2) - \frac{1}{2}I \Big] \cdots \Big[ (i_{\ell-1},i_\ell) - \frac{1}{2}I \Big], 
\end{align}
where $\ell = 2,4,...,2\lfloor n/2 \rfloor$, $B_0 = C_0 = I$, and subtracting $\frac{1}{2}I$ in $C_\ell$ is deliberate to make the operator basis traceless and thus orthonormal (cf. Eq.~(\ref{eq:transposition})). Since they are able to span relative phases for $S_n$ irreps corresponding to two-row Young diagrams, all three bases mentioned above are equivalent in the sense that one can be linearly represented by another. 

The above theorem uncovers the correspondence between locality and relative phases in relation to achieving $\text{SU}(2)$-symmetric $k$-designs. In particular, we show that in Appendix~\ref{sec:CQAk} whether Eq.~\eqref{eq:TkCQA} converges to $\text{SU}(2)$-symmetric $k$-designs can be reduced to the so-called \emph{moment problem} in algebraic geometry. To be precise, we consider the group $\text{CQA}^{(k)}$ in Eq.~\eqref{eq:CQA-kth-order}, incorporating $k$-th order YJM elements. Then, we show in Theorem~\ref{Thm:CQAkQubit} that, on qubits where the dominance relation (Lemma~\ref{lemma: total-ordering2}) of $S_n$ irreps becomes a total ordering, $\operatorname{CQA}^{(k)}$ containing $\{ P_l\}$ up to $l=\lfloor k/2 \rfloor$ admits a unique solution to the moment problem in Eq.~\eqref{eq: sum-of-power-poly} which corresponds exactly to desired Wick contractions, analogous to Eq.~\eqref{eq:Exapnsion1}, of irrep labels. 

Moreover, we prove in Theorem \ref{Thm:CQAk} that by incorporating $k$-th order YJM elements into CQA, it also forms $\text{SU}(d)$-symmetric unitary $k$-designs on general qudits ($d > 2$). For example, let $k = 1$.  Due to the restricted universality, any $M \in \operatorname{Comm}_1(\mathrm{CQA}) \subset \operatorname{End}(\mathcal{H})$ commutes with $S\mathcal{U}_\times$. Moreover, by definition, 
\begin{align}\label{eq:k-th-YJM}
	M X_i = X_i M \Rightarrow M (X_{i_1} \cdots X_{i_r}) = (X_{i_1} \cdots X_{i_r}) M.
\end{align}
Then, by the Okounkov--Vershik theorem \cite{Onishchik1990,Tolli2009}, $M$ commutes with arbitrary diagonal matrices including relative phases. Therefore, $M \in \operatorname{Comm}_1 (\mathcal{U}_\times)$ and CQA is an exact 1-design. For $k = 2$, the actions of first- and second-order YJM elements should be reformulated (in the form of tensor product representations of Lie algebra) as
\begin{align}
	X_i \otimes I + I \otimes X_i, \quad (X_k X_l) \otimes I + I \otimes (X_k X_l).
\end{align}
We prove in Lemma \ref{lemma:CQAk} that they are sufficient to generate tensor product representations $(X_{i_1} \cdots X_{i_r}) \otimes I + I \otimes (X_{i_1} \cdots X_{i_r})$ of arbitrarily higher order YJM elements. Again by the Okounkov--Vershik approach, the tensor product representations now commute with $M \in \operatorname{Comm}_2(\operatorname{CQA}) \subset \operatorname{End}(\mathcal{H}^{\otimes 2})$, indicating  $M \in \operatorname{Comm}_2(\mathcal{U}_\times)$, as in \eqref{eq:k-th-YJM}. For $k > 2$, we employ YJM elements up to $k$-th order to generate higher order tensor product representations and follow a similar argument to reach the conclusion. 

Finally, although it is desirable to reduce the locality, we can show that it is impossible to find an ensemble $\mathcal{E}$ composed of unitaries with {bounded locality}  that converges to a unitary $k$-design under $\text{SU}(d)$ symmetry for arbitrarily large $k$. This implies that local circuit models cannot converge to Haar randomness under  $\text{SU}(d)$ symmetry. From the above discussion, it is clear that in order to achieve higher order moments of the Haar distribution under $\text{SU}(d)$ symmetry, it is necessary to incorporate more relative phase factors into the ensemble. This unavoidably requires higher-order products of YJM elements and increases the locality. One may consider alternative ways to craft diagonal phase matrices such as using $S_n$ characters or center elements. However, we show in Theorem \ref{Thm:DesignLocality} that regardless of the approach taken, the locality must scale at least as $\Omega(\log p(n,d))$, where $p(n,d)$ is the number of all inequivalent $S_n$ irreps  from an $n$-qudit system related to the famous Hardy--Ramanujan asymptotic partition formula \cite{Ramanujan1918, Uspensky1920}. This lower bound can be tightened to $2 \lfloor n/2 \rfloor = \Theta(n)$ in the most explicit case of qubits ($d = 2$) to replenish all necessary relative phases by Theorem \ref{Thm:basis}. 


\section{Convergence of CQA dynamical models to $\text{SU}(d)$-symmetric $k$-designs}\label{sec: cqa-construction}

For many physical and practical applications, explicit local circuit models are desirable, even if they may not form a group. These local circuit models may produce distributions that  approximate unitary $k$-designs arbitrarily well after a certain number of applications of the local gates. Here, two fundamental questions arise: i) whether such an ensemble exists for a certain $k$, and ii) how fast (in what circuit depth) the ensemble  converges to a $k$-design, if possible. For quantum circuits without conservation laws where 2-local unitaries are able to achieve universality, the answer to the first question is straightforward: arbitrary $k$-designs can be achieved by many different 2-local random circuit models which hold importance in various contexts, including geometrically local, brickwork, and all-to-all interaction models and so forth \cite{Znidaric2008,Brown_2010,harrow2016local,Haferkamp2021,harrow2023approximate,Haferkamp2024linear}. However, the situation for the case with symmetry remains little understood and constitutes an important but inimical open problem (see the discussions in e.g.,~Refs.~\cite{kong2022near,marvian2023nonuniversality}). In this section, we address this open problem by introducing explicit local circuit ensembles that can converge to high-order unitary $k$-designs under  $\text{SU}(d)$ symmetry and, further, studying their convergence time. Although we have not been able to fully prove the convergence time scaling, we thoroughly discuss the mathematical obstacles in generalizing several classic approaches \cite{Diaconis1988, Knabe1988,Nachtergaele1996,Znidaric2008,Brown_2010,Varju2015,harrow2016local,Meckes2019,Haferkamp2021} that have been successfully used to understand the convergence in the case without symmetry, and report numerical results that strongly suggest worse convergence time scalings.


\subsection{Convergence of CQA dynamics and circuits}

We now study the convergence of the CQA ensemble and its variants defined in Section \ref{sec: main_results}. Recall that in each step of the random walk on the quantum circuit, the \emph{CQA ensemble} $\mathcal{E}_{\mathrm{CQA}}$ is defined by (i) sampling an element from the time evolution $\exp(-i\sum_{k,l} \beta_{kl} X_k X_l)$ of second order YJM elements, (ii) sampling an element from the time evolution $\exp(-it \tau_j)$ of the SWAP $\tau_j = (j,j+1)$ with $j$ being randomly selected from $1,...,n-1$. Pragmatically, using  $\text{SU}(d)$-symmetric random 4-local unitaries also fulfills the task, while it is worth mentioning that YJM elements exhibit many nice mathematical properties essential to our proofs and we shall prove the results for CQA ensemble at first. 

A basic property of YJM elements (as well as $\tau_j$) is that they only admit integer eigenvalues \cite{Jucys1974,Murphy1981,Okounkov1996,Tolli2009}. Therefore, it suffices to take the parameters $\beta_{kl}$ and $t$ from $[0,2\pi]$ instead of from the entire $\mathbb{R}$ when considering unitary time evolutions. The $k$-fold moment operator
corresponding to {one step} of the random walk is 
\begin{align}\label{eq:symmetry-protected}
	T_k^{\mathcal{E}_{\mathrm{CQA}}} = T_k^{\mathrm{YJM}} \Big( \frac{1}{n-1} \sum_{1 \leq j \leq n-1} T_k^{\tau_j} \Big) T_k^{\mathrm{YJM}},
\end{align}
with $T_k^{\tau_j}$ and $T_k^{\mathrm{YJM}}$ being the $k$-fold moment operators twirled by the time evolutions of $\tau_j = (j,j+1)$ and $\sum_{k,l} \beta_{kl} X_k X_l$, respectively (which form compact subgroups as parameters are taken from $[0,2\pi]$ uniformly).

To show that $\mathcal{E}_{\mathrm{CQA}}$ approaches an $\text{SU}(d)$-symmetric unitary $k$-design, we need to verify that the unit eigenspace $W_{k,\mathcal{E}_{\mathrm{CQA}}}^{\lambda = 1}$ of the operator $T_k^{\mathcal{E}_{\mathrm{CQA}}}$ equals $\operatorname{Comm}_k(\mathrm{CQA})$, which has previously been shown in Section \ref{sec:ExactDesign} to be equal to $\operatorname{Comm}_k(\mathcal{U}_\times)$, i.e., the commutant of $\text{SU}(d)$-symmetric unitaries under Haar distribution. We prove this by induction. Since each term in  $T_k^{\mathcal{E}_{\mathrm{CQA}}}$ is a Hermitian projection,
\begin{align}
	T_k^{\mathcal{E}_{\mathrm{CQA}}} M = M \quad\Leftrightarrow\quad T^{\tau_j}_k M = T_k^{\mathrm{YJM}} M = M
\end{align}
for any $j$. For the base case $2$-design, this is equivalent to 
\begin{align}\label{eq:LieBrackets}
	\begin{aligned}
		& [M, \tau_j \otimes I + I \otimes \tau_j] = 0, \\
		& [M, X_k X_l \otimes I + I \otimes X_k X_l] = 0,
	\end{aligned}
\end{align}
for $k, l,j < n$. Hence, $M$ commutes with the Lie algebra generators of the two-fold tensor product representation of the group CQA and it follows that 
\begin{align}
	W_{k= 2,\mathcal{E}_{\mathrm{CQA}}}^{\lambda = 1} \subset  \operatorname{Comm}_{k = 2}(\mathrm{CQA}) \subset  \operatorname{Comm}_{k = 2}(\mathcal{U}_\times).
\end{align}
Since the reverse direction $W_{k=2,\mathcal{E}_{\mathrm{CQA}}}^{\lambda = 1} \supset \operatorname{Comm}_{k = 2}(\mathcal{U}_\times)$ trivially holds by definition, we conclude the equivalence result $W_{k=2,\mathcal{E}_{\mathrm{CQA}}}^{\lambda = 1} = \operatorname{Comm}_{k = 2}(\mathcal{U}_\times)$ for $k=2$. For larger $k$, the result is proved similarly  by verifying that the Lie brackets in \eqref{eq:LieBrackets} also vanish for higher order tensors; the proof details are left to Lemma \ref{lemma:Ensemble2}. 
For the CQA local random circuit $\mathcal{E}_{\mathcal{V}_4}$, it is immediate to check that commuting with these  $\text{SU}(d)$-symmetric 4-local unitaries implies commuting with $\tau_j$ and YJM elements from CQA, which means that $\mathcal{E}_{\mathcal{V}_4}$ converges to  $\text{SU}(d)$-symmetric unitary $k$-designs as stated in \eqref{eq:V4Commutant}. We now summarize the conclusion as follows, with detailed proofs provided in Appendix \ref{sec:CQAEnsemble}.
\begin{theorem}
	(Informal) Repeated applications of $\mathcal{E}_{\mathrm{CQA}}$ or the  $\text{SU}(d)$-symmetric 4-local random circuit $\mathcal{E}_{\mathcal{V}_4}$ converge to $\text{SU}(d)$-symmetric unitary $k$-designs  for all $k < n(n-3)/2$.
\end{theorem}

As a reminder, the bound on $k$ arises from the same reasoning discussed in the previous subsection. One may ask if simpler constructions -- in particular, circuit models involving only 2-local unitaries -- exist. It is proved in Ref.~\cite{MarvianSUd} that, under $\text{SU}(d)$ symmetry, 2-local unitaries cannot even approximate unitary 2-designs for qudits with local dimension $d \geq 3$. This can be explicitly verified by our numerical methods developed later. To achieve convergence to a 2-design, the unit eigenspace of the 2-fold moment operator of the 2-local ensemble must be identical to $\operatorname{Comm}_2(\mathcal{U}_\times)$. It is not necessary to check the entire eigenspace; e.g., we identify irreducible sectors $S^\lambda$ with $\lambda = (3, 2, 1)$ from the direct sum of the Hilbert space of 6 qutrits under $\text{SU}(3)$ symmetry (Eq.~\eqref{eq:SchurWeyl}) and explicitly observe the inconsistency between the dimension of the unit eigenspace of any 2-local ensemble and that of $\operatorname{Comm}_2(\mathcal{U}_\times)$. More counterexamples can be found on irreps  with $\lambda = \lambda^T$ where $\lambda^T$ denotes the conjugate Young diagram of $\lambda$ (see more details in Refs.~\cite{Marin1,Marin2} as well as Appendices \ref{sec:CQAEnsemble} and \ref{sec:Approximate}).

By Schur-Weyl duality, up to a global phase, a 2-local $\text{SU}(d)$-symmetric unitary can be represented by $\exp(-it \tau)$ using a certain SWAP $\tau$. Uniformly sampling the parameter $t$ thus provides a way to sample from the Haar distribution of these 2-local $\text{SU}(d)$-symmetric unitaries. For the simpler qubit case ($d = 2$) with $\text{SU}(2)$ symmetry, 2-local unitaries are known to be sufficient for generating $k$-designs (see Proposition \ref{prop:DesignForQubits}).
There are various different 2-local circuit architectures aiming to capture different types of locality, including:
\begin{itemize}
	\item(1D local circuits) In each step of the random walk, we only sample $\exp(-it \tau_j)$ for an arbitrary $j = 1,...n-1$, defining an ensemble $\mathcal{E}_{\operatorname{eSWAP}}$. If one allows $j = 1$ and applies $\tau_n = (1,n)$, the ensemble is said to admit the periodic boundary condition. 
 
    \item (Brickwork circuits) We apply $\exp(-it \tau_1) \otimes \exp(-it \tau_3) \otimes \cdots$ and then $\exp(-it \tau_2) \otimes \exp(-it \tau_4) \otimes \cdots$ alternately for the random walk. This is known as the \emph{brickwork random circuits} under $\text{SU}(2)$ symmetry. 
 
	\item (All-to-all interaction circuits) In each step of the random walk, we sample $\exp(-it(i,j))$ with arbitrary $1 \leq i,j \leq n$. This constitutes the \emph{all-to-all interaction random circuits} under $\text{SU}(2)$ symmetry. 
\end{itemize}
We subsequently study the convergence time with respect to these architectures.


\subsection{On convergence time scaling}\label{sec: cqa-convergence}

\begin{figure*}[ht]
	\centering
    \includegraphics[width=0.45\textwidth]{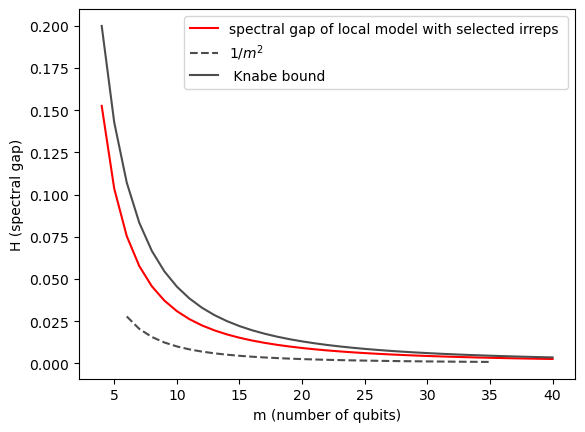}
    \includegraphics[width=0.45\textwidth]{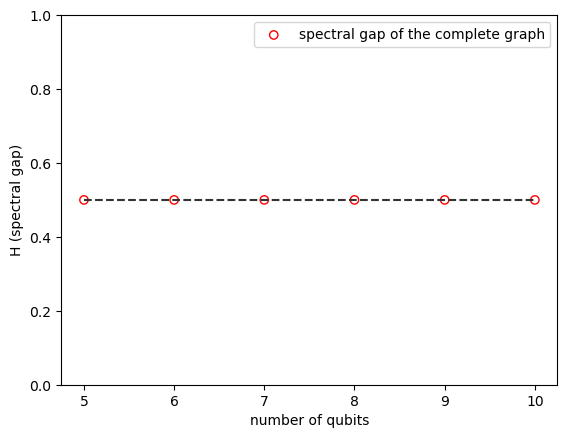}
	
	\caption{{(Left panel)} A direct numerical calculation based on Knabe's local gap method. Here $m$ is the number of qubits defined with the ``bulk Hamiltonian'' of $T_2^{\mathcal{E}_{\operatorname{eSWAP}}}$ on selected irreps  with $\lambda_1 = (m, 0)$, $\lambda_2 = (m-1, 1)$, $\lambda_3 = (m-2, 2)$ and $\lambda_4 = (m-1, 1)$. Our model exhibits spectral gaps that are strictly upper bounded by the Knabe bounds $6 /(m+1)(m+2)$ and seem to asymptotically converge to the Knabe bounds. Therefore, unlike for random circuit models without symmetry, the  Knabe method is inconclusive in determining the convergence time in the case with SU($2$) symmetry. 
    {(Right panel)} For all-to-all interaction circuit models with SU($2$) symmetry, the spectral gaps on irreps  as earlier discussed with respect to 2-designs remain constant, whereas the gap even increases without symmetry  \cite{Harrow2design2009,Znidaric2008,Haferkamp2021}. In conclusion, the theory of many-body spectral gap alone is inadequate for understanding the convergence time of random circuit models with  $\text{SU}(d)$ symmetry, indicating the need for novel approaches. }
	\label{Knabe-simulation}
\end{figure*}

As discussed earlier, we can determine the convergence time of the ensembles by bounding the infinity norm of $T_k^{\mathcal{E}_{\mathrm{CQA}}} - T_k^{\mathcal{U}_\times}$ or, equivalently, the spectral gap of the Hamiltonian $H \vcentcolon = n(I - T_k^{\mathcal{E}_{\mathrm{CQA}}} )$. In the quantum many-body theory literature,  various approaches for bounding the spectral gap of Hamiltonians including Knabe bounds \cite{Knabe1988,Gosset2016} and Nachtergaele's martingale method \cite{Nachtergaele1996,Cirac2006}  have been extensively studied. In Appendix \ref{sec:Approximate}, we establish an explicit quadratic bound on the convergence time to a 1-design, $p \geq (n-1)(2n\log d + \log (1/\epsilon))$, for $\mathcal{E}_\mathrm{CQA}$, using properties of YJM elements from representation theory. Unfortunately, although both Knabe's local gap and Nachtergaele's martingale method can be used to establish the convergence times of local random circuits to polynomial-designs for the symmetry-free case \cite{harrow2016local,Haferkamp2021}, there are fundamental obstacles to applying them to our case for $k \geq 2$ because of the decomposition of the Hilbert space \eqref{eq:ChargeDecomposition} with respect to $\text{SU}(d)$ symmetry. Intuitively speaking, the decomposition obstructs expressing a ground state of $H$ simply as a tensor product such as $\ket{\psi_{\sigma,d}}^{\otimes n}$ in Eq.~\eqref{eq:permutation}. Moreover, terms from \eqref{eq:symmetry-protected} such as $T_k^{\mathrm{YJM}} T_k^{\tau_r} T_k^{\mathrm{YJM}}$, and  $T_k^{\mathrm{YJM}} T_k^{\tau_s} T_k^{\mathrm{YJM}}$ do not commute even when $\tau_r$ and $\tau_s$ act on different qudits due to the intertwining with YJM elements.

Here, we briefly explain the limitations of these methods in our symmetric case, with a comprehensive discussion deferred to Appendices \ref{sec:Approximate} and \ref{sec:qudits}. Concurrently, we introduce our numerical methods. We first analyze the case of the 1D local random circuit $\mathcal{E}_{\operatorname{eSWAP}}$ with periodic boundary conditions defined above. Determining the second largest eigenvalue of the positive semidefinite operator 
\begin{align}\label{eq:T2SWAP}
	T_2^{\mathcal{E}_{\operatorname{eSWAP}}}  = \frac{1}{n} \sum_{1 \leq j \leq n} T_2^{\tau_j}
\end{align}
is equivalent to determining the {spectral gap} $\Delta(H)$  of 
\begin{align}\label{eq:1DHamiltonian}
	H \vcentcolon = n(I - T_2^{\mathcal{E}_{\operatorname{eSWAP}}} ) = \sum_i (I -  T_2^{(i,i+1)} ) = \sum_i P_i.
\end{align}
Based on this observation, we can potentially apply the method introduced by Knabe \cite{Knabe1988}, originally devised to estimate the spectral gap of 1D quantum spin chains with {periodic boundary conditions}. We now briefly introduce the method. Define the \emph{bulk Hamiltonian} as comprising all $P_j, \ldots, P_{m+j-1}$ terms in Eq.~\eqref{eq:1DHamiltonian}: $h_{m,j} = \sum_{i = j}^{m + j - 1} P_i$. By definition, for any $j$ and $j'$, $h_{m,j}$ and $h_{m,j'}$ are similar because $h_{m,j'}$ can be transformed from $h_{m,j}$ simply by tensor products of permutations $\sigma \in S_n$. Hence, 
\begin{align}
	\Delta(h_{m,j}) = \Delta(h_{m,j'}), \quad \forall j,j',
\end{align}
and we say that the system is \emph{permutation invariant}. An improvement of Knabe's local gap theory \cite{Gosset2016} indicates that 
\begin{align}\label{eq:KnabeBound}
	\Delta(H) \geq \frac{5(m^2 + 3m + 2)}{6(m^2 + 2m -3)} \Big( \Delta(h_{m,j}) - \frac{6}{(m+1)(m+2)} \Big).
\end{align}
To obtain a valid lower bound on the gap, we need to find a certain $m \geq 2$ such that $\Delta(h_{m,j}) > \frac{6}{(m+1)(m+2)}$. Let $m = 2$. it suffices to compute the gap of 
\begin{align}
	P_1 + P_2 = 2I - T_2^{(1,2)} - T_2^{(2,3)}.
\end{align}
Using the identity $\tau^2 = I$ for any SWAP $\tau$, and hence $e^{-it \tau} = \cos t \tau - i\sin t \tau$, we obtain
\begin{align}\label{eq:Tktau}
	& T_2^{\tau} = \frac{1}{2\pi} \int_0^{2\pi} (e^{-it \tau})^{\otimes 2} \otimes (e^{it \tau})^{\otimes 2} dt = \frac{1}{8} \Big(3 IIII + 3\tau\tau\tau\tau \notag \\
	&  + I\tau I\tau + I\tau\tau I + \tau I I\tau + \tau I\tau I - I I \tau\tau - \tau\tau I I \Big),
\end{align}
where we omit the tensor product symbols for simplicity. It is well known in $S_n$ representation theory that for any $S_n$ irrep $S^\lambda$, the matrix representation of each {adjacent transposition} $(j,j+1)$ can be explicitly read off from the so-called \emph{Young orthogonal form} \cite{Fulton1997,Tolli2009}.  With this, we can explicitly express Eq.~\eqref{eq:Tktau} in its matrix form restricted to irreps and then evaluate the gap numerically.
For example, when $n = 3$, we can compute by hand to obtain $\Delta(P_1 + P_2) = {3}/{8}$. Perhaps surprising at first glance, this result holds for arbitrary large $n$ because of the \emph{$S_n$ branching rule} \cite{Sagan01,Goodman2009}, which states that any $S_n$ irrep is a direct sum of $S_{n-1}$ irreps and so forth. In our case, the Young orthogonal forms of $(1,2), (2,3) \in S_n$ are just repetitions of those appearing in $S_3$. Therefore, the gap for $P_1 + P_2$ containing $(1,2)$ and $(2,3)$ is the same for arbitrary $n$ (for more details, refer to Appendix \ref{sec:Approximate}). Following a similar principle, we scale up our numerical computation for local bulk Hamiltonians with larger $m$. The results are plotted on the left panel of Fig.~\ref{Knabe-simulation}, according to which the local gap is below the threshold demanded in \eqref{eq:KnabeBound} and asymptotically approaches the threshold. This suggests that the local gap threshold method is not applicable to $\text{SU}(d)$-symmetric local random circuits with { periodic boundary conditions}. 
Incidentally, the martingale method \cite{Nachtergaele1996,harrow2016local} designed for 1D system with {open} boundary conditions is also ruled out because its applicability would imply a constant gap for $P_1 + \cdots + P_{m-1}$ for large $m$, which contradicts the asymptotic behavior observed numerically. 

The calculation for the all-to-all interaction circuit model is analogous to the above: we apply Eq.~\eqref{eq:Tktau} and evaluate the gap using Young orthogonal forms. To be precise, let
\begin{align}\label{eq:Complete}
		H \vcentcolon = \sum_{1 \leq i < j \leq n} (I - T_2^{(i,j)}) = \sum_{1 \leq i < j \leq n}  P_2^{(i,j)}.
\end{align}
Given any collection of $m$ qubits, let $h_{i_1,...,i_m} = \sum_{i < j \in \{i_1,...,i_m\}} P_k^{(i,j)}$ with local gap $\Delta(h_{i_1,...,i_m})$. Analogous to Knabe's original derivation of 1D local gap threshold, we have
\begin{align}\label{eq:LocalGap}
	\Delta(H) \geq 1 + \frac{n-2}{m-2}(\Delta(h_{i_1,...,i_m}) - 1).
\end{align}
If $\Delta(h_{i_1,...,i_m}) \geq 1$, $\Delta(H) \geq 1$ for all $n \geq m$. Otherwise, when $\gamma_m < 1$, which is the case according to our numerical results (see the right panel of Fig.~\ref{Knabe-simulation}), the bound would decrease to a non-positive number and thus become invalid. Based on these facts, we make the following conjecture and leave rigorous mathematical verification for future work:
\begin{conjecture}
	The spectral gap of the Hamiltonian $\sum_{i=1}^{n-1} (I -  T_k^{(i,i+1)} )$ scales as $\Theta(1/n^2)$. As a result, the 1D local random circuit model {with open boundary} converges in $\Theta(n^4 + n^3\log(1/\epsilon))$ steps to an $\epsilon$-approximate $\text{SU}(2)$-symmetric 2-design. The all-to-all interaction random circuit model converges in $\Theta(n^3 + n^2\log(1/\epsilon))$ steps to an $\epsilon$-approximate $\text{SU}(2)$-symmetric 2-design. 
\end{conjecture}

Our numerical results for  $\text{SU}(2)$-symmetric 1D local random circuits can also be used to estimate the spectral gap $\Delta(\nu^{bw},k=2)$ of the brickwork model using the so-called \emph{detectability lemma} \cite{Detectability2009,Detectability2016,harrow2016local,Haferkamp2021}, which states that
\begin{align}
	\Delta(\nu^{bw},2) \geq 1 - \frac{1}{\Delta(H) / 4+1}
\end{align}
if the Hamiltonian $H = \sum_i P_i$ is frustration-free and the $P_i$ are orthogonal projections, which holds for \eqref{eq:1DHamiltonian}. Our previous analysis thus indicates that the brickwork model converges in $O(n^3 + n^2\log(1/\epsilon))$ steps.

To summarize, both the Knabe and Nachtergaele methods are used to establish that the spectral gap of the underlying random circuit model is at least some constant without normalization. However, our numerical simulations  provide compelling evidence that the spectral gap of random circuits with SU($2$) symmetry is asymptotically subconstant without normalization. For general qudits with a larger local dimension $d > 2$, the utilization of the CQA ensemble with YJM elements explicitly violates assumptions such as the commutativity of certain bulk Hamiltonians or their ground state projections, rendering the application of these methods \cite{Nachtergaele1996,harrow2016local,Haferkamp2021} impractical due to the intricate behavior of the commutant under  $\text{SU}(d)$ symmetry, as mentioned in Section \ref{sec:PreliminariesSymmetricDesign}. The mathematical details can be found in Appendix \ref{sec:Approximate}. 


\section{Discussion}

In this paper, we have presented an in-depth study of the convergence of local quantum circuits and evolutions under $\text{SU}(d)$ symmetry. 
In particular, we have developed a systematic framework of mathematical approaches based on the CQA group, using which we have resolved the open problem of whether local circuit models can exactly form or converge to high-order $k$-designs in the presence of $\text{SU}(d)$ symmetry by explicit constructions. Our results bridge numerous important areas in mathematics and physics, including $S_n$ representation theory, $k$-designs and many-body spectral gap, and significantly sharpen the recently understood incompatibility between universality and locality in the presence of continuous symmetries \cite{MarvianNature,marvian2023nonuniversality}. More specifically, we now understand that although unbounded locality is necessary to reach arbitrarily high-order designs, merely 4-local ensembles are capable of achieving polynomial-order designs, which are sufficiently high for applications.

Moreover, this study strengthens our understanding of the fundamental discrepancy between quantum information processing with and without symmetries and conservation laws. In particular, for the key problem of analytically proving convergence time scales, we have systematically discussed how several classic methods that have been successful in cases without symmetry cease to work. Specifically, our analysis on CQA ensembles suggests that the spectral gaps with respect to the infinity norm exhibit a power-law-decaying behavior, in contrast to local circuit models without symmetry which have constant spectral gaps, leading to polynomially slower convergence. Such behaviors preclude the applicability of classic methods for analytically proving bounds on the spectral gap and convergence rate, including the Knabe's local gap and Nachtergaele's martingale method, suggesting the need for radically new analytical approaches to address this challenge. This difficulty also originates from the decomposition of irreps  (and the lack of good branching rules). We leave a more rigorous study of the gaps and convergence times as important future work.

Also of interest for future work is to extend the study to encompass different types of symmetries by incorporating additional mathematical techniques, especially $\text{U}(1)$, which is of natural physical importance. It is worth noting that the subconstant spectral gap feature and the aforementioned mathematical difficulties are expected to carry over to $\text{U}(1)$ and continuous symmetries in general, which indicate that continuous symmetries induce fundamental discrepancies in the properties of dynamics with locality. 

Furthermore, given the extensive interest in random circuit models and the importance of symmetries, our models, techniques and results are anticipated to find broad applications. As mentioned, we have explored various applications to areas including quantum information scrambling, covariant quantum error correction, and quantum machine learning in a companion paper \cite{SUd-k-Design2023Application}, and more may be found in quantum information and physics and contexts.

\emph{Note added}
We note that a recent paper \cite{Marvian3local} posted after the release of this work demonstrates that the group $\mathcal{V}_3$ of 3-local $\text{SU}(d)$-symmetric unitaries also satisfies $S\mathcal{U}_\times \subsetneqq \mathcal{V}_3$, which optimizes the necessary locality for the problems of $\text{SU}(d)$-symmetric universality addressed in \cite{Zheng2021SpeedingUL} and k-designs in this work.


\begin{acknowledgments}

We acknowledge helpful discussions with Gregory S. Bentsen, Bill Fefferman, Laimei Nie, Shengqi Sang, Sergii Strelchuk, Brian Swingle, Yunfei Wang, and Pei Zeng, among others. J.L. is supported in part by International Business Machines (IBM) Quantum through the Chicago Quantum Exchange, and the Pritzker School of Molecular Engineering at the University of Chicago through AFOSR MURI (FA9550-21-1-0209). J.L. is also a co-founder of SeQure, a startup working on AI and cryptography. L.J. acknowledges support from the ARO (W911NF-23-1-0077), ARO MURI (W911NF-21-1-0325), AFOSR MURI (FA9550-19-1-0399, FA9550-21-1-0209), NSF (OMA-1936118, ERC-1941583, OMA-2137642), NTT Research, Packard Foundation (2020-71479), and the Marshall and Arlene Bennett Family Research Program. This material is based upon work supported by the U.S. Department of Energy, Office of Science, National Quantum Information Science Research Centers. This research used resources of the Oak Ridge Leadership Computing Facility, which is a DOE Office of Science User Facility supported under Contract DE-AC05-00OR22725. Z.-W.L. is supported in part by a startup funding from YMSC, Tsinghua University, and NSFC under Grant No.~12475023.
    
\end{acknowledgments}

\bibliography{CQA.bib}

\widetext
\appendix

\begingroup
\titleformat{\section}[block]{\large\bfseries\filcenter}{}{0pt}{}
\section*{Appendix}
\endgroup

\setcounter{tocdepth}{2}

\startcontents[sections]
\titlecontents{section}[0pt]{\vspace{5mm}}{\thecontentslabel\hspace{1em}}{}{\titlerule*[1pc]{.}\contentspage}
\titlecontents{subsection}[1.5em]{}{\thecontentslabel\hspace{1em}}{}{\titlerule*[1pc]{.}\contentspage}
\printcontents[sections]{}{1}{\section*{}\vspace{-10mm}}


\section{Mathematical preliminaries}

We introduce some basic notions and facts from $S_n$ representation theory as well as our CQA model to lay the foundation for later mathematical proofs. We also refer interested readers to Refs.~\cite{Fulton1997,Sagan01,Goodman2009,Tolli2009,Zheng2021SpeedingUL} for more systematic presentations on these topics.

\subsection{Miscellaneous facts about $S_n$ representation theory}\label{sec:SnTheory}

Irreducible representations (irreps) of the symmetric group $S_n$ permuting $n$ nodes are in one-to-one correspondence with the so-called \emph{Young diagrams}. For instance, for $S_6$, the following two Young diagrams stand for the \emph{trivial} and the \emph{standard} representation, respectively:
\begin{align*}
	\ytableausetup{boxsize=1.25em} \ydiagram{6}, \qquad \ydiagram{5,1},
\end{align*}
the direct sum of which is more familiar as the six-dimensional \emph{defining} representation under which each $\sigma \in S_6$ permutes the components of vectors from $\mathbb{R}^6$. 

\begin{definition}\label{def:YoungDiagram}
	Formally, let $\lambda = (\lambda_1,\ldots,\lambda_r)$ be a collection of positive integers such that $\lambda_i \geq \lambda_{i+1}$ and $\sum_i \lambda_i = n$. Then $\lambda$ is called a \emph{partition} of the integer $n$, denoted by $\lambda \vdash n$. Obviously, $\lambda$ defines a Young diagram abstractly and the $S_n$ irrep corresponding to this Young diagram is always denoted as $S^\lambda$. The dimension of this irrep is given by the \emph{hook-length formula}:
	\begin{align}
		\dim S^\lambda = \frac{n!}{\prod_{(x,y) \in \lambda} h_{x,y} },
	\end{align}
    where $(x,y)$ specifies a box from $\lambda$ by its \emph{row} and \emph{column numbers}, and the \emph{hook-length} $h(x,y)$ counts the number of all boxes to the right of or below $(x,y)$ plus itself.
\end{definition}

Given an arbitrary $S_n$ irrep $S^\lambda$, there is a canonical way to label a basis, called the \emph{Gelfand--Tsetlin (GZ) basis} or \emph{Young--Yamanouchi basis}, on the representation space, using \emph{standard Young tableau} $T$, which are defined by filling into each box of $\lambda$ a positive integer from $1,2,\ldots,n$ in an increasing order from left to right and top to bottom. For instance, the standard representation of $S_6$ mentioned above is five-dimensional with five basis vectors labeled as
\begin{align*}
	\ytableaushort{12345, 6}, \quad  \ytableaushort{12346, 5}, \quad \ytableaushort{12356, 4} \quad \ytableaushort{12456, 3}, \quad \ytableaushort{13456, 2}.
\end{align*}
When we study the group of $\text{SU}(d)$-symmetric unitaries in the main text, the Young basis $\{\ket{\alpha_T}\}$ labeled by standard tableaux is implicitly used and a detailed treatment can be found in Appendix \ref{sec:SnCommutant}. 

\begin{definition}\label{def:YJM}
	For $1 < k \leq n$, the \textit{Young--Jucys--Murphy element}, or YJM element for short, is defined as a (formal) sum of \emph{transpositions} or SWAPs 
	\begin{align}\label{eq:A-YJM}
		X_i = (1,i) + (2,i) + \cdots + (i-1,i).
	\end{align}
	We set $X_1 = 0$ as a convention. 
\end{definition} 

The YJM element is a central concept used in our work developed by Young \cite{Young1977}, Jucys \cite{Jucys1974} and Murphy \cite{Murphy1981} and later used by Okounkov and Vershik \cite{Okounkov1996}. Under any $S_n$ representation, it may be more comprehensible to treat $X_i = (1,i) + (2,i) + \cdots + (i-1,i)$ as the sum of matrix representations of these transpositions, or we can say that the representation is extended to the \emph{group algebra} 
\begin{align}
	\mathbb{C}[S_n] = \left\{ \sum_i c_i \sigma_i; \sigma_i \in S_n \right\}
\end{align}
consisting of formal finite linear combinations of $S_n$ group elements. By the Wedderburn theorem \cite{Sagan01,Goodman2009}, $\mathbb{C}[S_n]$ is isomorphic with the direct sum of all inequivalent $S_n$ irreps $\bigoplus_\lambda \mathrm{1}_{\dim S^\lambda} \otimes (S^\lambda)$ with multiplicities equal to their dimension. It also provides a perspective for the discussion of $k$-designs through the lens of Lie groups and Lie algebras in Appendix \ref{sec:A-ExactDesign}.

Let us consider coordinate differences $x-y$ of boxes from a Young diagram $\lambda$. Given any standard tableau $T$ of $\lambda$, its \emph{content vector} is defined by rearranging them with respect to the order of boxes determined by the tableau. For instance, the content vectors of the above five standard tableaux are listed as follows:
\begin{align*}
	(0,1,2,3,4,-1), \quad (0,1,2,3,-1,4), \quad (0,1,2,-1,3,4), \quad (0,1,-1,2,3,4), \quad (0,-1,1,2,3,4).
\end{align*} 
An important feature of YJM elements is their special actions under the Young basis as revealed by content vectors: 
\begin{enumerate}
	\item They are diagonal matrices under the Young basis (even each single transposition $(i,j)$ from Eq.~\eqref{eq:A-YJM} may not be diagonal).
	
	\item The diagonal entry of $X_i$ under the Young basis vector $\ket{\alpha_T}$ corresponding to standard tableau $T$ is just the $i$-th component of the content vector.
\end{enumerate}
On the irrep $S^{(5,1)}$, 
\begin{align}\label{eq:YJMExample}
	\begin{aligned}
		& X_1 = \begin{pmatrix} 0 & 0 & 0 & 0 & 0 \\ 0 & 0 & 0 & 0 & 0 \\ 0 & 0 & 0 & 0 & 0 \\  0 & 0 & 0 & 0 & 0 \\	0 & 0 & 0 & 0 & 0 \end{pmatrix}, \quad\ \ \ 
		X_2 = \begin{pmatrix} 1 & 0 & 0 & 0 & 0 \\ 0 & 1 & 0 & 0 & 0 \\ 0 & 0 & 1 & 0 & 0 \\  0 & 0 & 0 & 1 & 0 \\	0 & 0 & 0 & 0 & -1 \end{pmatrix}, \quad 
		X_3 = \begin{pmatrix} 2 & 0 & 0 & 0 & 0 \\ 0 & 2 & 0 & 0 & 0 \\ 0 & 0 & 2 & 0 & 0 \\  0 & 0 & 0 & -1 & 0 \\	0 & 0 & 0 & 0 & 1 \end{pmatrix}, \\
		& X_4 = \begin{pmatrix} 3 & 0 & 0 & 0 & 0 \\ 0 & 3 & 0 & 0 & 0 \\ 0 & 0 & -1 & 0 & 0 \\  0 & 0 & 0 & 2 & 0 \\	0 & 0 & 0 & 0 & 2 \end{pmatrix}, \quad
		X_5 = \begin{pmatrix} 4 & 0 & 0 & 0 & 0 \\ 0 & -1 & 0 & 0 & 0 \\ 0 & 0 & 3 & 0 & 0 \\  0 & 0 & 0 & 3 & 0 \\	0 & 0 & 0 & 0 & 3 \end{pmatrix}, \quad
		X_6 = \begin{pmatrix} -1 & 0 & 0 & 0 & 0 \\ 0 & 4 & 0 & 0 & 0 \\ 0 & 0 & 4 & 0 & 0 \\  0 & 0 & 0 & 4 & 0 \\	0 & 0 & 0 & 0 & 4 \end{pmatrix}.
	\end{aligned}
\end{align}
In summary, Young basis vectors $\ket{\alpha_T}$, standard tableaux $T$ and content vectors $\alpha_T$ are in one-to-one correspondence and uniquely determine the matrix representations of YJM elements. We will introduce and apply other remarkable properties of YJM elements in Appendices \ref{sec:CQAk} and \ref{sec:Approximate} when deriving our new results.

As mentioned when introducing our numerical method in Section \ref{sec: cqa-construction}, the matrix representation of each \emph{adjacent transposition} $(i,i+1)$ can be explicitly read off in the Young basis by the \emph{Young orthogonal form}: Let $r = \alpha_T(i+1) - \alpha_T(i)$ be the \emph{axial distance} and let $(i,i+1) \cdot T$ denote the tableau defined by exchanging integers $i, i +1$ from $T$. It is easy to check that as long as $r \neq \pm 1$, $(i,i+1) \cdot T$ is still a standard Young tableau. Then
\begin{align}\label{eq:YoungOrthogonal}
	(i,i+1) \ket{\alpha_T} = \frac{1}{r} \ket{\alpha_T} + \sqrt{1 - \frac{1}{r^2}} \ket{\alpha_{(i,i+1) \cdot T}}, \quad  (i,i+1) \ket{\alpha_{(i,i+1) \cdot T}} = \sqrt{1 - \frac{1}{r^2}} \ket{\alpha_T} - \frac{1}{r} \ket{\alpha_{(i,i+1) \cdot T}}.  
\end{align} 
On the irrep $S^{(5,1)}$, 
\begin{align}
	\renewcommand\arraystretch{1.25}
	\begin{aligned}
		& (1,2) = \begin{pmatrix} 1 & 0 & 0 & 0 & 0 \\ 0 & 1 & 0 & 0 & 0 \\ 0 & 0 & 1 & 0 & 0 \\  0 & 0 & 0 & 1 & 0 \\	0 & 0 & 0 & 0 & -1 \end{pmatrix}, \quad\quad 
		(2,3) = \begin{pmatrix} 1 & 0 & 0 & 0 & 0 \\ 0 & 1 & 0 & 0 & 0 \\ 0 & 0 & 1 & 0 & 0 \\  0 & 0 & 0 & -\frac{1}{2} & \frac{\sqrt{3}}{2} \\	0 & 0 & 0 & \frac{\sqrt{3}}{2} & \frac{1}{2} \end{pmatrix}, \quad 
		(3,4) = \begin{pmatrix} 1 & 0 & 0 & 0 & 0 \\ 0 & 1 & 0 & 0 & 0 \\ 0 & 0 & -\frac{1}{3} & \frac{2\sqrt{2}}{3} & 0 \\  0 & 0 & \frac{2\sqrt{2}}{3} & \frac{1}{3} & 0 \\	0 & 0 & 0 & 0 & 1 \end{pmatrix}, \\
		& (4,5) = \begin{pmatrix} 1 & 0 & 0 & 0 & 0 \\ 0 & -\frac{1}{4} & \frac{\sqrt{15}}{4} & 0 & 0 \\ 0 & \frac{\sqrt{15}}{4} & \frac{1}{4}  & 0 & 0 \\  0 & 0 & 0 & 1 & 0 \\ 0 & 0 & 0 & 0 & 1 \end{pmatrix}, \
		(5,6) = \begin{pmatrix}  -\frac{1}{5} & \frac{2\sqrt{6}}{5} & 0 & 0 & 0 \\  \frac{2\sqrt{6}}{5} & \frac{1}{5} & 0 & 0 & 0 \\ 0 & 0 & 1 & 0 & 0 \\  0 & 0 & 0 & 1 & 0 \\	0 & 0 & 0 & 0 & 1 \end{pmatrix}.
	\end{aligned}
\end{align}
Having access to the matrix representations of all adjacent transpositions allows us to numerically calculate the matrix representing any permutation $\sigma \in S_n$. There are also classical or quantum \emph{$S_n$-fast Fourier transform} methods designed for such tasks \cite{Clausen_1993,Maslen_1998}. A detailed illustration for the usage of Young orthogonal form in our numerical computation can be found in Appendix \ref{sec:Approximate}.


\begin{definition}\label{def:CycleType}
	We say that a permutation $\sigma \in S_n$ is of \emph{cycle type} $\lambda = (\lambda_1,\ldots,\lambda_r)$ where $\lambda \vdash n$ corresponds to a partition or Young diagram, if it is decomposed into cycles of lengths $\lambda_1,\ldots,\lambda_r$. 
\end{definition}

For instance, $\sigma = (134)(56) \in S_6$ is of cycle type $\lambda = (3,2,1)$. The trivial permutation $\operatorname{id} \in S_n$ is of type $(1,\ldots,1)$. Transpositions or SWAPs $(i,j)$ are just 2-cycles, so products of transpositions such as $(i,j)(k,l) \cdots (s,t)$ are of type $(2,2,\ldots,2,1,\ldots,1)$. 

\begin{definition}\label{def:PartitionFunction}
	Let $p(n)$ denote the {number of partitions} of $n$. It equals the number of all inequivalent $S_n$ irreps, as well as the number of possible cycle types in $S_n$. Analogously, we define $p(n,d)$ as the {number of partitions of $n$ with at most $d$ parts}, i.e., the number of all Young diagrams of $n$ boxes with at most $d$ rows. 
\end{definition}

We encounter $p(n)$ and $p(n,d)$ in the main text when discussing the difficulty of computing frame potentials in the presence of $\text{SU}(d)$ symmetry. They also appear when we study the locality required to achieve arbitrary $k$-designs (Appendices \ref{sec:SnCommutant} and \ref{sec:Locality}). Due to the celebrated work of Ramanujan and Hardy \cite{Ramanujan1918} and Uspensky, \cite{Uspensky1920}, 
\begin{align}\label{eq:Ramanujan}
	p(n) \sim \frac{e^{\sqrt{\pi^2 2n /3}}}{4n\sqrt{3}}, n \to \infty.
\end{align}
There has been further study on this \cite{Rademacher1938,Erdos1942}, and various useful bounds on $p(n)$ have been found later, such as \cite{Maroti2003, Wladimir2009}
\begin{align}
	\frac{e^{2\sqrt{n}}}{an} < p(n) < e^{b\sqrt{n}}.
\end{align}
If $d = 2$, $p(n,2) = \lfloor \frac{n}{2} \rfloor + 1$. However, there are no closed-form formulas for these partition functions in general. 


\begin{proposition}
	Let $c_\mu \in \mathbb{C}[S]$ be the sum of all $\sigma \in S_n$ with cycle type $\mu$. Considering all possible Young diagrams of size $n$, the collection $\{ c_\mu \}_{\mu \vdash n}$ forms a basis for the \emph{center} $Z(\mathbb{C}[S_n])$ consisting of all elements that commute with $\mathbb{C}[S_n]$. 
\end{proposition}  

By definition, $c_\mu$ commutes with any $\sigma \in S_n$. By the Wedderburn theorem \cite{Sagan01,Goodman2009}, its matrix representation, still denoted by $c_\mu$ for simplicity, under any $S_n$ irrep $S^\lambda$ is just a scalar. As a result, the representation of $Z(\mathbb{C}[S_n])$ consists of scalar matrices within any $S_n$ irrep. Being a basis of $Z(\mathbb{C}[S_n])$ means being a basis capable of spanning all scalar matrices, called \emph{relative phase factors} when we study $k$-design with symmetry, respecting the direct sum $\bigoplus S^\lambda$ of all inequivalent $S_n$ irreps. In Appendices \ref{sec:Locality} and \ref{sec:CQAk}, the diagonal phase matrix $e^{-ic_\mu}$ helps to replenish relative phase factors for the group of $\text{SU}(d)$-symmetric unitaries. 

Besides the basis $\{ c_\mu \}$ defined above, we still have the following two kinds of center bases:

\begin{theorem}\label{thm:CenterBases}
	The following two collections also constitute bases for $Z(\mathbb{C}[S_n])$:
	\begin{enumerate}
		\item Consider the $S_n$ group character
		\begin{align}\label{eq:Character}
			\chi_\lambda (\sigma) = \operatorname{tr}_\lambda \sigma
		\end{align}
		defined by taking the trace of $\sigma \in S_n$ restricted to the irrep $S^\lambda$. Then
		\begin{align}\label{eq:OrthogonalProjection}
			\Pi_\mu \vcentcolon = \frac{\dim S^\mu}{n!} \sum_{\sigma \in S_n} \bar{\chi}_\mu(\sigma) \sigma
		\end{align}
		is a projection exclusively into the irrep $S^\mu$. The collection $\{\Pi_\mu \}$ is an orthonormal basis. 
		
		\item Consider the YJM elements $X_i$. For any $\mu = (\mu_1,\ldots,\mu_r) \vdash n$, we set
		\begin{align}
			X_\mu = \sum_{2 \leq i_1 \neq i_2 \neq \cdots \neq i_r \leq n} X_{i_1}^{\mu_1 - 1} X_{i_2}^{\mu_2 - 1} \cdots X_{i_r}^{\mu_r - 1}
		\end{align} 
		The collection $\{X_\mu\}$ is also a basis for $Z(\mathbb{C}[S_n])$ \cite{Jucys1974,Murphy1981,Tolli2009}.
	\end{enumerate}
\end{theorem}

A basis element $c_\mu$ is defined by summing over all permutations of a given cycle type $\mu$, and $\Pi_\mu$ even requires doing this over the whole symmetric group with $n!$ elements. By Theorem \ref{Thm:basis} in the main text, and also the discussion in Appendices \ref{sec:Locality} and \ref{sec:CQAk}, a more constructive way to build basis elements with a desired locality is to employ the YJM elements, which turns out to enable $2k$-local CQA to form an exact $k$-design for any constant $k$ unconditionally in $n$.


In Appendix \ref{sec:CQA2}, we also need to compute the character $\chi_\lambda (\sigma)$ explicitly for some $\sigma \in S_n$, so we briefly introduce the method here. We first present the following proposition which is crucial to the proofs in Appendix \ref{sec:Locality}:

\begin{proposition}\label{prop:Integer}
	For any $\sigma \in S_n$, $\chi_\lambda(\sigma) \in \mathbb{Z}$. That is, $S_n$ characters are integer-valued.
\end{proposition}

One can prove a variety of similar facts using Galois theory for general finite groups. For our purpose, we simply note that there is the so-called \emph{Young's natural representation} which is a \emph{non-unitary} representation of $S^\lambda$ under which each $\sigma$ is expressed as matrices with integer entries \cite{Sagan01}. As trace is invariant under matrix similarity, $\chi_\lambda (\sigma) = \operatorname{tr}_\lambda \sigma \in \mathbb{Z}$ in general. 

It is also well known that permutations $\sigma$ and $\sigma'$ with the same cycle type are conjugate to each other, so $\chi_\lambda (\sigma) = \chi_\lambda (\sigma')$ and hence we only care about the value of $\chi_\lambda$ for a given cycle type $\mu$. The so-called \emph{Frobenius character formula} \cite{Goodman2009} expresses $S_n$ character values as coefficients of a power series. The coefficients can be formally computed by, e.g., contour integrals using the residue theorem. However, closed-form formulas only exist for very few simple cases \cite{Ingram1950,Roichman1996}. For instance, the characters for 2-cycles (transpositions/SWAPs) are
\begin{align}\label{eq:CharacterValue}
	& \frac{\chi_\lambda(i,j)}{\dim S^\lambda} = \frac{2}{n(n-1)} \sum_i \Big( \binom{\lambda_i}{2} - \binom{\lambda_i'}{2} \Big) = \frac{1}{n(n-1)} \sum_i \big[ (\lambda_i - 1)(\lambda_i - 1 + i) - i(i-1) \big], 
\end{align}
where $\lambda'$ denotes the conjugate of $\lambda$, e.g.,
\begin{align*}
	\lambda = \ytableausetup{boxsize=1.25em,aligntableaux=center} \ydiagram{4,2,1} \qquad \lambda' = \ydiagram{3,2,1,1}
\end{align*}
If $\lambda_i < 2$, the corresponding binomial coefficient is set to zero.


Let us relate the above character formula to some techniques involving YJM elements. Restricted to any irrep $S^\lambda$, we can associate the following invariants: 
\begin{align}
	P_l = \left(\sum_i X_i\right)^l, l = 1,2,\ldots
\end{align}
where $\sum_i X_i$ is the summation of all YJM elements. Let us check its matrix form under the Young basis $\{\ket{\alpha_T}\}$:
\begin{align}
	\left(\sum_i X_i\right) \ket{\alpha_T} = \sum_i \alpha_T(i) \ket{\alpha_T},
\end{align}
where the $\alpha_T$ are the content vectors. Obviously, for any fixed Young diagram $\lambda$, the sum of all components of any of its content vector $\alpha_T$ is simply equal to the sum of all coordinate differences, and we denote it as $\alpha_\lambda$. Then
\begin{align}\label{eq:P_l}
	P_l \ket{\alpha_T} = (\alpha_\lambda)^l \ket{\alpha_T} 
\end{align}
for all standard tableaux or Young basis vectors of the Young diagram $\lambda$.  

Let $\operatorname{tr}_\lambda$ denote the trace within $S^\lambda$ (it is just the $S_n$ character in Eq.~\eqref{eq:Character}). When $l = 1$, we note that
\begin{align}\label{eq:P_l2}
	\alpha_\lambda = \frac{\operatorname{tr}_\lambda (P_l) }{\dim S^\lambda} = \frac{\operatorname{tr}_\lambda \sum_i X_i}{\dim S^\lambda} = \frac{n(n-1)}{2} \frac{\operatorname{tr}_\lambda (i,j)}{\dim S^\lambda} = \frac{n(n-1)}{2} \frac{\chi_\lambda(i,j)}{\dim S^\lambda},
\end{align}
which gives another way to compute the character value of 2-cycles by summing all components from the content vector. The method using YJM elements and content vectors to express general $S_n$ characters can be found in Ref.~\cite{Lassalle2008}.


\begin{definition}\label{def:Dominance}
	Given two partitions $\lambda = (\lambda_i), \mu = (\mu_i) \vdash n$. We say that $\lambda$ \emph{dominates} $\mu$, denoted by $\lambda \unrhd \mu$, if for all $j > 0$, $\sum^j_i \lambda_i \geq \sum^j_i \mu_i$. 
\end{definition}

For instance, we have
\begin{align}
	(6) \unrhd (5,1) \unrhd (4,2) \unrhd (4,1^2) 
\end{align}
where $(4,1^2)$ is the abbreviation of $(4,1,1)$. The dominance relation is not \emph{totally ordered}, e.g., we cannot compare $(4,1^2)$ and $(3,3)$. However, in the case of qubits ($d = 2$), only two-row Young diagrams need to be considered (see Appendix \ref{sec:SchurWeyl}) and partitions $\lambda = (\lambda_1, \lambda_2)$ with $\lambda_1 \geq \lambda_2$ clearly give rise to a total ordering. 

\begin{lemma}\label{lemma: total-ordering2}
	For any two unequal partitions  $\lambda, \mu \vdash n$, if $\lambda \unrhd \mu$, then $\alpha_\lambda > \alpha_\mu$.
\end{lemma}
\begin{proof}
	We prove this lemma by induction. Suppose that the statement holds for $n-1$. Given unequal $\lambda, \mu \vdash n$ with $\lambda \unrhd \mu$, there should be some $i$ such that $\lambda_i > \mu_i$, where $\lambda_i$ and $\mu_i$ are the lengths of the $i$-th rows of $\lambda$ and $\mu$, respectively. If $\lambda_i > \lambda_{i+1}$ and $\mu_i > \mu_{i+1}$, then we discard the right-hand side boxes on the $i$-th rows of $\lambda$ and $\mu$. The resultant Young diagrams, denoted by $\lambda'$ and $\mu'$, still satisfy the relation $\lambda' \unrhd \mu'$. Then, by the induction hypothesis, $\alpha_{\lambda'} > \alpha_{\mu'}$. On the other hand, the content of the discarded box from $\lambda$ is larger than that from $\mu$ by definition, hence we conclude that $\alpha_\lambda > \alpha_\mu$.
	
	Suppose that $\lambda_i = \lambda_{i+1} = \cdots = \lambda_r$ or $\mu_i = \mu_{i+1} = \cdots = \mu_s$. Then we are only allowed to discard the right-hand side boxes of $\lambda_r$ and $\mu_s$ to ensure that $\lambda'$ and $\mu'$ are well-defined Young diagrams. Even when $r \neq s$, the dominance relation still holds because $\lambda_r > \mu_s$. By the same argument as above, we complete the proof. 
\end{proof}

By Eq.~\eqref{eq:P_l2}, the above lemma says that $\frac{\chi_{\lambda}(1,2)}{\dim S^\lambda}$ is strictly increasing with respect to the dominance order of $\lambda$. Lots of counterexamples occur when this order fails to hold: e.g., $(3,3)$ and $(4,1^2)$.


Let us end this subsection with some explicit analysis on the dimension of $S_n$ irreps of two-row Young diagrams. In this circumstance, the hook-length formula in.\ref{def:YoungDiagram} can be further simplified as  \cite{Ingram1950,Sagan01}
\begin{align}\label{eq:two-row-dim}
	\dim S^\lambda = d_\lambda = \binom{n}{r} - \binom{n}{r-1} = \frac{n - 2r +1}{n - r + 1}\binom{n}{r}.
\end{align}
For binomial coefficients, we have another two useful bounds (assume that $n = 2m$) \cite{Gallier2011}:
\begin{align}
	\frac{2^n}{\sqrt{\pi(\frac{n}{2} + \frac{1}{3})}} \leq \binom{n}{n/2} \leq \frac{2^n}{\sqrt{\pi(\frac{n}{2} + \frac{1}{4})}}, \quad e^{-(\frac{n}{2} - r)^2/(n-r+1)} \leq \binom{n}{r} \Big/ \binom{n}{\frac{n}{2}} \leq e^{-(\frac{n}{2} - r)^2/(n-r)}.
\end{align}
Therefore, the ratio of $\dim S^\lambda$ to the dimension of the entire Hilbert space is
\begin{align}\label{eq:dim-Comparision}
	\frac{1}{\sqrt{\pi(\frac{n}{2} + \frac{1}{3})}} \frac{n - 2r +1}{n - r + 1}\binom{n}{r} \Big/ \binom{n}{\frac{n}{2}} \leq \frac{\dim S^\lambda}{2^n} = \frac{1}{2^n} \frac{n - 2r +1}{n - r + 1}\binom{n}{r} \leq \frac{1}{\sqrt{\pi(\frac{n}{2} + \frac{1}{4})}} \frac{n - 2r +1}{n - r + 1}\binom{n}{r} \Big/ \binom{n}{\frac{n}{2}}.
\end{align}

A lower bound for general $d$-row Young diagrams is also useful \cite{Mishchenko1996,Giambruno2015}. 
Suppose the first row $\lambda_1$ and $\lambda_1'$ of $\lambda$ and its conjugate is upper bounded by $\frac{n}{\alpha}$ for some $\alpha > 1$. Then, 
\begin{align}\label{eq:dim-Comparision2}
	\dim S^\lambda \geq \frac{\alpha^n}{n^{d(d+2)/2}}.
\end{align}
We will use these results for specific cases in Appendix \ref{sec:CQA2}.


\subsection{Schur--Weyl duality and CQA architecture}\label{sec:SchurWeyl}

We now provide a brief review on $S_n$-Convolutional Quantum Alternating Ans{\"a}tze and the group CQA proposed in Ref.~\cite{Zheng2021SpeedingUL} which can be shown to form exact unitary $k$-designs with $\text{SU}(d)$ symmetry. It also motivates the definition of CQA ensembles used in forming approximate $k$-designs.

For quantum systems, there is a discrete set of translations corresponding to permuting the qudits as well as a continuous notion of translation corresponding to spatial rotations by elements of $\text{SU}(d)$. To be precise, let $V$ be a $d$-dimensional complex Hilbert space with orthonormal basis $\{e_1,\ldots,e_d\}$. The tensor product space $V^{\otimes n}$ admits two natural representations: the \textit{tensor product representation} $\pi_{\operatorname{SU}(d)}$ of $\text{SU}(d)$, acting as
\begin{align}\label{eq:SUd}
	\pi_{\operatorname{SU}(d)}(g) (e_{i_1} \otimes \cdots \otimes e_{i_n}) \vcentcolon = g \cdot e_{i_1} \otimes \cdots \otimes g \cdot e_{i_n}, 
\end{align}
where $g \cdot e_{i_k}$ is the fundamental representation of $\text{SU}(d)$, and the \textit{permutation representation} $\pi_{S_n}$ of $S_n$ acting as
\begin{align}\label{eq:Sn}
	\pi_{S_n}(\sigma) (e_{i_1} \otimes \cdots \otimes e_{i_n}) \vcentcolon =
	e_{i_{\sigma^{-1}(1)}} \otimes \cdots \otimes e_{i_{\sigma^{-1}(n)}}.
\end{align}
We treat $\mathcal{H} = V^{\otimes n}$ as the Hilbert space of an $n$-qudit system. Schur--Weyl duality states that the action of $\text{SU}(d)$ and $S_n$ on $V^{\otimes n}$ jointly decompose the space into irreducible representations of both groups in the form 
\begin{align}\label{eq:A-SchurWeyl}
	V^{\otimes n} = \bigoplus_\lambda W_\lambda \otimes S^\lambda.
\end{align}
Again, $\lambda$ denotes a Young diagram. In this setting, it corresponds not only to a unique $S_n$ irrep $S^\lambda$, but also an $\text{SU}(d)$ irrep $W_\lambda$ \cite{Goodman2009,Tolli2009}. It should be noted that within an $n$-qudit system, only irreps corresponding to $\lambda$ ranging over {Young diagrams of size $n$ with at most $d$ rows} can be found in the decomposition. 

We denote by $\mathbbm{1}_{m_{\operatorname{SU}(d),\mu}} \cong S^\mu, \mathbbm{1}_{ m_{S_n,\lambda}} \cong W_\lambda$ the multiplicity spaces of $\text{SU}(d)$ and $S_n$ irreps, respectively. Then
\begin{align}
	& \pi_{\operatorname{SU}(d)} \cong \bigoplus_\mu W_\mu \otimes \mathbbm{1}_{m_{\operatorname{SU}(d),\mu}}, \quad  \pi_{S_n} \cong \bigoplus_\lambda \mathbbm{1}_{ m_{S_n,\lambda}} \otimes S^\lambda,
\end{align}
where $m_{\operatorname{SU}(d),\mu}=\dim S^\mu$ and $m_{S_n,\lambda} = \dim W_\lambda$.

An operator $A$ acting on the system being $\text{SU}(d)$-symmetric/invariant means that
\begin{align}
	\pi_{\operatorname{SU}(d)}(g) A = A \pi_{\operatorname{SU}(d)}(g) \text{ or } g^{\otimes n} A = A g^{\otimes n}.
\end{align}
One can check by Eqs.~\eqref{eq:SUd} and \eqref{eq:Sn} that these permutation actions clearly commute with $g^{\otimes n}$. Furthermore, Schur--Weyl duality and the double commutant theorem \cite{Goodman2009,Tolli2009} confirm that $\text{SU}(d)$-symmetric operators are exactly built from permutations in the symmetric group $S_n$. That is, they can be expressed as linear combinations, such as $\sum c_i \sigma_i$, of permutations.

Decomposing the entire space into $\text{SU}(d)$ irreps is a conventional practice in physics. Quantum states living in these subspaces are actually permutation-invariant or $S_n$-symmetric. Since our focus is on quantum circuits with $\text{SU}(d)$ symmetry, we should decompose the entire Hilbert space with respect to $S_n$ irreps (for more details, see Refs.~\cite{Tolli2009,krovi,Zheng2021SpeedingUL}). As a reminder, although the entire Hilbert space is decomposed into smaller subspaces, one should not expect that related problems including computing the ground state energy of $\text{SU}(d)$-symmetric Hamiltonian or constructing a $\text{SU}(d)$-symmetric random quantum circuit, would become easier. There are two reasons in general:
\begin{enumerate}
	\item There are various inequivalent $S_n$ irreps from the decomposition to deal with, and the total number is $p(n,d)$, which scales at most superpolynomially (see \eqref{eq:Ramanujan}) with $n$ and has no closed-form formula for evaluation. 
	
	\item Even for qubits with $d = 2$, using the hook-length formula from Definition \ref{def:YoungDiagram}, we know that (cf.~Eq.~\eqref{eq:two-row-dim})
	\begin{align}
		\dim S^{(m,m)} = \frac{(2m)!}{(m+1)!\, m!} = \frac{2^m}{m+1} \prod_{k=1}^m \frac{2k-1}{k} > \frac{2^m}{m+1} 
	\end{align}
    for the $S_n$ irrep of Young diagram $\lambda = (m,m)$ on a $2m$-qubit system. One can find other examples with exponentially large subspaces respecting the $\text{SU}(d)$ symmetry \cite{Giambruno2015}, which still cause difficulties when approaching the problem.
\end{enumerate}

We now introduce the mathematical definition of the $S_n$-CQA ansatz:

\begin{definition}\label{def:CQA}
	The $S_n$-CQA ansatz is defined as
	\begin{align}\label{eq:CQAansatz}
		\begin{aligned}
			\cdots \exp(-i \sum_{k,l} \beta_{kl} X_k X_l ) \exp(-i\gamma H_S) 
			\exp(-i \sum_{k,l} \beta'_{kl} X_k X_l) \exp(-i\gamma' H_S) \cdots,
		\end{aligned}
	\end{align}
	where $X_k X_l$ are products of YJM elements which are 4-local and still diagonal under the Young basis (see the example given in \eqref{eq:YJMExample}). The Hamiltonian $H_S$ is defined as the summation of adjacent transpositions $\sum_{i = 1}^{n-1} (i,i+1)$. 
\end{definition}

One can also set $k \leq l$ in the above definition because YJM elements are commutative with each other. Moreover, let us define the group generated by alternating exponentials from \eqref{eq:CQAansatz}:
\begin{align}
	\mathrm{CQA} = \Big\langle \exp(-i \sum_{k,l} \beta_{kl} X_k X_l), \quad \exp(-i\gamma H_S) \Big\rangle.
\end{align}
Obviously, CQA is contained in the \emph{group of $\text{SU}(d)$-symmetric unitaries}. To define this group, let $\text{U}(S^\lambda)$ denote the unitary group acting on the representation space $S^\lambda$, i.e., $\text{U}(S^\lambda) \cong \operatorname{U}(\dim S^\lambda)$. A typical element $g$ from the group of $\text{SU}(d)$-symmetric unitaries is then a collection of unitaries:
\begin{align}\label{eq:GroupElements}
	g = \bigoplus_\lambda U_\lambda^{\oplus m_{S_n, \lambda}},
\end{align}
where $U_\lambda \in \operatorname{U}(S^\lambda)$ and $\lambda$ range over all Young diagrams of size $n$ with at most $d$ rows. For simplicity, we omit the multiplicities and denote this group by  
\begin{align}
	\text{either } \bigoplus_\lambda \operatorname{U}(S^\lambda) \ \text{ or } \ \mathcal{U}_\times.
\end{align}
Equivalent copies of $S_n$ irreps pose no extra difficulties in computing $k$-fold channel/$k$-th moment operator twirled through $\text{SU}(d)$-symmetric unitaries in Appendix \ref{sec:LR}, but it is one of the obstacles when we calculate the frame potential/dimension of the commutant of $\mathcal{U}_\times$ in Appendix \ref{sec:FramePotential}.

On the other hand, by restricting the phase factors to be 1 on each $S^\lambda$, we have the special unitary group $\text{SU}(S^\lambda)$ as well as $\bigoplus_\lambda \operatorname{SU}(S^\lambda) = S\mathcal{U}_\times$ consisting of $\text{SU}(d)$-symmetric unitaries with unit determinant on each $S_n$ irrep block, i.e., unitaries with trivial relative phase factors with respect to each irrep. We also define $\mathcal{V}_4$ to be the group generated by $\text{SU}(d)$-symmetric 4-local unitaries. It is demonstrated in Ref.~\cite{Zheng2021SpeedingUL} that
\begin{align}
	S\mathcal{U}_\times \subsetneqq \mathrm{CQA} \subsetneqq \mathcal{V}_4 \subsetneqq \mathcal{U}_\times, 
\end{align}
establishing a theoretical guarantee for searching the ground state energy of the frustrated 2D Heisenberg model using the $S_n$-CQA ansatz, because relative phase factors can be ignored when we measure the expectation value in the experiment. 

With a focus on locality, 2-local unitaries are sufficient for universality as well as generating designs. After imposing the $\text{SU}(d)$ symmetry however, it has recently been shown in Refs.~\cite{Marin1,Marin2,MarvianSUd} that when $d \geq 3$, 2-local $\text{SU}(d)$-symmetric unitaries cannot even generate $S\mathcal{U}_\times$. The group CQA accomplishes the generation of $S\mathcal{U}_\times$ by incorporating 4-local $\text{SU}(d)$-symmetric unitaries. In the following appendices, we will present mathematical details for using CQA to generate unitary $\text{SU}(d)$-symmetric $k$-designs in both exact and approximate senses.


\section{Characterizing designs by commutant under group representation}\label{sec:CommutantTheory}

To study whether an ensemble forms a $k$-design, major approaches include computing the frame potential of the ensemble or analyzing the commutant algebra in the representation space. In this appendix, we illustrate these strategies in detail, establish their mathematical relationship, and finally move on to determining the commutant of the group $\mathcal{U}_\times$ of $\text{SU}(d)$-symmetric unitaries (Theorem \ref{Thm:Commutant}) and discussing the limitation of considering frame potentials in the presence of $\text{SU}(d)$ symmetry. These perspectives connect various approaches for characterizing unitary designs such as the tensor product expander \cite{HarrowTEP08,HarrowTEP09,harrow2016local,Haferkamp2021} and the frame potential \cite{RobertsChaos2017,junyu2017chaos,hunter2019unitary,junyu2020chargescrambler,brian2022linear}, based on which we derive our results on $\text{SU}(d)$-symmetric unitary designs in Appendices \ref{sec:A-ExactDesign} and \ref{sec:AppDesign}. 


\subsection{Quantum $k$-fold channel}\label{sec:k-channel}

We first provide the definition of a $k$-fold channel \cite{Dankert2026PRA,harrow2016local,hunter2019unitary} or the $k$-th moment (super-)operator associated with a compact group $G$.

\begin{definition}
	Given a compact group $G$ with Haar measure $\mu$ and a unitary representation $\rho$ on the concerned Hilbert space $\mathcal{H}$. For any operator $M \in \operatorname{End}(\mathcal{H}^{\otimes k})$, the \emph{$k$-fold channel} twirled by the Haar measure $\mu$ over $G$ acting on $M$ is given by 
	\begin{align}\label{eq: tpe}
		T_k^G(M) = \int_G d\mu(g) \rho^{\otimes k}(g) M (\rho^{\otimes k}(g))^\dagger = \int_G dU U^{\otimes k} M U^{\dagger \otimes k},
	\end{align}
	where we denote the matrix representations of group elements simply by $U$ and $V$ on the right-hand side of the above equation. Despite its integral form, as a super-operator, $T_k^G(\cdot)$ is merely a linear map acting on $\operatorname{End}(\mathcal{H}^{\otimes k})$ and can be reformulated as the \emph{$k$-th moment (super)-operator}:
	\begin{align}
		 T_k^G = \int_G U^{\otimes k } \otimes \bar{U}^{ \otimes k} dU.
	\end{align}
	Replacing $G$ by an arbitrary ensemble $\mathcal{E}$, $T_k^{\mathcal{E}}$ can be analogously defined, which provides a basis for the study of (approximate) $k$-designs.
\end{definition}

In later contexts, when we write $T_k^G$ for certain a compact group $G$, the integral is automatically understood to be carried out over the Haar measure.
Since the Haar measure is left-invariant, $T_k^G(M)$ commutes with the $k$-fold tensor product representation $\rho^{\otimes k}$ of $G$:
\begin{align}
	V^{\otimes k} T_k^G(M) V^{\dagger \otimes k} = \int_G (VU)^{\otimes k} M (UV)^{\dagger \otimes k} dU = T_k^G(M).
\end{align}
Putting it another way, $T_k^G$ projects $M$ into the \emph{commutant algebra} 
\begin{align}
	\operatorname{Comm}_k(G) \vcentcolon = \{M \in \operatorname{End}(\mathcal{H}^{\otimes k}); U^{\otimes k} M = M U^{\otimes k} \}, 
\end{align}
i.e., the subspace of all operators that commute with the tensor product representation $\rho^{\otimes k}$. The operator $T_k^G$ is surjective since if $M \in \operatorname{Comm}(M)$, by definition we have $T_k^G(M) = M$. Hence it is a \emph{projector} from $\operatorname{End}(\mathcal{H}^{\otimes k})$ onto $\operatorname{Comm}_k(G)$. Then we obtain the following identity either by the invariance of the Haar measure or by the property of projection: 
\begin{align}
	T_k^G(T_k^G(M)) = \int_G dU dV (VU)^{\otimes k} M (UV)^{\dagger \otimes k} = \int_G d(UV) (VU)^{\otimes k} M (UV)^{\dagger \otimes k} = T_k^G(M),
\end{align}
which further implies that $T_k^G$ has eigenvalues either 0 or 1.

For the common case of unitary designs without any symmetry assumptions, $G = \operatorname{U}(d^n) \equiv \operatorname{U}(N)$ with $\rho^{\otimes k}$ is given by the $k$-fold tensor products of fundamental representation of $\text{U}(N)$. Therefore, by the Schur--Weyl duality and the double commutant theorem (cf. Schur--Weyl duality on an $n$-qudit system), the commutant algebra is isomorphic to the representation of the symmetric group algebra $\mathbb{C}[S_k]$ which permutes elements from $\mathcal{H}^{\otimes k}$. In the presence of $\text{SU}(d)$ symmetry, the group of interest is replaced by $\mathcal{U}_\times$ defined in Section \ref{sec:SchurWeyl} and we denote by $T_k^{\mathcal{U}_\times}$ the  corresponding $k$-th moment operator. 

To establish the generation of $k$-designs with $S_n$-CQA, we will later analyze $T_k^{\mathrm{CQA}}$ and $T_k^{\mathcal{E}_{\mathrm{CQA}}}$ twirled by CQA and the ensemble $\mathcal{E}_{\mathrm{CQA}}$ and compare them with $T_k^{\operatorname{Haar}}$ in Appendices \ref{sec:CQA2} and \ref{sec:Approximate}, respectively.


\subsection{Approximate generation of unitary $k$-designs }\label{sec:Definitions}

The viewpoint that $T_k^G$ is a projector onto the commutant of $G$ provides a foundation for the characterization of approximate unitary $k$-designs. We adopt the following strong definition \cite{harrow2016local,harrow2023approximate} (see also e.g.,~Refs.~\cite{Dankert2026PRA,Harrow2design2009,Liu_2018,hunter2019unitary,Haferkamp2021,Gao2022,vanDam2002,Low2010} for various other definitions as well as comparison of operator norms): 

\begin{definition}\label{def:AppDesign}
	Given a compact group $G$, an ensemble of unitaries $\mathcal{E}$ is called an \emph{$\epsilon$-approximate unitary $k$-design with respect to  $G$} if the following matrix inequality holds in the sense of complete positivity (i.e., $A \leq_{\mathrm{cp}} B$ means $B - A$ is completely positive):
	\begin{align}
		(1 - \epsilon) T_k^G \leq_{\mathrm{cp}} T_k^{\mathcal{E}} \leq_{\mathrm{cp}} (1 + \epsilon) T_k^G. 
	\end{align} 
	We denote by $c_{\mathrm{cp}}(\mathcal{E}, k)$ the smallest constant $\epsilon$ achieving the above bound.
\end{definition}

\begin{remark}
	There are various other conditions for the definition of approximate $k$-designs in the literature, including the following:
	\begin{enumerate}
		\item The induced $2$-norm of the difference of $k$-th moment operators satisfies
		\begin{align}
			\Vert T_k^{\mathcal{E}} - T_k^G \Vert_{2 \to 2} \leq \epsilon.
		\end{align}
		We denote by $g(\mathcal{E}, k)$ the smallest constant $\epsilon$ achieving the above bound. Viewing super-operators $T_k^{\mathcal{E}}, T_k^G$ as ordinary operators, the induced $2$-norm is exactly the infinity norm that we have used in the main text. When the operator is Hermitian and positive semidefinite, $g(\mathcal{E}, k)$ is simply the largest eigenvalue of $T_k^{\mathcal{E}} - T_k^G$.
		
		\item The diamond norm of the difference of $k$-th moment operators satisfies
		\begin{align}
			\Vert T_k^{\mathcal{E}} - T_k^G \Vert_{\diamond} \leq \epsilon.
		\end{align}
		We denote by $c_\diamond(\mathcal{E}, k)$ the smallest constant $\epsilon$ achieving the above bound.
	\end{enumerate}
\end{remark}

\begin{lemma}
	These conditions are related by the following inequalities:
	\begin{align}
			& \frac{c_{\mathrm{cp}}(\mathcal{E}, k)}{N^{2k}} \leq c_\diamond(\mathcal{E}, k) \leq 2c_{\mathrm{cp}}(\mathcal{E},k),  \\
			&\frac{g(\mathcal{E},k)}{2N^k} \leq c_{\mathrm{cp}}(\mathcal{E}, k) \leq N^{2k}g(\mathcal{E},k), \label{ineq:cp}
			\\ 
            &\frac{g(\mathcal{E}, k)}{N^k} \leq c_\diamond(\mathcal{E}, k) \leq N^k g(\mathcal{E},k).
	\end{align}
\end{lemma}
\begin{proof}
	For the ordinary case $G = \operatorname{U}(N)$ without symmetry, we refer interested readers to Ref.~\cite{harrow2023approximate} for comprehensive proofs. We only prove the second inequality in  \eqref{ineq:cp} for the general case; the others can be simply obtained by properties of the Schatten norms and induced norms regardless of the kind of super-operators that are being considered (for more details, also see Refs.~\cite{Low2010,watrous_2018}). 
	
	Let us denote the Choi-Jamiolkowski representations of $T_k^{\mathcal{U}_\times}, T_k^{\mathcal{E}}$ by
	\begin{align}
		& J(T_k^{\mathcal{U}_\times}) = T_k^{\mathcal{U}_\times} \otimes \mathrm{id}_{\mathcal{H}^{\otimes 2k}} \big( \operatorname{vec}(I_{\mathcal{H}^{\otimes k} }) \otimes \operatorname{vec}(I_{\mathcal{H}^{\otimes k}}) \big), \\
		& J(T_k^{\mathcal{E}}) = T_k^{\mathcal{E}} \otimes \mathrm{id}_{\mathcal{H}^{\otimes 2k}} \big(\operatorname{vec}(I_{\mathcal{H}^{\otimes k} }) \otimes \operatorname{vec}(I_{\mathcal{H}^{\otimes k}}) \big),
	\end{align}
	respectively, where $\mathrm{id}_{\mathcal{H}^{\otimes 2k}}$ is the identity map acting on the operator space of $\mathcal{H}^{\otimes k}$, while $I_{\mathcal{H}^{\otimes k} }$ is the ordinary identity matrix with $\operatorname{vec}(I_{\mathcal{H}^{\otimes k} })$ being its vectorization form. It is well known that
	\begin{align}
		(1 + \epsilon)T_k^{\mathcal{U}_\times} - T_k^\mathcal{E} \geq_{\mathrm{cp}} 0 \quad \Leftrightarrow \quad
		(1 + \epsilon)J(T_k^{\mathcal{U}_\times}) - J(T_k^\mathcal{E}) \geq 0,
	\end{align}
	where the second inequality is defined in the sense of  positive semidefiniteness.
	
	Our first step is to explicitly solve the eigenpairs of $J(T_k^{\mathcal{U}_\times})$. Based on this, we study the eigenspaces of $J(T_k^{\mathcal{E}})$ to bound $c_{\mathrm{cp}}(\mathcal{E},k)$. Let 
	\begin{align}
		\mathcal{H} \cong \bigoplus_{\lambda } \mathbbm{1}_{m_\lambda} \otimes S^\lambda
	\end{align}
	denote the decomposition of the Hilbert space $\mathcal{H}$ with respect to the representation of $G$ and multiplicities. In our case, $G = \mathcal{U}_\times$ and $S^\lambda$ refers to the $S_n$ irreps. We further decompose the tensor product 
	\begin{align}\label{eq:kfoldDecomposition}
		\mathcal{H}^k \cong \Big( \bigoplus_{\lambda } \mathbbm{1}_{m_\lambda} \otimes S^\lambda \Big)^{\otimes k} \cong \bigoplus_{\stackrel{r \leq k,\lambda_1 \neq \cdots \neq \lambda_r}{d_1 + \cdots d_r = k}} \Big( \bigoplus_{\mu_{\lambda_1}} \mathbbm{1}_{m_{\mu_{\lambda_1}}} \otimes Q^{\mu_{\lambda_1}} \Big)_{d_1} \otimes \cdots \otimes \Big( \bigoplus_{\mu_{\lambda_r}} \mathbbm{1}_{m_{\mu_{\lambda_r}}} \otimes Q^{\mu_{\lambda_r}} \Big)_{d_r},
	\end{align}
	where $\big( \bigoplus_{\mu_{\lambda_i}} \mathbbm{1}_{m_{\mu_{\lambda_r}}} \otimes Q^{\mu_{\lambda_i}} \big)_{d_i}$ is obtained by Schur-Weyl duality when we decompose the $d_r$-fold tensor product of $\mathbbm{1}_{m_{\lambda_r}} \otimes S^{\lambda_r}$. As a caveat, arranging 
	\begin{align}
		\underbrace{\mathbbm{1}_{m_{\lambda_1}} \otimes S^{\lambda_1}}_{d_1 \text{ copies}} ,\ldots,\underbrace{\mathbbm{1}_{m_{\lambda_1}} \otimes S^{\lambda_1}}_{d_r \text{ copies}}
	\end{align}
	in different orders when taking tensor products yields isomorphic copies. They are all absorbed into the multiplicities $\mathbbm{1}_{m_{\mu_{\lambda_1}}},\ldots,\mathbbm{1}_{m_{\mu_{\lambda_r}}}$ above.   
	
	Taking an orthonormal basis with respect to the decomposition in \eqref{eq:kfoldDecomposition}, we consider the following maximally entangled state:
	\begin{align}
		\begin{aligned}
			\frac{1}{\sqrt{N^k}} \operatorname{vec}(I_{\mathcal{H}^{\otimes k} }) = & \frac{1}{\sqrt{N^k}} \sum_{\stackrel{r \leq k,\lambda_1 \neq \cdots \neq \lambda_r}{d_1 + \cdots d_r = k}}  \Big( \sum_{\mu_{\lambda_1}} \sqrt{m_{\mu_{\lambda_1}} \dim Q^{\mu_{\lambda_1}}}  \ket{\Phi_{m_{\mu_{\lambda_1}}}} \ket{\Phi_{Q^{\mu_{\lambda_1}}}} \Big) \otimes \\
			& \cdots \otimes \Big( \sum_{\mu_{\lambda_1}} \sqrt{m_{\mu_{\lambda_r}} \dim Q^{\mu_{\lambda_1}}}  \ket{\Phi_{m_{\mu_{\lambda_r}}}} \ket{\Phi_{Q^{\mu_{\lambda_r}}}} \Big),
		\end{aligned}
	\end{align}
	where $\ket{\Phi_{m_{\mu_{\lambda_i}}}}$ and $\ket{\Phi_{Q^{\mu_{\lambda_i}}}}$ are maximally entangled states defined on the corresponding subspaces.
	
	To compute $J(T_k^{\mathcal{U}_\times})$, we note that
	\begin{align}\label{eq:JHaar}
		\begin{aligned}
			\int_{\mathcal{U}_\times} (U^{\otimes k} \otimes I_{\mathcal{H}^{\otimes k}}) 
			\ket{\Phi_{m_{\mu_{\lambda_1}}}} \ket{\Phi_{Q^{\mu_{\lambda_1}}}} & \otimes \cdots \otimes \ket{\Phi_{m_{\mu_{\lambda_r}}}} \ket{\Phi_{Q^{\mu_{\lambda_r}}}} \\
			& \bra{\Phi_{m_{\mu'_{\lambda_1'}}}} \bra{\Phi_{Q^{\mu'_{\lambda_1'}}}} \otimes \cdots \otimes \bra{\Phi_{m_{\mu'_{\lambda_r'}}}} \bra{\Phi_{Q^{\mu'_{\lambda_r'}}}}
			U^{\dagger \otimes k} \otimes I_{\mathcal{H}^{\otimes k}})  dU
		\end{aligned}
	\end{align}
	vanishes unless $\lambda_i = \lambda_i'$ and $\mu_{\lambda_i} = \mu_{\lambda_i}'$, for all $i$, according to the Schur orthogonality, which we will introduce in Appendix \ref{sec:LR}. In that case, the integral is given by
	\begin{align}
		\frac{1}{N^k} \sum_{\stackrel{r \leq k,\lambda_1 \neq \cdots \neq \lambda_r}{d_1 + \cdots d_r = k}}  \Big( \sum_{\mu_{\lambda_1}} \ket{\Phi_{m_{\mu_{\lambda_1}}}} \bra{\Phi_{m_{\mu_{\lambda_1}}}}
		\frac{m_{\mu_{\lambda_1}}}{\dim Q^{\mu_{\lambda_1}} } I_{Q^{\mu_{\lambda_1}}}^{\otimes 2} \Big) \otimes \cdots \otimes 
		\Big( \sum_{\mu_{\lambda_r}} \ket{\Phi_{m_{\mu_{\lambda_r}}}} \bra{\Phi_{m_{\mu_{\lambda_r}}}}
		\frac{m_{\mu_{\lambda_r}}}{\dim Q^{\mu_{\lambda_r}} } I_{Q^{\mu_{\lambda_r}}}^{\otimes 2} \Big). 
	\end{align}
	The identity matrices $I_{Q^{\mu_{\lambda_i}}}$ arise from integrating over the Haar measure of $\mathcal{U}_\times$. The maximally entangled state $\ket{\Phi_{Q^{\mu_{\lambda_i}}}} \bra{\Phi_{Q^{\mu_{\lambda_i}}}}$ is thus averaged through (the decomposition of) $U^{\otimes k}$. 	
	
	This result also yields an eigenbasis for $J(T_k^{\mathcal{U}_\times})$ simply given by 
	\begin{align}\label{eq:JEigenbasis}
		\ket{\Phi_{m_{\mu_{\lambda_1}}}} \ket{v_{Q^{\mu_{\lambda_1}}}} \otimes \cdots \otimes \ket{\Phi_{m_{\mu_{\lambda_r}}}} \ket{v_{Q^{\mu_{\lambda_r}}}},
	\end{align} 
	where $\{\ket{v_{Q^{\mu_{\lambda_i}}}}\}$ is an arbitrary orthonormal basis on the irrep subspace $Q^{\mu_{\lambda_i}}$. The eigenvalues are just
	\begin{align}
		\frac{1}{N^k} \frac{m_{\mu_{\lambda_1}}}{\dim Q^{\mu_{\lambda_1}} } \cdots \frac{m_{\mu_{\lambda_r}}}{\dim Q^{\mu_{\lambda_r}} } \geq \frac{1}{N^{2k}}
	\end{align}
	because the dimension of any (tensor product)  subspace from the decomposition in \eqref{eq:kfoldDecomposition} cannot exceed $\dim \mathcal{H}^{\otimes k} = N^k$.
	
	We now study $J(T_k^{\mathcal{E}})$. The integral \eqref{eq:JHaar} over a general ensemble $\mathcal{E}$ may not be zero. Nevertheless, $J(T_k^{\mathcal{E}})$ still acts on the subspace $S$ spanned by \eqref{eq:JEigenbasis}. Suppose that
	\begin{align}
		\epsilon N^{-2k} \geq \Vert J(T_k^{\mathcal{U}_\times}) - J(T_k^{\mathcal{E}}) \Vert_\infty, 
	\end{align}
	i.e.,
	\begin{align}
		\epsilon \lambda_i\big( J(T_k^{\mathcal{U}_\times}) \big) \geq \Vert J(T_k^{\mathcal{U}_\times}) - J(T_k^{\mathcal{E}}) \Vert_\infty \geq \lambda_{\max} \big( J(T_k^{\mathcal{U}_\times}) - J(T_k^{\mathcal{E}}) \big), 
	\end{align}
	for any eigenvalue $\lambda_i\big( J(T_k^{\mathcal{U}_\times}) \big)$ restricted to the subspace $S$. By our previous argument, this implies that
	\begin{align}
		\begin{aligned}
			& \epsilon J(T_k^{\mathcal{U}_\times}) \geq J(T_k^{\mathcal{U}_\times}) - J(T_k^{\mathcal{E}})
			\quad\Leftrightarrow\quad (1 + \epsilon )J(T_k^{\mathcal{U}_\times}) - J(T_k^\mathcal{E}) \geq 0.   
		\end{aligned}
	\end{align}
	Therefore, 
	\begin{align}
		c_{\mathrm{cp}}(\mathcal{E},k) \leq N^{2k}\Vert J(T_k^{\mathcal{U}_\times}) - J(T_k^{\mathcal{E}}) \Vert_\infty 
		\leq N^{2k}\Vert J(T_k^{\mathcal{U}_\times}) - J(T_k^{\mathcal{E}}) \Vert_2
		\leq N^{2k} \Vert T_k^{\mathcal{U}_\times} - T_k^{\mathcal{E}} \Vert_{2 \to 2} = N^{2k} g(\mathcal{E}, k),
	\end{align}
	concluding the proof.
\end{proof}

\begin{lemma}\label{lemma:EnsembleHermitian}
	Recall that $T_k^{\mathcal{E}}$ is merely a linear map on $\operatorname{End}(\mathcal{H}^{\otimes k})$. With further conditions on the measure $\nu$ of $\mathcal{E}$ being specified, $T_k^{\mathcal{E}}$ satisfies the following properties:
	\begin{enumerate}
		\item If $\nu$ is left-invariant, then $T_k^{\mathcal{E}}$ is a projector onto $\operatorname{Comm}_k(\mathcal{E})$.
		
		\item If $\nu$ is invariant under inverse, i.e.,
		\begin{align}
			\int_\mathcal{E} d\nu(g) f(g^{-1}) = \int_\mathcal{E} d\nu(g) f(g)
		\end{align}
	    for any function $f$ defined on $\mathcal{E}$, then $T_k^{\mathcal{E}}$ is Hermitian.
	\end{enumerate}
	In particular, when $\mathcal{E}$ is taken as the restricted Haar measure over some compact subgroup of $G$, both of the above properties hold.
\end{lemma}    

We have discussed the first property in Appendix \ref{sec:k-channel}. The second property is also straightforward by using $U^{-1} = U^\dagger$ for unitary representation:
\begin{align}
	T_k^{\mathcal{E} \dagger} = \int_S V^{\dagger \otimes k} \bar{V}^{\dagger \otimes k} dV = \int_S (V^{-1})^{\otimes k} M (\bar{V}^{-1})^{ \otimes k} dV = T_k^\mathcal{E}. 
\end{align} 
Most ensembles encountered in the literature, such as in Refs.~\cite{Harrow2design2009,harrow2016local,harrow2023approximate,Haferkamp2021}, as well as our CQA ensemble defined in Appendix \ref{sec:CQAEnsemble}, induce Hermitian $k$-fold channels that can be diagonalized with an operator norm equal to the largest absolute value  of their eigenvalues. This fact is consistently used in our study. The notion of the frame potential applies more generally to the non-Hermitian case, which we also use in Appendix \ref{sec:Locality} to demonstrate that ensembles with a constant (bounded) locality can never even approximately generate $\text{SU}(d)$-symmetric $k$-designs with arbitrarily large $k$.

Besides, by applying the bi-invariance of the Haar measure and the Fubini theorem, which holds for well-behaved measures including  restricted Haar measures on compact subgroups, we see that $T_k^{\mathcal{E}}$ commutes with $T_k^G$: 
\begin{align}
	\begin{aligned}
		T^{\mathcal{E}}_k T^G_k(M) & = \int_\mathcal{E} \int_G V^{\otimes k}  U^{\otimes k} M U^{\dagger \otimes k} V^{\dagger \otimes k} dU dV = \int_G U^{\otimes k} M U^{\dagger \otimes k} dU = T^G_k(M) \\
		& = \int_\mathcal{E} \int_G U^{\otimes k} V^{\otimes k} M V^{\dagger \otimes k} U^{\dagger \otimes k} dU dV \\
		& = \int_G  \int_\mathcal{E} U^{\otimes k} V^{\otimes k} M V^{\dagger \otimes k} U^{\dagger \otimes k} dV dU = T^G_k T^\mathcal{E}_k(M). 
	\end{aligned}
\end{align}
By Lemma \ref{lemma:EnsembleHermitian}, $T_k^G$ is always diagonalizable. Assuming that $T_k^\mathcal{E}$ is Hermitian and hence diagonalizable, they can be {simultaneously diagonalized}. For example, 
\begin{align}\label{eq:EigenvaluesExample}
	T^G_k = \begin{pmatrix} 1 & & & & & \\ & 1 & & & & \\ & & & 0 & & \\ & & & & 0 & \\ & & & & & 0 \end{pmatrix}, \quad T^\mathcal{E}_k = \begin{pmatrix} 1 & & & & & \\ & 1 & & & & \\ & & & \lambda & & \\ & & & & \mu & \\ & & & & & \nu \end{pmatrix}.
\end{align} 
Obviously, the eigenspace corresponding to the unit eigenvalue of $T_k^G$ is exactly $\operatorname{Comm}_k(G)$ whose eigenvectors, by definition, also commute with the restricted representation on the ensemble $\mathcal{E}$. Therefore, 
\begin{align}\label{eq:AboveFact}
	\operatorname{Comm}_k(G) \subset \operatorname{Comm}_k(\mathcal{E}).
\end{align}
as instantiated in \eqref{eq:EigenvaluesExample}. It is now clear that only when $0 \leq \vert \lambda \vert, \vert \mu \vert, \vert \nu \vert < 1$, the convolution of  $T^\mathcal{E}_k$ converges to $T_k^G$ and thus forms an approximate $k$-design with respect to $G$ in the sense of Definition \ref{def:AppDesign}. Within this framework, evaluating the upper bound of the \emph{second largest absolute eigenvalue} of  $\lambda,\mu,\nu$ helps  determine the convergence speed of $\mathcal{E}$ to unitary $k$-designs. The case in which $T^\mathcal{E}_k$ is non-Hermitian can be addressed by calculating the frame potential as introduced later.

As a basic application of this method, suppose that $\mathcal{E}$ is taken as a compact subgroup of $G$, e.g, a one-parameter subgroup, equipped with the Haar measure inherited from that of $G$. Then $T^\mathcal{E}_k$ is also a projector and the eigenvalues $\lambda$, $\mu$, and $\nu$ exemplified above is either 0 or 1. Therefore, $T^\mathcal{E}_k = T^G_k$ if and only if $\operatorname{Comm}_k(\mathcal{E}) = \operatorname{Comm}_k(G)$, which further indicates the following simple but important conclusion.

\begin{fact}\label{Thm:GroupDesign}
	If the unitary ensemble $\mathcal{E}$ is given by a compact Lie subgroup of $G$ with restricted Haar measure, then $\mathcal{E}$ either forms an exact unitary $k$-design or it can never generate a unitary $k$-design in the approximate sense, meaning that it cannot generate a unitary $k$-design with arbitrary precision in terms of any measure defined in Definition \ref{def:AppDesign} or converge to a unitary $k$-design.
\end{fact}

As an immediate and insightful example, one-parameter subgroups generally do not even form an approximate $k$-design when the ambient group $G$ is of large dimension. However, ensembles consisting of various one-parameter subgroups may fulfill the task. With further conditions being specified, we verify this in Appendix \ref{sec:CQAEnsemble} for ensembles motivated by CQA. 


\subsection{Random walks on compact groups}\label{sec:RandomWalk}

We now review the relationship between unitary $k$-designs and \emph{random walks on compact groups} \cite{Diaconis1988,Applebaum2014,Meckes2019}, which would elaborate on the strategy of computing the second largest eigenvalue with more deep insights from probability and representation theory of groups. To begin with, let us define the convergence of measures over groups:

\begin{definition}
	Suppose that $\{\mu_p\}$ is a sequence of probability densities/measures over a compact group $G$. Then it \emph{converges in the weak star topology},  or simply \emph{converges weakly}, to the Haar measure $\mu$, denoted by $\mu_p \xrightarrow{w} \mu$, if for any continuous (and automatically bounded) function $f$ defined on $G$,
	\begin{align}\label{eq:MeasureConvergence}
		\lim_{p \to \infty} \int_G f(g) d\mu_p(g)  = \int_G f(g) d\mu(g), \ \text{ or equivalently, } \ \lim_{p \to \infty} \mathbb{E}_{\mu_p}(f) = \mathbb{E}_{\mu}(f) .
	\end{align}
    Note that the limit is considered independently for each single function $f$. Requiring \emph{uniform convergence} for all $f$ is somewhat too strong for continuous compact groups \cite{Diaconis1988,Meckes2019}. 
\end{definition}

In order to check whether $\mu_p \xrightarrow{w} \mu$, we can use the following Lévy continuity theorem generalized from classical Euclidean space to compact groups, which translates the convergence of expectations by \emph{Fourier transformation} to the convergence of certain operators acting on irreducible representations of $G$.  

\begin{theorem}[Lévy continuity theorem \cite{Applebaum2014}] 
	Given any irrep $\pi$ of a compact group $G$ and any density function $\nu$, the operator
	\begin{align}
		\hat{\nu}(\pi) \vcentcolon = \int_G \pi(g) d\nu(g)
	\end{align}
    acting on the representation space of $\pi$ is called the Fourier transform or characteristic function of $\nu$. The convergence of $\mu_n \to \mu$ defined above is equivalent to the convergence of matrix entries $\hat{\mu}_n(\pi)_{ij} \to \hat{\mu}(\pi)$ for all inequivalent $G$ irreps.
\end{theorem}

Let us check the Fourier transformation of the Haar measure $\mu$:
\begin{align}
	\hat{\mu}(\pi) = \int_G \pi(g) d\mu(g) = \begin{cases} 1 & \pi \text{ is the trivial representation} \\ 0 & \text{otherwise}
	\end{cases}.
\end{align}
This is due to the so-called \emph{Schur orthogonality}, which we formally introduce and use in Appendix \ref{sec:LR}. Since $\hat{\mu}_n(\pi) \equiv 1$ on the trivial representation, the operator norm 
\begin{align}
	\Big\Vert \bigoplus_\lambda \hat{\mu}_n(\pi) - \bigoplus_\lambda \hat{\mu}(\pi) \Big\Vert
\end{align}
evaluated over all inequivalent $G$ irreps or the second largest absolute value of eigenvalues of $\bigoplus_\lambda \hat{\mu}_n(\pi)$ (if it has a discrete spectrum) determines whether and how fast $\mu_n$ converges to $\mu$. 

\begin{remark}
	A \emph{random walk} on the group $G$ is simply a sequence $\{S_n\}$ of \emph{random variables} $S_n = X_1 \cdots X_n$ for which the $X_i$ are independent random variables with values in the group $G$ distributed according to the same density $\nu$. This induces a sequence of densities $\{\nu^{\ast n}\}$, with which one can examine the convergence properties via the previous theorem. 
	
	On the other hand, when studying unitary $k$-designs, we define $k$-th moment operator
	\begin{align}
		T_k^{\mathcal{E}} = \int_\mathcal{E} V^{\otimes k} \otimes \bar{V}^{\otimes k} dV
	\end{align} 
	of an unitary ensemble $\mathcal{E}$ and compare it with $T_k^{G}$. The ensemble is sampled multiple times, imitating a random walk on a quantum circuit. We also note that the tensor product $V^{\otimes k} \otimes \bar{V}^{\otimes k}$ from the integral can  in principle be further decomposed with respect to the irreps of $G$, thus we can interpret $T_k^{\mathcal{E}}$ as a \emph{truncated} Fourier transform of the measure $\nu$ prescribed in $\mathcal{E}$. 
	
	Consequently, the formation of unitary $k$-designs is weaker compared to the convergence of measures. For instance, $\text{SU}(N)$ is an exact $k$-design to $\text{U}(N)$ for arbitrary $k$, but $\text{SU}(N) \neq $ $\text{U}(N)$ so one cannot say that the Haar measure of $\text{SU}(N)$ converges to that of $\text{U}(N)$. Even when comparing the integral with respect to these measures, there are mismatches: let $\det$ denote the determinant function, then
	\begin{align}
		\int_{\operatorname{SU}(N)} \det V dV = 1 \neq  \int_{\operatorname{U}(N)} \det U dU = 0.
	\end{align}
	The formal reason is that there is no guarantee that $V^{\otimes k} \otimes \bar{V}^{\otimes k}$ encompasses all inequivalent irreps even when $k \to \infty$, e.g., the 1-dimensional representation $\det$ given by taking the determinant \cite{Fulton1997,Goodman2009}. However, unitary $k$-designs are more practical and relevant for quantum computation,  where we focus on the conjugate actions of unitaries on density matrices as $U^{\otimes k} \rho U^{\dagger \otimes k}$.
\end{remark}


\subsection{Frame potential, spectral form factors, and $k$-invariance}\label{sec:FramePotential}

We now demonstrate that, for the characterization of $k$-design properties, the perspective of defining $k$-fold channels and computing their second largest eigenvalues is closely related to the following two concepts: the \emph{spectral form factor} $R^{\mathcal{E}}_{2k}$ and the \emph{frame potential} $F^{(k)}_{\mathcal{E}}$, which are widely used in recent physics literature \cite{Gross2006,RobertsChaos2017,junyu2017chaos,hunter2019unitary,junyu2020chargescrambler,brian2022linear}. The spectral form factor is defined as
\begin{align}\label{eq: form-factor}
	R^{\mathcal{E}}_{2k} \vcentcolon = \int_{\mathcal{E}} dU |\operatorname{tr}(U)|^{2k}.
\end{align}
Using the facts that traces interchange with integrals and the identity $\operatorname{tr} (U \otimes V) = \operatorname{tr} U \operatorname{tr} V$, we obtain
\begin{align}
	R^{\mathcal{E}}_{2k} = \operatorname{tr} \int_{\mathcal{E}} U^{\otimes k } \otimes \bar{U}^{ \otimes k}  dU = \operatorname{tr} T_k^{\mathcal{E}}. 
\end{align}
In particular, when $\mathcal{E} = G$, $R^G_{2k} = \operatorname{tr} T_k^G$ measures precisely the {dimension} of the commutant $\operatorname{Comm}_k(G)$ explained previously. For the case without symmetry, let $G = U(d^n) \equiv U(N)$. It is well-known by Schur--Weyl duality that $R^{G}_{2k} = k!$ when $k < d^n = N$ \cite{Goodman2009,Tolli2009}. In the most general setting of arbitrarily large $k$, $R^{G}_{2k}$ is proved to be equal to the number of permutations having no \emph{increasing subsequence} of length greater than $d^n$ \cite{Rains1998}, which relates to the so-called \emph{increasing subsequence problem} from combinatorics \cite{Sagan01}.

The frame potential measures the 2-norm distance between a given ensemble and the Haar-random unitary: 
\begin{align}
	F^{(k)}_{\mathcal{E}} = \int_{\mathcal{E}} dU dV \Vert \operatorname{tr}(UV^\dagger) \Vert^{2k}.
\end{align}
Comparing to the spectral form factor, the frame potential is defined for more general choices of ensembles:

\begin{proposition}\label{prop: FramePotential-Commutant}
	Given an arbitrary ensemble $\mathcal{E}$,
	\begin{align}
		F^{(k)}_{\mathcal{E}} = \operatorname{tr} (T_k^{\mathcal{E} \dag} T_k^{\mathcal{E}} ),
	\end{align}
    which is simply the squared 2-norm of $T_k^{\mathcal{E}}$. If $T_k^{\mathcal{E}}$ is Hermitian, $F^{(k)}_{\mathcal{E}} = \operatorname{tr} ( (T_k^{\mathcal{E}})^2 )$. When $\mathcal{E}$ is compact subgroup, 
	\begin{align}
		F^{(k)}_{\mathcal{E}} = R^{\mathcal{E}}_{2k}.
	\end{align}
    In either cases, the frame potential is lower-bounded by $F^{(k)}_G = R^G_{2k}$.
	\begin{proof}
		Similarly to how we derived $R^{\mathcal{E}}_{2k} = \operatorname{tr} T_k^{\mathcal{E}}$  above, 
\begin{align}
			F_{\mathcal{E}}^{(k)} & = \int_{U,V \in \mathcal{E}} \vert \operatorname{tr}(U^\dagger V) \vert^{2k} dV dU = \int_{U,V \in \mathcal{E}} \operatorname{tr} ((U^\dagger V)^{\otimes k} \otimes (\overline{U^\dagger V})^{\otimes k}) dV dU \notag \\
			& = \operatorname{tr} \int_{U,V \in \mathcal{E}} (U^\dagger V)^{\otimes k} \otimes (\overline{U^\dagger V})^{\otimes k} dV dU = \operatorname{tr} \int_{U \in \mathcal{E}} \int_{V \in \mathcal{E}} \Big( U^{\dagger \otimes k} \otimes \bar{U}^{\dagger \otimes k} \Big)  \Big( V^{\otimes k} \otimes \bar{V}^{\otimes k} \Big) dV dU \\
			& = \operatorname{tr} \big( T_k^{\mathcal{E} \dag} T_k^{\mathcal{E}} \big). \notag
		\end{align}	
		When $\mathcal{E}$ is taken as a compact subgroup of $G$ with the restricted Haar measure, $T_k^{\mathcal{E}} = T_k^{\mathcal{E} \dag}$ becomes a projector by Lemma \ref{lemma:EnsembleHermitian}. Then,
	\begin{align}
			F_{\mathcal{E}}^{(k)} = \operatorname{tr} \big( (T_k^{\mathcal{E}})^2 \big) = \operatorname{tr}T^{\mathcal{E}}_k = R^{\mathcal{E}}_{2k}.
		\end{align}
			Finally, regardless of whether or not $T_k^{\mathcal{E}}$ is Hermitian, $T_k^{\mathcal{E} \dagger} T_k^{\mathcal{E}}$ is always diagonalizable with nonnegative eigenvalues. Let $W_{k,\mathcal{E}}^{\lambda = 1}$ denote its unit eigenspace, for any $M \in \operatorname{Comm}_k(G)$,  
		\begin{align}
			\begin{aligned}
				& U^{\otimes k} M = U^{\otimes k} M, U^{\dag \otimes k} M = U^{\dag \otimes k} M \\
				\implies\quad & T_k^{\mathcal{E} \dagger} T_k^{\mathcal{E}} (M) = M \\
				\implies \quad & F^{(k)}_G = R^G_{2k} = \operatorname{tr}(T_k^G) = \dim \operatorname{Comm}_k(G) \leq \dim W_{k,\mathcal{E}}^{\lambda = 1} \leq F_{\mathcal{E}}^{(k)}.
			\end{aligned}
		\end{align}
	    This concludes the proof. 
	\end{proof}
\end{proposition}

Note that the last statement, $F^{(k)}_G \leq F_{\mathcal{E}}^{(k)}$, can be verified directly using the bi-invariance of Haar measure \cite{RobertsChaos2017,junyu2017chaos,hunter2019unitary,junyu2020chargescrambler,brian2022linear} when comparing the real Haar randomness with that assigned by the ensemble $\mathcal{E}$. Our method incorporates insights from commutant theory. Its usefulness will be further demonstrated in Appendices \ref{sec:Locality} and \ref{sec:CQAEnsemble}.

We now introduce the notion of $k$-invariance \cite{junyu2020chargescrambler}, which characterizes how invariant the ensemble is under the Haar-random unitary. For a given ensemble $\mathcal{E}$, its $k$-invariance $I_{\mathcal{E}}^{(k)}$ is defined by
\begin{align}
	I_{\mathcal{E}}^{(k)} = F_{\mathcal{E}}^{(k)} - F_{\tilde{\mathcal{E}}}^{(k)},
\end{align}
where $\tilde{\mathcal{E}}$ is obtained from averaging $\mathcal{E}$ over the Haar measure
\begin{align}
	\tilde{\mathcal{E}} = \left\{\int_{G} d W\left(W U W^{\dagger}\right): U \in \mathcal{E}\right\}.
\end{align}
By employing methods similar to those used above,
\begin{align}\label{eq:k-invariance}
	F^{(k)}_{\tilde{\mathcal{E}}} & = \operatorname{tr}\left( \int_{\tilde{\mathcal{E}}} d \tilde{U} \int_{\tilde{\mathcal{E}}} d \tilde{V} \left( \tilde{U}^{\dagger \otimes k} \otimes \bar{\tilde{U}}^{\dagger \otimes k}\right) \left( V^{\otimes k} \otimes \bar{V}^{\otimes k} \right) \right) \notag \\
	& = \operatorname{tr} \left( \int_{\mathcal{E}} dU \int_{G} dW  \int_{\mathcal{E}} dV \int_{G} dX  \left((WU^\dag W^\dag)^{\otimes k} \otimes  (\overline{WU^\dag W^\dag})^{\otimes k} \right) \Big( (XVX^\dag)^{\otimes k} \otimes (\overline{XVX^\dag})^{\otimes k} \Big) \right) \\
	& = \operatorname{tr}\Big( \big( T^G_{k} T^{\mathcal{E} \dag}_k T^G_{k} \big) \big( T^G_{k} T^{\mathcal{E}}_k T^G_{k} \big) \Big) = \operatorname{tr}\Big( T^G_{k} T^{\mathcal{E} \dag}_k T^{\mathcal{E}}_k T^G_{k} \Big) = \operatorname{tr}( T^G_{k})  \notag.
\end{align}
Obviously, the $k$-invariance $I^{(k)}_{\mathcal{E}} \geq 0$. If $\mathcal{E}$ is an exact $k$-design, $I^{(k)}_{\mathcal{E}} = 0$. We call any ensemble for which $I^{(k)}_{\mathcal{E}} = 0$  $k$-invariant. 

\begin{remark}
	With the introduction of the commutant, the spectral gap of the $k$-fold channel, and the frame potential characterizations of ensembles provided above, we can now discuss their relationship. We assume that $T_k^{\mathcal{E}}$ is Hermitian with nonnegative eigenvalues, i.e., is positive semidefinite. This assumption holds for all CQA ensembles defined in Appendix \ref{sec:CQAEnsemble}. In  exotic scenarios in which $T_k^{\mathcal{E}}$ is non-Hermitian, we consider the operator $T_k^{\mathcal{E} \dagger} T_k^{\mathcal{E}}$ instead. 
	
	With this assumption, in the language of commutant theory, we evaluate the \emph{second largest eigenvalue} $\lambda$ of $T_k^{\mathcal{E}}$. It has been shown in Refs.~\cite{HarrowTEP08,HarrowTEP09,harrow2016local} that, for a random walk with $p$ steps (or a random circuit of depth $p$) to achieve an $\epsilon$-approximate $k$-design, i.e.,~achieve $\Vert (T_k^{\mathcal{E}})^p - T_k^G \Vert_{\diamond} \leq \epsilon$, the smallest $p$ needed is 
	\begin{align}
		\tilde{p} = \frac{1}{\log \frac{1}{\lambda}} \log \frac{N^{2k}}{\epsilon}.
	\end{align}
    Since for large $x$ it holds that $x \leq (x+1) \log (x+1)$, we have
    \begin{align}\label{eq:InequalitySecondEigenvalue}
    	\frac{1}{\log \frac{1}{\lambda}} = \frac{1}{\log \frac{1 - \lambda}{\lambda} + 1} \leq \frac{\frac{1 - \lambda}{\lambda} + 1}{\frac{1 - \lambda}{\lambda}} = \frac{1}{1- \lambda}\quad\implies\quad  \tilde{p} \leq  \frac{1}{1- \lambda} \log \frac{N^{2k}}{\epsilon}.
    \end{align}
    Therefore, a polynomial \emph{spectral gap} between the first and second largest eigenvalues of $T_k^\mathcal{E}$ guarantees an efficient random circuit scheme (also see Refs.~\cite{vanDam2002,Low2010}). 
    
    On the other hand, suppose that we consider frame potential $F^{(k)}_{\mathcal{E}}(p)$ for each $p$. Then, the inequality
    \begin{align}\label{eq:InequalityFramePotential}
    	\Vert (T_k^{\mathcal{E}})^p - T_k^G \Vert_{\diamond}^2 \leq N^{2k} \Big(F^{(k)}_{\mathcal{E}}(p) - F^{(k)}_G \Big)
    \end{align}
    can be applied to bound the difference under the diamond norm \cite{hunter2019unitary,junyu2020chargescrambler,brian2022linear}. By Proposition \ref{prop: FramePotential-Commutant}, the knowledge of both the second largest eigenvalue $\lambda$ and $F^{(k)}_G = \dim \operatorname{Comm}_k(G)$ is sufficient for bounding 
    \begin{align}\label{eq:FramePotential-Eigenvalue}
    	F^{(k)}_{\mathcal{E}}(p) = \operatorname{tr}( (T_k^{\mathcal{E}})^{2p}) = \sum_i \lambda_i^{2p},
    \end{align}
    where the $\lambda_i$ denote eigenvalues of $T_k^{\mathcal{E}}$. Conversely, with knowledge of $F^{(k)}_{\mathcal{E}}(p)$ for all $p \in \mathbb{N}$ one can uniquely determine these eigenvalues $\lambda_i$  through the so-called \emph{moment problem} studied in number theory and algebraic geometry \cite{schmudgen2020ten}. Even though it is generally impossible to explicitly solve $\lambda_i$  in Eq.~\eqref{eq:FramePotential-Eigenvalue}, this consideration unifies the concepts of frame potential, commutant, operator traces and eigenvalues in the context of characterizing $k$-design properties.
\end{remark}


\subsection{Commutant of the group of $\text{SU}(d)$-symmetric unitaries}\label{sec:SnCommutant}

In later sections where we prove our main results on unitary $k$-design under $\text{SU}(d)$ symmetry, we will always adopt the approach using $S_n$ representation theory and analyzing the commutant and eigenvalues, instead of computing the frame potential. At the end of this section, we explain the potential difficulty of working with the frame potential in the presence of $\text{SU}(d)$ symmetry, and describe the commutant of $\mathcal{U}_\times$ with an explicit spanning set, which is core to our computation of $T_{k=2}^{\mathrm{CQA}}(M)$ for the analysis of several related applications in Ref.~\cite{SUd-k-Design2023Application}.

To lay a basis for the proofs, we first consider the commutant algebra $\operatorname{Comm}_k(\operatorname{U}(d^n))$ for a generic $n$-qudit system $H = V^{\otimes n}$ with no symmetry. It is well-known by Schur--Weyl duality and the double commutant theorem \cite{Goodman2009,Tolli2009} that
\begin{align}
	\operatorname{Comm}_k(\operatorname{U}(d^n)) = \operatorname{span}\{ \sigma, \sigma \in S_k \},
\end{align}
where the $\sigma$ should be understood as permutations on the $k$-fold tensor product $\mathcal{H}^{\otimes k}$. To express $\sigma$ explicitly, note that any $M \in \operatorname{Comm}_k(\operatorname{U}(N))$ is just an element from $\operatorname{End}(\mathcal{H}^{\otimes k})$ which has a standard basis given by tensor products of matrix units, i.e., matrices $E_{ij}$ with unit entry in the position $(i,j)$ and zero entries elsewhere. Under the computational basis $\{ \ket{i} \}_{i=1}^N$,
\begin{align}
	E_{ij} = \ket{i} \bra{j}.
\end{align}
Then it is straightforward to check by definition that
\begin{align}\label{eq:Permutation}
	\sum_{i,j} E_{ii} \otimes E_{jj}, \quad \sum_{i,j} E_{ij} \otimes E_{ji}, \text{ and } \sum_{i,j,k,r,s} E_{is} \otimes E_{jk} \otimes E_{kr} \otimes E_{rj} \otimes E_{si}
\end{align}
correspond to the identity matrix, the transposition $(1,2)$ on the first two indices, and the permutation $(15) (234)$, which swaps the first and fifth indices while translating the second, third, and forth indices cyclically, respectively. Note that we define these operators on $\mathcal{H}^{\otimes k}$ for arbitrary $k$, and the cumbersome tensor products with identity matrix $I$ in the above expressions are omitted. A general permutation $\sigma$ of cycle type $\mu = (\mu_1,\ldots,\mu_r) \vdash k$ (Definition \ref{def:CycleType}) can be written out following this procedure: when there is a basis vector label $i$ appearing as a covariant (contravariant) index of some matrix unit from the tensor product, it should be assigned again as a contravariant (covariant) index. Besides using the computational basis, permutations can also be expanded by (generalized) Pauli matrices with nice properties, which is useful for various applications \cite{SUd-k-Design2023Application}. These expansions are all important for the study of unitary $k$-designs \cite{Oliveira2design2007a,Oliveira2design2007b,Znidaric2008,Harrow2design2009,Brown_2010,harrow2016local,RobertsChaos2017,Gross2021}.

We now discuss the more involved case of $\operatorname{Comm}_k(\mathcal{U}_\times)$. Recall that, as we study $\text{SU}(d)$-symmetric quantum circuits,  by Schur--Weyl duality, the entire Hilbert space $\mathcal{H} = V^{\otimes n}$ of qudits decomposes into irreps $S^\lambda$ of $S_n$ with multiplicities $m_{S_n,\lambda}$ (Appendix \ref{sec:SchurWeyl}):
\begin{align}
	\mathcal{H} \cong \bigoplus_\lambda \mathbbm{1}_{m_{S_n,\lambda}} \otimes S^\lambda.
\end{align}
In Appendix \ref{sec:SnTheory}, we have introduced the spanning of these irreps by the Young--Yamanouchi basis. We change from the computational basis $\{\ket{i}\}$ to the Young basis $\{\ket{\alpha_T, m}\}$ with $\alpha_T$ labeling a basis vector and $m$ recording the irrep multiplicity by Schur transform \cite{Tolli2009,krovi,Zheng2021SpeedingUL}, and redefine the matrix unit as 
\begin{align}
	E_{(\alpha_T, m), (\alpha_{T'}, m')} = \ket{\alpha_T, m} \bra{\alpha_{T'}, m'}.
\end{align}
It turns out that the commutant $\operatorname{Comm}_k(\mathcal{U}_\times)$ is spanned by ``permutations'' generalized from 
\eqref{eq:Permutation} using the Young--Yamanouchi basis. As a simple but enlightening example, we have
\begin{align}\label{eq:GeneralPermutation}
	\sum_{T_1,T_2} E_{(\alpha_{T_1}, m_1),(\alpha_{T_1}, m_1')} \otimes E_{(\alpha_{T_2}, m_2),(\alpha_{T_2}, m_2')}
\end{align}
generalized from $\sum_{i,j} E_{ii} \otimes E_{jj}$. However, \eqref{eq:GeneralPermutation} no longer represents the identity matrix because the summation is taken within two $S_n$ irreps labeled by the Young diagrams of $T_1$ and $T_2$, but not over the entire space $\mathcal{H}^{\otimes k}$. Besides, the multiplicity indices can vary arbitrarily as there is {no need} to require $m_1 = m_1'$ or $m_2 = m_2'$. We only write covariant and contravariant basis vector labels in pairs. Moreover, when $k = 2$ or $k=5$,
\begin{align}\label{eq:GeneralPermutation2}
	\sum_{T_1,T_2} & E_{(\alpha_{T_1}, m_1), (\alpha_{T_2}, m_2)} \otimes E_{(\alpha_{T_2}, m_2'),(\alpha_{T_1}, m_1')}, \\
	\sum_{T_1,T_2,T_3,T_4,T_5} & E_{(\alpha_{T_1}, m_1), (\alpha_{T_5}, m_5)} \otimes E_{(\alpha_{T_2}, m_2), (\alpha_{T_3}, m_3)} \\
	& \otimes E_{(\alpha_{T_3}, m_3'), (\alpha_{T_4}, m_4)} \otimes E_{(\alpha_{T_4}, m_4'), (\alpha_{T_2}, m_2')} \otimes E_{(\alpha_{T_5}, m_5'), (\alpha_{T_1}, m_1')}. \notag
\end{align}
(\ref{eq:GeneralPermutation}) and (\ref{eq:GeneralPermutation2}) show how to generalize examples of permutations in \eqref{eq:Permutation}. With all these preparations, we prove the following theorem.

\begin{theorem}\label{Thm:Commutant}
The commutant $\operatorname{Comm}_k(\mathcal{U}_\times)$ is spanned by the collection of all generalized permutations
\begin{align}
	\sum E_{(\alpha_{T_1}, m_1), (\alpha_{T_2}, m_2)} \otimes E_{(\alpha_{T_3}, m_3), (\alpha_{T_4}, m_4)} \otimes \cdots \otimes E_{(\alpha_{T_{2k-1}}, m_{2k-1} ), (\alpha_{T_{2k}}, m_{2k} )},
\end{align}
where the basis vector labels come in pairs and there are no restrictions on multiplicity indices such as Eqs.~\eqref{eq:GeneralPermutation} and \eqref{eq:GeneralPermutation2}.
\end{theorem}
\begin{proof}
	By definition, matrix representations $U$ of group elements $g \in \mathcal{U}_\times$ are just collections of unitary matrices acting on inequivalent $S_n$ irrep blocks with identical copies on the multiplicity spaces (Eq.~\eqref{eq:GroupElements}). So the conjugation action of $\mathcal{U}_\times$ on any simple tensor product of matrix units is given by 
	\begin{align}
		(U_1 E_{(\alpha_{T_1}, m_1), (\alpha_{T_2}, m_2)} U_2^\dagger) \otimes (U_3 E_{(\alpha_{T_3}, m_3), (\alpha_{T_4}, m_4)} U_4^\dagger) \otimes \cdots \otimes (U_{2k-1} E_{(\alpha_{T_{2k-1}}, m_{2k-1} ), (\alpha_{T_{2k}}, m_{2k} )} U_{2k}^\dagger) 
	\end{align}
	where the $U_i$ are unitaries acting on decomposed $S_n$ irreps spanned by $\{ \ket{\alpha_{T_i}} \}$. A generic element $M \in \operatorname{End}(\mathcal{H}^{\otimes k})$ is a linear combination of simple tensor products of matrix units:
	\begin{align}\label{eq:SimpleTensors}
		M = \sum_{T_i,M_i} c_{T_i,M_i} E_{(\alpha_{T_1}, m_1), (\alpha_{T_2}, m_2)} \otimes E_{(\alpha_{T_3}, m_3), (\alpha_{T_4}, m_4)} \otimes \cdots \otimes E_{(\alpha_{T_{2k-1}}, m_{2k-1}), (\alpha_{T_{2k}}, m_{2k})} ,
	\end{align}
    as they form a standard basis for $\operatorname{End}(\mathcal{H}^{\otimes k})$. Belonging to $\operatorname{Comm}_k(\mathcal{U}_\times)$ means being invariant under the conjugate action of an arbitrary $U = \rho(g)$. 
	
	We first take the $U_i$ as arbitrary diagonal phase matrices. Then the conjugate action accounts for scalar products with phase factors on each simple tensor. Since these simple tensors are linearly independent, being invariant under phase change implies such invariance for individual simple tensors from \eqref{eq:SimpleTensors}, which leads to the requirements on coupling covariant and contravariant basis vector indices. For instance, when $k = 2$, phase-change-invariant simple tensors are of the following forms:
	\begin{align}
		(U_1 E_{(\alpha_{T_1}, m_1), (\alpha_{T_1}, m_1')} U_1^\dagger) \otimes (U_2 E_{(\alpha_{T_2}, m_2), (\alpha_{T_2}, m_2')} U_2^\dagger)
	\end{align}
    or
    \begin{align}
    	(U_1 E_{(\alpha_{T_1}, m_1), (\alpha_{T_2}, m_2)} U_2^\dagger) \otimes (U_2 E_{(\alpha_{T_2}, m_2'), (\alpha_{T_1}, m_1')} U_1^\dagger).
    \end{align}
    With this example, let $U_1$ be a matrix that exchanges arbitrary rows and columns as
	\begin{align}
		U_1 = \begin{pmatrix} 1 & 0 & 0 \\ 0 & 0 & 1 \\ 0 & 1 & 0		\end{pmatrix},
	\end{align}
	while  $U_2$ is set to be the identity matrix. Conjugated by these kind of $\mathcal{U}_\times$ group elements, the basis vector label $\alpha_{T_1}$ varies arbitrarily inside the $S_n$ irrep acted on by $U_1$. Therefore, being invariant indicates that we should take summations over $T_1, T_2$, yielding  
	\begin{align}
		\sum_{T_1,T_2} E_{(\alpha_{T_1}, m_1),(\alpha_{T_1}, m_1')} \otimes E_{(\alpha_{T_2}, m_2),(\alpha_{T_2}, m_2')}
	\end{align}
	and
	\begin{align}
		\sum_{T_1,T_2} E_{(\alpha_{T_1}, m_1), (\alpha_{T_2}, m_2)} \otimes E_{(\alpha_{T_2}, m_2'),(\alpha_{T_1}, m_1')}.
	\end{align}
	However, these considerations do not affect the choices of multiplicity labels $m_i$, which is why they are assigned arbitrarily.
	
	We still need to prove that the operators $M$ spanned with the above requirements commute with all other unitaries from $\mathcal{U}_\times$. This is done by considering the Lie algebra $\mathfrak{g} = \mathfrak{L}(\mathcal{U}_\times)$ consisting of 
	\begin{align}
		E \otimes I \otimes \cdots \otimes I + I \otimes E \otimes \cdots \otimes I + \cdots I \otimes \cdots \otimes I \otimes E \in \operatorname{End}(\mathcal{H}^{\otimes k}),
	\end{align}
    where $E$ is an anti-Hermitian matrix respecting the decomposition of $\mathcal{H}^{\otimes k}$ under $\text{SU}(d)$ symmetry. We expand $E$ by matrix units and examine the commutativity. Using the example
    \begin{align}
    	M = \sum_{T_1,T_2} E_{(\alpha_{T_1}, m_1), (\alpha_{T_2}, m_2)} \otimes E_{(\alpha_{T_2}, m_2'), (\alpha_{T_1}, m_1')},
    \end{align}
    we have
    \begin{align}
    	(E_{(\alpha_T, m), (\alpha_{T'}, m')} \otimes I) M =  \sum_{T_2} E_{(\alpha_T, m), (\alpha_{T_2}, m_2)} \otimes E_{(\alpha_{T_2}, m_2'), (\alpha_{T'}, m')} = M (I \otimes E_{(\alpha_T, m), (\alpha_{T'}, m')})
    \end{align}
    by contracting the same tensor indices. Analogously, $(I \otimes E_{(\alpha_T, m), (\alpha_{T'}, m')}) M = M (E_{(\alpha_T, m), (\alpha_{T'}, m')} \otimes I)$ and hence $M$ commutes with the Lie algebra elements. The general case for arbitrary $k$ can be similarly deduced.
\end{proof}

\begin{remark}
    With this theorem, let us try to compute the frame potential $F^{(k)}_{\mathcal{U}_\times}$ for $G = \mathcal{U}_\times$. The simplest case is $k = 1$, where $\operatorname{Comm}_k( \mathcal{U}_\times)$ is spanned by
	\begin{align}
		 \sum_T E_{(\alpha_T, m), (\alpha_T, m')}
	\end{align}
    for each $S_n$ irrep appearing in the decomposition and with arbitrary copies. For $n$-qubit systems, Schur--Weyl duality indicates that there are $p(n,2) = \lfloor \frac{n}{2} \rfloor + 1$ inequivalent $S_n$ irreps corresponding to two-row Young diagrams in the decomposition. Moreover, the multiplicity of the irrep $S^\lambda$ with $\lambda = (n-r,r)$ is $n-2r + 1$. Therefore, for qubits,
    \begin{align}
    	F^{(1)}_{\mathcal{U}_\times} = \dim \operatorname{Comm}_{k = 1}\left( \mathcal{U}_\times\right) = \sum_{0 \leq r \neq \leq \lfloor n/2 \rfloor} (n-2r + 1)^2.
    \end{align}
    When $k = 2$, 
    \begin{align}
    	\begin{aligned}
    		F^{(2)}_{\mathcal{U}_\times} = \dim \operatorname{Comm}_{k = 2}\left( \mathcal{U}_\times\right) = & (n+1)^4 + 2\sum_{1 \leq r \leq \lfloor n/2 \rfloor} (n-2r + 1)^4 \\
    		& + 2\sum_{0 \leq r \neq s \leq \lfloor n/2 \rfloor} (n-2r + 1)^2 (n-2s + 1)^2,
    	\end{aligned}
    \end{align}
    where the first two terms on the right-hand side appear when the four indices of basis vectors $\alpha_T$ from \eqref{eq:GeneralPermutation} are all selected from an equivalent $S_n$ irrep. As the trivial irrep ($r = 1$) of $S_n$ is 1-dimensional with no freedom for exchanging indices, we separate it from the second term. The third term counts the number of cases in which two pairs of indices are chosen from two inequivalent $S_n$ irreps. Then the general case follows. 
    
    For qudits, the number of inequivalent $S_n$ irreps is $p(n,d)$, which denotes the number of Young diagrams with $n$ boxes and at most $d$ rows (Definition \ref{def:PartitionFunction}). Unfortunately, there is no closed-form formula for $p(n,d)$, except the asymptotic formula due to Ramanujan, Hardy, and Uspensky \cite{Ramanujan1918,Uspensky1920} and  bounds later developed in Refs.~\cite{Maroti2003,Wladimir2009}. Consequently, it may be infeasible to calculate the frame potential $F^{(k)}_{\mathcal{U}_\times}$ for the group of $\text{SU}(d)$-symmetric unitaries, let alone comparing it with $F^{(k)}_{\mathcal{E}}$ in \eqref{eq:InequalityFramePotential}. To circumvent this obstacle to some extent, we will work exclusively with the commutant algebra $\operatorname{Comm}_k(\mathcal{U}_\times)$ when presenting our main results.
\end{remark}


\section{Exact $k$-designs with $\text{SU}(d)$ symmetry from CQA}\label{sec:A-ExactDesign}

In this appendix, we provide details of our results on local ensembles forming exact $k$-designs in the presence of $\text{SU}(d)$ symmetry. We prove that, for a general $n$-qudit system, the group CQA generated by unitary time evolutions of SWAPs and second-order YJM elements, which are local, is an exact $\text{SU}(d)$-symmetric $k$-design  for $k$ up to at least $O(n^2)$ (Theorem \ref{Thm:CQA2}). We also prove that  to extend this result for arbitrary large $k$, it is necessary to incorporate non-local gates (Theorem \ref{Thm:DesignLocality}), which stands in clear contrast to the symmetry-free case where 2-local unitaries are always enough to approximate Haar randomness to arbitrary precision \cite{Dankert2026PRA,Harrow2design2009}. At the price of increasing locality, we also provide an explicit way to construct ensembles that achieve $k$-designs of arbitrary order by employing $k$-th order YJM elements (Theorem \ref{Thm:CQAk}). The proofs in this section are largely based on group representation theory \cite{Fulton1997,Goodman2009} and the Okounkov--Vershik approach \cite{Okounkov1996,Tolli2009}.


\subsection{Comparison between the $k$-fold channels of $S\mathcal{U}_\times$ and $\mathcal{U}_\times$}\label{sec:LR}

It is straightforward to see that for arbitrary $N$ and $k$, $\text{SU}(N)$ is an exact $k$-design for $\text{U}(N)$:
\begin{align}\label{eq:A-CommonDesign}
	T_k^{\operatorname{SU}(N)} = \int_{\operatorname{SU}(N)} V^{\otimes k} \otimes \bar{V}^{\otimes k} d\nu = \int_{\operatorname{U}(N)} U^{\otimes k} \otimes \bar{U}^{\otimes k} d\mu = T_k^{\operatorname{U}(N)},
\end{align}
where $\nu$ is the restricted Haar measure  on $\text{SU}(N)$ restricted from $\mu$. Apparently, any \emph{global phase} from $U \in \operatorname{U}(N)$ would get canceled after taking the tensor product $U \otimes \bar{U}$ and hence the above integrals are identical. However, this is in general not true for the $T_k^{S\mathcal{U}_\times }$ and $T_k^{\mathcal{U}_\times}$ --- the $k$-fold channels twirled over the groups $S\mathcal{U}_\times$ and $\mathcal{U}_\times$, respectively. Intuitively speaking, with the entire Hilbert space $\mathcal{H} = V^{\otimes n}$ of $n$-qudits being decomposed into various $S_n$ irreps, the integrands in $T_k^{S\mathcal{U}_\times}$ and $T_k^{\mathcal{U}_\times}$ become intricate with tensor products of various submatrices defined on these irreps (see Appendix \ref{sec:SchurWeyl}). In this subsection, we rigorously discuss when $S\mathcal{U}_\times$ fails to form an exact design with respect to $\mathcal{U}_\times$, which paves the way for understanding the more complicated cases involving YJM elements and CQA later.

To motivate our statement, let us consider the simplest nontrivial case where there are only two inequivalent $S_n$ irreps when we decompose the system: $S^{\lambda_1}$ and $S^{\lambda_2}$ of dimension $d_1$ and $d_1$, respectively. Then, $g \in S\mathcal{U}_\times$ is represented as (cf.~\eqref{eq:GroupElements} and Theorem \ref{Thm:Commutant})
\begin{align}
	\begin{pmatrix} V_1 & \\ & V_2 \end{pmatrix} = V_1 \oplus V_2, 
\end{align}
where $V_i \in \operatorname{SU}(d_i)$ and we write $U_i = e^{-i\phi} V_i \in \operatorname{U}(d_i)$ if $g$ is taken from $\mathcal{U}_\times$. Then the integrand
\begin{align}
	g^{\otimes k} \otimes \bar{g}^{\otimes k} = (V_1 \oplus V_2)^{\otimes k} \otimes (\bar{V}_1 \oplus \bar{V}_2)^{\otimes k}  
\end{align}
can be expanded as the following direct sums of tensor products:
\begin{align}\label{eq:SUDesign}
	\big( V_1^{\otimes r} \otimes V_2^{\otimes k - r} \otimes \bar{V}_1^{\otimes s} \otimes \bar{V}_2^{\otimes k - s} \big) \oplus \cdots \oplus \big( V_2^{\otimes r'} \otimes V_1^{\otimes k - r'} \otimes \bar{V}_2^{\otimes s'} \otimes \bar{V}_1^{\otimes k - s'} \big).
\end{align}
As a result, $T_k^{\text{S}\mathcal{U}_\times }$, as well as $T_k^{\mathcal{U}_\times}$, is just a direct sum of integrals with these tensor products:
\begin{align}
	\begin{aligned}
		\Big( \int_{S\mathcal{U}_\times} V_1^{\otimes r} \otimes V_2^{\otimes k - r} \otimes \bar{V}_1^{\otimes s} & \otimes \bar{V}_2^{\otimes k - s} d\nu_1 d\nu_2 \Big) \oplus \cdots \\
		& \cdots \oplus \Big( \int_{S\mathcal{U}_\times} V_2^{\otimes r'} \otimes V_1^{\otimes k - r'} \otimes \bar{V}_2^{\otimes s'} \otimes \bar{V}_1^{\otimes k - s'} d\nu_1 d\nu_2 \Big).
	\end{aligned}
\end{align}
Different from Eq.~\eqref{eq:A-CommonDesign}, each integral is evaluated on both $\text{SU}(d_1)$ and $\text{SU}(d_2)$ under our assumption that the space decomposes as $S^{\lambda_1} \oplus S^{\lambda_2}$. We now show by representation theory that one can always find certain $r$ and $s$ such that
\begin{align}\label{eq:Counterexample}
	\int_{S\mathcal{U}_\times} V_1^{\otimes r} \otimes V_2^{\otimes k - r} \otimes \bar{V}_1^{\otimes s} \otimes \bar{V}_2^{\otimes k - s} d\nu_1 d\nu_2 \neq 0 = \int_{\mathcal{U}_\times} U_1^{\otimes r} \otimes U_2^{\otimes k - r} \otimes \bar{U}_1^{\otimes s} \otimes \bar{U}_2^{\otimes k - s} d\mu_1 d\mu_2.
\end{align}
There are lots of similar counterexamples from other direct sum components from \eqref{eq:SUDesign}. In conclusion,
$T_k^{S\mathcal{U}_\times } \neq T_k^{\mathcal{U}_\times}$ in general. 

To handle integrals over various compact groups, we first rearrange $U_1^{\otimes r} \otimes U_2^{\otimes k - r} \otimes \bar{U}_1^{\otimes s} \otimes \bar{U}_2^{\otimes k - s}$ into $(U_1^{\otimes r} \otimes \bar{U}_1^{\otimes s}) \otimes (U_2^{\otimes k - r} \otimes \bar{U}_2^{\otimes k - s})$ and then apply the Fubini theorem to evaluate
\begin{align}\label{eq:Rearrangement}
	\Big( \int U_1^{\otimes r} \otimes \bar{U}_1^{\otimes s} d\mu_1 \Big) \otimes \Big( \int U_2^{\otimes k - r} \otimes \bar{U}_2^{\otimes k - s} d\mu_2 \Big).
\end{align}
Such rearrangements are invertible for tensor products and hence, to obtain the evidence in Eq.~\eqref{eq:Counterexample}, we just need to check whether the above integral vanishes on $S\mathcal{U}_\times$ or $\mathcal{U}_\times$.

When we integrate over $\mathcal{U}_\times$, the answer is immediate: unless $r = s$, 
\begin{align}
	\int_{\operatorname{U}(d_1)} (U_1^{\otimes r} \otimes \bar{U}_1^{\otimes s}) d\mu_1 = \int_{\operatorname{SU}(d_1)} \frac{1}{2\pi} \int_0^{2\pi} e^{-i(r-s)\phi} (V_1^{\otimes r} \otimes \bar{V}_1^{\otimes s}) d\phi d\nu_1
\end{align}
equals zero because $\int_0^{2\pi} e^{-i(r-s)\phi} = 0$ (cf. Lemma \ref{lemma:VariableChange}). 

When we integrate over $S\mathcal{U}_\times$, setting $r = s$ yields the same result as for $\mathcal{U}_\times$. That is, counterexamples occur when $r \neq s$. The crucial observation for identifying them is that each integrand, such as $V_1^{\otimes r} \otimes \bar{V}_1^{\otimes s}$, is an element from the tensor product of the fundamental and conjugate representations of $\text{SU}(d_1)$ and it can be further decomposed into a direct sum of irreps:
\begin{align}\label{eq:LRdecomposition}
	V_1^{\otimes r} \otimes \bar{V}_1^{\otimes s} \sim \bigoplus_\alpha V_\alpha.
\end{align}
Recall that we have applied this observation to interpret $T_k^{\mathcal{E}}$ as a truncated Fourier transform of its associated measure $\nu$ in Appendix \ref{sec:RandomWalk}. On the other hand, the so-called \emph{Schur orthogonality} \cite{Fulton1997,Goodman2009} asserts that for any compact group $G$ with irreps $\pi_\alpha$ and $\pi_\beta$, 
\begin{align}
	\int_G \pi_\alpha(g)_{ij} \overline{\pi_\beta(g)}_{kl} = \begin{cases} 0, & \pi_\alpha \ncong \pi_\beta \\ \frac{1}{\dim \pi_1} \delta_{ij} \delta_{kl}, & \pi_\alpha \cong \pi_\beta \end{cases}.
\end{align}
To apply these facts, we first expand high order tensor products like $V_1^{\otimes r} \otimes \bar{V}_1^{\otimes s}$ by \eqref{eq:LRdecomposition}. Then
\begin{align}
	\int_{\operatorname{SU}(d_1)} (V_1^{\otimes r} \otimes \bar{V}_1^{\otimes s}) d\nu_1 = \bigoplus_\alpha  \int_{\operatorname{SU}(d_1)} V_\alpha d\nu_1 = \bigoplus_\alpha \int_{\operatorname{SU}(d_1)} V_\alpha \otimes \boldsymbol{1} d\nu_1, 
\end{align}
where $\boldsymbol{1}$ is simply the 1-dimensional unit scalar given by the trivial representation of $\operatorname{SU}(d_1)$. As long as the decomposition of \eqref{eq:LRdecomposition} bears the trivial representation, the integral is nonzero by Schur orthogonality which leads to a counterexample.

\begin{example}
	As a concrete example, suppose that $d_1 \leq k$ and $d_2 = 1$. Let $r = d_1$ and $s = 0$. Then,
	\begin{align}
		(V_1^{\otimes r} \otimes \bar{V}_1^{\otimes s}) \otimes (V_2^{\otimes k - r} \otimes \bar{V}_2^{\otimes k - s}) = V_1^{\otimes d_1} \otimes \boldsymbol{1}^{\otimes 2k - d_1}.
	\end{align}
	Moreover, it can be shown by the  \emph{Littlewood--Richardson rule} illustrated in the following lemma that for any $d$, the $d$-fold tensor product representation of $\text{SU}(d)$ contains the trivial representation (e.g., $d = 2,3$ familiar in quantum angular momentum and quark theory). Therefore,
	\begin{align}\label{eq:TrivialExample}
		\int_{\operatorname{SU}(d_1)} V_1^{\otimes d_1} d\nu_1 \otimes \int_{\operatorname{SU}(1)} \boldsymbol{1}^{\otimes 2k - d_1} d\nu_2 \neq 0 = \int_{\operatorname{U}(d_1)} U_1^{\otimes d_1} d\mu_1 \otimes \int_{\operatorname{U}(1)} e^{-i (k-d_1)\phi} e^{ik\phi} d\mu_2.
	\end{align}
	Note that the 1-dimensional representation of $\text{U}(1)$ is given by $e^{-i\phi}$. 
\end{example}

This inconsistency occurs when we consider arbitrary $n$-qudit systems. To explain the reason, we recite the following basic fact from $S_n$ representation theory: {when $n \geq 5$, except the 1-dimensional trivial representation and sign representation, there is no irrep with dimension lower than $n-1$} \cite{Rasala1977,Sagan01,Sellke2020}. When, e.g.,  $n = 4$, the statement fails as there is a two dimensional irrep of Young diagram $\lambda = (2,2)$ (by the hook-length formula in Definition \ref{def:YoungDiagram}).

Assume that there are $\geq 5$ qudits in the system. Recall that, by Schur--Weyl duality, only the $S_n$ irreps corresponding to Young diagrams of size $n$ and at most $d$ rows appear when decomposing the system with $\text{SU}(d)$ symmetry.
\begin{enumerate}
	\item[(a)] In common cases in which $d < n$, the above counterexample cannot happen for $k < n-1$ because there is no $S_n$ irrep, except the trivial one, that has dimension no larger than $k < n-1$. We also prove in the following lemma that it is impossible to find {any other} kinds of counterexamples in this case.
	
	\item[(b)]  When  $d \geq n$, the inconsistency always exists because we can select $V_1$ and $V_2$ from the trivial and sign representations, respectively. Even though  the corresponding representation spaces are 1-dimensional, the phase factors are integrated independently and the integral vanishes like the right-hand side of Eq.~\eqref{eq:TrivialExample}.
\end{enumerate}

\begin{lemma}\label{lemma:LR}
	Let $r \neq s \leq k < d$ with $r,s,k$, and $d$ defined before. The tensor product $V^{\otimes r} \otimes \bar{V}^{\otimes s}$ for $V \in \operatorname{SU}(d)$ cannot be decomposed into the trivial representation. Therefore, the integration of \eqref{eq:LRdecomposition} always vanishes over $\text{SU}(d)$. Trivial representations only appear when $k \geq d$. 
\end{lemma}
\begin{proof}
	The proof is based on the method of recording $\text{U}(d)$ irreps by Young diagrams and results from the Littlewood–Richardson rule. We refer interested readers to Refs.~\cite{Fulton1997,Goodman2009} for more details. For our purpose, it suffices to assume the following facts: 
	\begin{enumerate}
		\item The fundamental representation of $\text{U}(d)$ is denoted by the one-box Young diagram $(1)$.
		
		\item The conjugate representation corresponds to the Young diagram $(1^{d-1})$ of $d - 1$ boxes in one column.
		
		\item The trivial representation is given by the Young diagram $(1^d)$ of $d$ boxes in one column.
	\end{enumerate}
	Other $\text{U}(d)$ irreps are also expressed by Young diagrams under certain conditions. Especially, the Littlewood–Richardson rule indicates that the Young diagram of each irrep arises from the decomposition of $U_1^{\otimes r} \otimes \bar{U}_1^{\otimes s}$ contains $r + s(d - 1)$ boxes (but may be allocated with different rows or columns). As long as this number can be divided by $d$, the trivial representation (restricted back to $\text{SU}(d)$) appears in the decomposition.
	
	Therefore, for $k < d$, we consider the following two cases:
	\begin{enumerate}
		\item $r > s$: then $r + s(d - 1) = (r - s) + sd$. Since $r, s \leq k < d$, $d$ cannot divide $r - s$.
		
		\item $r < s$: then $r + s(d - 1) = rd + (s - r)(d - 1)$. Assume that $d$ divides $(s - r)(d - 1)$. Since the greatest common divisor of $d-1$ and $d$ is just 1, this requires $s - r$ to be divided by $d$, which is impossible.
	\end{enumerate}
	When $k \geq d$, there are several ways to produce the trivial representation from tensor product as in the previous example.  
\end{proof}	

With the above argumentation, we now analyze the real situation where matrix representations of elements from $S\mathcal{U}_\times$ and $\mathcal{U}_\times$ consist of various submatrix blocks from inequivalent $S_n$ irreps with possible equivalent copies as 
\begin{align}
	\bigoplus_\lambda V_\lambda^{\oplus m_{S_n, \lambda}}, \quad \bigoplus_\lambda U_\lambda^{\oplus m_{S_n, \lambda}}.
\end{align}
Submatrices from equivalent copies of any irreps are integrated simultaneously. Inequivalent submatrices are arranged as before and they are integrated independently. Then the integrals of paired fundamental and conjugate representations on one $S_n$ sector ($r = s$ in our previous illustration) are always the same for both $S\mathcal{U}_\times$ and $\mathcal{U}_\times$; inconsistency may appear otherwise for $S_n$ blocks of specific dimensions violating Lemma \ref{lemma:LR}. Combining with the Fact below \eqref{eq:AboveFact}, we conclude the following.

\begin{theorem}\label{Thm:SU}
	Suppose that $n\geq 5$ and $d < n$. If $k < n-1$, then the group $S\mathcal{U}_\times$ with restricted Haar measure forms an exact $k$-design with respect to $\mathcal{U}_\times$. For larger $k$, $S\mathcal{U}_\times$ and $\mathcal{U}_\times$ have different commutants and thus $S\mathcal{U}_\times$ cannot even converge to $k$-designs with respect to $\mathcal{U}_\times$.
\end{theorem}


\subsection{Comparison between the $k$-fold channels of CQA and $\mathcal{U}_\times$}\label{sec:CQA2}

In Appendix \ref{sec:LR}, we have showed that $T_k^{S\mathcal{U}_\times } = T_k^{\mathcal{U}_\times}$ only when $d < n$ and $k < n-1$. The group CQA, on the other hand, possesses accessibility to part of the phase factors due to the employment of YJM elements (Definition \ref{def:YJM}), which would alleviate this problem to some extent. We are now going to make this point clear. As a reminder, results from Refs.~\cite{Zeier2011,Zeier2015} point out that the commutant  $\operatorname{Comm}_2(H)$ of a Lie subgroup $H$ of any compact semisimple group $G$ is strictly larger than $\operatorname{Comm}_2(G)$ and hence $H$ can never be a unitary 2-design. However, the ambient Lie group $\mathcal{U}_\times$ considered here is not semisimple because they have a nontrivial center or, informally, elements with nontrivial phase factors (Theorem \ref{thm:CenterBases}), so the result is not applicable to our case.

Following the method from Appendix \ref{sec:LR}, we first check the integral defining $T_k^{\mathrm{CQA}}$ by representation theory. To begin with, we apply the main theoretical result of Ref.~\cite{Zheng2021SpeedingUL}: 
\begin{align}
	S\mathcal{U}_\times \subset \mathrm{CQA}.
\end{align}
Therefore, by the Fact below \eqref{eq:AboveFact} and \ref{Thm:SU}, 
\begin{align}
	\operatorname{Comm}_k\left(S\mathcal{U}_\times\right) \supset \operatorname{Comm}_k(\mathrm{CQA}) \supset \operatorname{Comm}_k\left(\mathcal{U}_\times\right),
\end{align}
and CQA constitutes an exact $k$-design with respect to $\mathcal{U}_\times$ for all $k < n-1$ and $d < n$. With nontrivial phase factors given by YJM elements, the statement now holds for larger $k$. To be precise, let us consider the Lie algebra of $\mathrm{CQA}$ which is proved in Ref.~\cite{Zheng2021SpeedingUL} to satisfy
\begin{align}
	\mathfrak{L}(\mathrm{CQA}) = \mathfrak{L} \left(S\mathcal{U}_\times\right) \oplus \mathfrak{z},
\end{align}
where $\mathfrak{z}$ is a subspace spanned by 
\begin{align}\label{eq:PhaseFactor}
	i c_\sigma = \bigoplus_\lambda i \frac{\operatorname{tr}_\lambda (\sigma)}{\dim S^\lambda} I_\lambda = \bigoplus_\lambda i \frac{\chi_\lambda (\sigma)}{\dim S^\lambda} I_\lambda
\end{align}
where $\sigma$ is any 2-cycle, 3-cycle and $(2,2)$-cycle (Definition \ref{def:CycleType}) and $I_\lambda$ is the identity matrix on the $S^\lambda$ irrep block. Intuitively, second-order YJM elements produce these cycles. Let $H$ denote the integral Lie group formed by taking exponentials of elements from $\mathfrak{z}$. Define the map 
\begin{align}
	f: S\mathcal{U}_\times \times H \rightarrow \mathrm{CQA} \text{ by } f(g,h) = gh.
\end{align}
Obviously, $f$ is a Lie group homomorphism because the $S\mathcal{U}_\times$ commute with $H$ by definition. Then it is natural to evaluate the integral defining $T_k^{\mathrm{CQA}}$ using $f$ as a ``change of variables''. However, $f$ is surjective but not an isomorphism. A more familiar case is that $\text{SU}(d) \times \operatorname{U}(1) \ncong \operatorname{U}(d)$. Despite this, we still have the following lemma.

\begin{lemma}\label{lemma:VariableChange}
	$\operatorname{Comm}_k(S\mathcal{U}_\times \times H) = \operatorname{Comm}_k(\mathrm{CQA})$. Therefore, if we denote by $\{c_i\}$ a basis of $\mathfrak{z}$, then
	\begin{align}\label{eq:A-TkCQA}
		T_k^{\mathrm{CQA}} & = T_k^{S\mathcal{U}_\times \times H} = \int_{S\mathcal{U}_\times \times H} f(g)^{\otimes k} \otimes \overline{f(g)}^{\otimes k} d\mu(g) \notag \\
		& = \int_{\gamma_j} \int_{S\mathcal{U}_\times} (e^{-i \sum_j \gamma_j c_j} V)^{\otimes k} \otimes (e^{i \sum_j \gamma_j c_j} \bar{V})^{\otimes k} dV d\gamma \\
		& = \int_{\gamma_j} (e^{-i \sum_j \gamma_j c_j})^{\otimes k} \otimes (e^{i \sum_j \gamma_j c_j} )^{\otimes k} d\gamma \int_{S\mathcal{U}_\times} V^{\otimes k} \otimes \bar{V}^{\otimes k} dV = T_k^{ H} T_k^{S\mathcal{U}_\times} =  T_k^{S\mathcal{U}_\times} T_k^{ H} . \notag
	\end{align}
\end{lemma}
\begin{proof}
	Since $f$ defined above is a group homomorphism, it formally endows the product group $S\mathcal{U}_\times$ a representation on $\mathcal{H} = V^{\otimes n}$. It is thus legal to define $\operatorname{Comm}_k(S\mathcal{U}_\times \times H)$ as the image  of the function $f$ which is exactly equal to $\operatorname{Comm}_k(\mathrm{CQA})$. Since we integrate with respect to the Haar measure of $S\mathcal{U}_\times$, $T_k^{S\mathcal{U}_\times \times H}$ is a Hermitian projector (Lemma \ref{lemma:EnsembleHermitian}) and equals $T_k^{\mathrm{CQA}}$. 
\end{proof}

Let us revisit the example of Eq.~\eqref{eq:TrivialExample} after adding YJM elements.

\begin{example}
	Suppose that $k = n-1$. The expansion of $V^{\otimes k} \otimes \overline{V}^{\otimes k}$ always contains the term $V_1^{\otimes n-1} \otimes \boldsymbol{1}^{\otimes n-1}$ where $V_1$ is an arbitrary unitary (of unit determinant) acting on the $(n-1)$-dimensional $S_n$ irrep block with identity $\boldsymbol{1}$ taken from the trivial irrep. Then 
	\begin{align}
		\int_{\operatorname{SU}(n-1)} V_1^{\otimes n-1} d\nu_1 \otimes \int_{\operatorname{SU}(1)} \boldsymbol{1}^{\otimes n-1} d\nu_2 \neq 0 = \int_{\operatorname{U}(n-1)} U_1^{\otimes n-1} d\nu_1 \otimes \int_{\operatorname{U}(1)} e^{i(n-1)\phi} d\nu_2
	\end{align}
    by Schur orthogonality and the Littlewood--Richardson rule. However, it can be seen in \eqref{eq:A-TkCQA} that the integral over CQA has one more term of phase factors $e^{-i \sum_{j=1}^N \gamma_j c_j}$ given by the Lie algebra $\mathfrak{z}$, which would ultimately eliminate the above inconsistency. 
	
	Let us integrate the phase factors provided by first order YJM elements consisting of merely 2-cycles. As introduced in Appendix \ref{sec:SnTheory}, $c_{\tau} = c_{\tau'}$ for any 2-cycles $\tau$ and $\tau'$. We also compute by Eq.~\eqref{eq:CharacterValue} that
	\begin{align}\label{eq:2CycleCharacter}
		\frac{\operatorname{tr}_{(n)} (\tau) }{\dim S^{(n)}} = 1 = -\frac{\operatorname{tr}_{1^{(n)}} (\tau) }{\dim S^{(1^n)} }, \quad \frac{\operatorname{tr}_{(n-1,1)} (\tau)}{\dim S^{(n-1,1)}} = \frac{n-3}{n-1} = -\frac{\operatorname{tr}_{(2,1^{(n-2)})} (\tau)}{\dim S^{(2,1^{n-2})}},
	\end{align}
    where $\lambda = (n), (1^n), (n-1,1)$, and $(2,1^{n-2})$ denote the trivial, sign, standard and conjugate representation of $S_n$ respectively (Definition \ref{def:YoungDiagram}).
     Let $I_\lambda$ denote the identity matrix on the $S^\lambda$ irrep block. The phase integral accompanied with $V_1^{\otimes n-1} \otimes I_{(n)}^{\otimes n-1}$ includes
	\begin{align}\label{eq:PhaseIntegral}
		\frac{1}{2\pi} \int_0^{2\pi} \Big( (e^{-i\frac{n-3}{n-1} \gamma I_{(n-1,1)}})^{\otimes n-1} \otimes (e^{i\gamma I_{(n)}})^{\otimes n-1} \Big) d\gamma = \frac{1}{2\pi} \int_0^{2\pi} e^{i2\gamma} I d\gamma = 0.
	\end{align}
\end{example}

\begin{theorem}\label{Thm:CQA2}
	The following holds for exact $k$-designs with $\text{SU}(d)$ symmetry:
	\begin{enumerate}
		\item For an $n$-qudit system with $n \geq 9$ and $d < n$, the group CQA generated by 4-local $\text{SU}(d)$-symmetric Hamiltonians forms an exact $\text{SU}(d)$-symmetric $k$-design with respect to $\mathcal{U}_\times$ for all $k < n(n-3)/2$. When $d \geq n$, the largest possible $k$ is precisely $2n-4$.
		
		\item For an $n$-qubit system ($d = 2$) with $n \geq 10$, CQA is an exact $\text{SU}(2)$-symmetric $k$-design for all $k < n(n-1)(n-5)/6$. 
	\end{enumerate}
\end{theorem}
\begin{proof}	
	The first claim is demonstrated by checking phase integrals as in the previous example. Recall from \eqref{eq:GroupElements} that any $g \in \mathrm{CQA}$ consists of various unitaries on $S_n$ irreps from the decomposition of the entire Hilbert space $\mathcal{H}$, so we expand $g^{\otimes k} \otimes \bar{g}^{\otimes k}$ as in \eqref{eq:SUDesign} before calculating the integral.  When $n \geq 9$, the third lowest dimension of these irrep blocks is $n(n-3)/2$ \cite{Rasala1977} and Lemma \ref{lemma:LR} indicates that 
    \begin{align}
	\int_{\text{SU}(S^\lambda)} V^{\otimes r} \otimes \bar{V}^{\otimes s} dV = \int_{\text{U}(S^\lambda)} U^{\otimes r} \otimes \bar{U}^{\otimes s} dU
    \end{align}
    when $r,s < n(n-3)/2$ but $\dim S^\lambda \geq n(n-3)/2$. Inconsistency arises when we integrate over even lower-dimensional irrep blocks, and we need to verify that the phase integral as in Eq.~\eqref{eq:PhaseIntegral} would remedy the problem.
	
	Since we assume that $d < n$, $S_n$ irreps with dimension lower than $n(n-3)/2$ are just the 1-dimensional trivial irrep and the $(n-1)$-dimensional standard irrep with its conjugate. With respect to these three irreps, let us denote by $\phi_i, i = 1,2,3$ the phase factors of 2-cycle given by YJM elements in CQA. Similarly, we take $\psi_i, i = 1,2,3$ as the phase factors of 3-cycle (or $(2,2)$-cycle). As a reminder,
	\begin{align}
		\phi_1 = \psi_1 = 1, \quad \phi_2 = \frac{n-3}{n-1} = -\phi_3.
	\end{align}
	Explicit formula for $\psi_2$ and $\psi_3$ can be found in Refs.~\cite{Ingram1950,Lassalle2008}. For now it suffices to know that $\psi_2 = \psi_3$. By definition, they are combined as 
	\begin{align}
		\begin{aligned}
			& \phi_1(r_1 - s_1) + \phi_2(r_2 - s_2) + \phi_3(r_3 - s_3), \\
			& \psi_1(r_1 - s_1) + \psi_2(r_2 - s_2) + \psi_3(r_3 - s_3),
		\end{aligned}
	\end{align}
    in an expansion of $g^{\otimes k} \otimes \bar{g}^{\otimes k}$, with $r_i$ and $s_i$ being the tensor folds of the aforementioned three irreps. Since $\sum_i r_i = \sum_i s_i = k$, we rewrite them as
    \begin{align}
    	\begin{aligned}
    		& (\phi_1 - \phi_3)(r_1 - s_1) + (\phi_2 - \phi_3)(r_2 - s_2), \\
    		& (\psi_1 - \psi_3) (r_1 - s_1) + (\psi_2 - \psi_3)(r_2 - s_2).
    	\end{aligned}
    \end{align}
    Note that $(\psi_1 - \psi_3) (r_1 - s_1) + (\psi_2 - \psi_3)(r_2 - s_2) = (\psi_1 - \psi_3) (r_1 - s_1)$ and it is only when $r_i = s_i$ for all $i$ that they yield trivial phase. Therefore, under the assistance of phase integrals, $T_k^{\mathrm{CQA}} = T_k^{\mathcal{U}_\times}$ for all $k = n(n-3)/2$. 
    
    As a reminder, the above proof requires $d < n$. Otherwise, the 1-dimensional sign representation always arises from the decomposition of $\mathcal{H}$ under $\text{SU}(d)$ symmetry. Then, the previous argument fails for $k = 2(n-4)$. Indeed, one can check by separately considering Lemma \ref{lemma:LR} and the phase integral that
    \begin{align}
    	\hspace{-1em} \int_{\mathrm{CQA}} W_{(n)}^{\otimes n-3} \otimes W_{(2,1^{(n-2)})}^{\otimes n-1} \otimes \bar{W}_{1^{(n)}}^{\otimes n-3} \otimes \bar{W}_{(n-1,1)}^{\otimes n-1} dW \neq 0 = 
    	\int_{\mathcal{U}_{\times,\U(1)}} U_{(n)}^{\otimes n-3} \otimes U_{(2,1^{(n-2)})}^{\otimes n-1} \otimes \bar{U}_{1^{(n)}}^{\otimes n-3} \otimes \bar{U}_{(n-1,1)}^{\otimes n-1} dU,
    \end{align}
    where for the phase integral, we have to use characters \cite{Ingram1950}
    \begin{align}
    	\frac{\operatorname{tr}_{(n-1,1)} (i,j)(k,l)}{\dim S^{(n-1,1)}} = \frac{n-4}{n-1} = \frac{\operatorname{tr}_{(2,1^{(n-2)})} (i,j)(k,l)}{\dim S^{(2,1^{n-2})}}, \quad \frac{\operatorname{tr}_{(n-1,1)} (i,j,k)}{\dim S^{(n-1,1)}} = \frac{n-5}{n-1} = \frac{\operatorname{tr}_{(2,1^{(n-2)})} (i,j,k)}{\dim S^{(2,1^{n-2})}},
    \end{align}
    for $(2,2)$-cycles and $3$-cycles given by YJM elements. The order is just $n-3 + n - 1 = 2n - 4$ and it is easy to check that $T_k^{\mathrm{CQA}} = T_k^{\mathcal{U}_\times}$ for any lower $k$.
    
	For the case of qubits, we argue by $S_n$ character formula on two-row Young diagrams (Schur--Weyl duality) that the bound can be improved to $n(n-1)(n-5)/6$. Our strategy is to count the number of inequivalent irrep blocks appearing in each specific term from the expansion. For instance, there are terms bearing two and three inequivalent irreps which we rearrange as 
    \begin{align}\label{eq:SUDesign2}
		(V_1^{\otimes r} \otimes \overline{V}_1^{\otimes s} \otimes (V_2^{\otimes k - r} \otimes \overline{V}_2^{\otimes k - s}), \quad  (V_1^{\otimes r_1} \otimes \overline{V}_1^{\otimes s_1} \otimes (V_2^{\otimes r_2} \otimes \overline{V}_2^{\otimes s_2}) \otimes (V_3^{\otimes k - r_1 - r_2} \otimes \overline{V}_3^{\otimes k - s_1 - s_2}).
	\end{align}
	The first type has been exemplified several times in the preceding contexts. We still use $\phi_1, \phi_2$ to denote the phase factor induced by any 2-cycle transposition for $V_1$ and $V_2$, respectively. Since two-row Young diagrams ($d = 2$ for qubits) are totally ordered, Lemma \ref{lemma: total-ordering2} implies that $\phi_1 \neq \phi_2$.
	
	As before, we only need to examine the situation when $r \neq s$, because when $r = s$, the integrals over $\mathcal{U}_\times$ and $S\mathcal{U}_\times$ are identical. For arbitrary $k$, the phase integral like Eq.~\eqref{eq:PhaseIntegral} must vanish since
	\begin{align}
		\phi_1(r - s) + \phi_2 (k - r - (k - s)) = (\phi_1 - \phi_2) (r - s) \neq 0.
	\end{align}
	Therefore, any possible inconsistency between $T_k^{\mathrm{CQA}}$ and $T_k^{\mathcal{U}_\times}$ should be spotted on expanded terms with at least three inequivalent irrep blocks. We deal with this case by making use of the phase factors produced by $\Big( \frac{2}{n(n-1)}\sum_{l=1}^n X_l \Big)^2$. It turns out that they are just powers of $\phi_i$ given by 2-cycles (see Eq.~\eqref{eq:P_l}). All these phase factors are combined as 
	\begin{align}
		\begin{aligned}
			& (\phi_1 - \phi_3)(r_1 - s_1) + (\phi_2 - \phi_3)(r_2 - s_2), \\
			& (\phi_1^2 - \phi_3^2) (r_1 - s_1) + (\phi_2^2 - \phi_3^2)(r_2 - s_2).
		\end{aligned}
	\end{align}
	Unless $r_i = s_i$, the above equations have no homogeneous solutions. Therefore, counterexamples only occur when at least four inequivalent irreps emerge. Let $D$ be the fourth lowest dimensions of these irreps. Then, Lemma \ref{lemma:LR} requires that $k \geq D$.
	To derive this fourth lowest dimension $D$, we compute by the hook-length formula in Definition \ref{def:YoungDiagram} to obtain that
	\begin{align}
		\dim S^{(n)} = 1,\quad \dim S^{(n-1,1)} = n-1, \quad \dim S^{(n-2,2)} = \frac{n(n-3)}{2}, \quad \dim S^{(n-3,3)} = \frac{n(n-1)(n-5)}{6}.
	\end{align}
	For $S_n$ irreps of two-row Young diagrams, their dimensions also satisfy the following identity (see the end of Appendix \ref{sec:SnTheory}):
	\begin{align}
		\dim S^{(n-r,r)} = \binom{n}{r} - \binom{n}{r-1}
	\end{align}
	where $0 \leq r \leq \lfloor n/2 \rfloor$ and $\binom{n}{-1} = 0$ by convention.The last step is verifying $D = \dim S^{(n-3,3)}$ when we only consider two-row Young diagrams and $n \geq 15$.
    
    When $n \geq 10$, $S^{(n-5,5)}, S^{(n-4,4)}$ arise from the decomposition of the physical Hilbert space under $\text{SU}(2)$ symmetry. A direct computation indicates that
    \begin{align}
	  \dim S^{(n-5,5)}, \dim S^{(n-4,4)} \geq \dim S^{(n-3,3)}.
    \end{align}
    As a caveat, only when $n \geq 12$, $S^{(n-6,6)}$ appears and we still have 
    \begin{align}
       \dim S^{(n-6,6)} \geq \dim S^{(n-3,3)}.
    \end{align}
    Similarly, $\dim S^{(n-7,7)} \geq \dim S^{(n-3,3)}$ for $n \geq 14$. To deal with the cases for larger $r$, we note that for $n \geq 16$
    \begin{align}
      \binom{n}{r} - \binom{n}{r-1} = \frac{n-2r+1}{n-r+1} \binom{n}{r} \geq \frac{1}{n+1} \binom{n}{r} \geq \frac{1}{n+1} \binom{n}{8} \geq \dim S^{(n-3,3)}. 
    \end{align}
    because the binomial coefficients are increasing. This concludes the proof.
\end{proof}

\begin{remark}
	We introduce a related result proved in Refs.~\cite{Marin1,Marin2,MarvianSU2,MarvianNature}, that the group
	\begin{align}
		\operatorname{eSWAP} = \langle e^{-i\theta_{rs} (r,s)} \rangle
	\end{align} 
	generated by all $n(n-1)/2$ transpositions contains $S\mathcal{U}_\times$ on qubits ($d = 2$). 
	\begin{enumerate}
		\item[(a)] This group achieves the restricted universality on qubits:
        \begin{align}
          S\mathcal{U}_\times \subsetneqq \operatorname{eSWAP} \subsetneqq \mathcal{V}_2 \subsetneqq \mathcal{U}_\times. 
          \end{align}
          Since eSWAP satisfies the restricted universality and contains 2-cycles, techniques from the above theorem can be applied to show that eSWAP forms an exact $k$-design for all $k < n(n-3)/2$ on an $n$-qubit system. However, the inclusion property on the LHS fails to hold for general qudits, for which we propose the framework of the 4-local CQA to solve the problem. 
		
		\item[(b)] As a reminder, eSWAP can be generated by arbitrary generating sets of SWAPs. We explain the point from the perspective of Lie algebra. Let $\tau_1$ and $\tau_2$ denote two SWAPs taken from the generating set. A simple calculation of Lie brackets shows 
        \begin{align}
           [[\tau_1, \tau_2], \tau_1] = 2\tau_1 \tau_2 \tau_1 - 2\tau_2. 
           \end{align}
        Therefore, the conjugate actions such as $\tau_1 \tau_2 \tau_1$ can be utilized to generate arbitrary SWAPs. As a result, in Appendix \ref{sec:CQAEnsemble}, we define the ensemble $\mathcal{E}_{\operatorname{eSWAP}}$ merely using unitary evolutions generated by of nearest-neighbor SWAPs $(j,j+1)$.
	\end{enumerate} 
    The group eSWAP fails to achieve restricted universality and $k$-designs when $d \geq 3$, as indicated in Refs.~\cite{Marin1,Marin2,MarvianSU2,MarvianNature,MarvianSUd}. We also provide numerical evidence in Section \ref{sec:CQAEnsemble} by restricting the $k$-th moment operator of eSWAP to certain $S_n$ irreps.
\end{remark}

As a reminder, it is still possible to refine the bound on $k$ provided above: we use phase factors given by the square of the sum of all YJM elements $(\sum_{l=1}^n X_l)^2$, but there are still other kinds of second-order YJM elements $X_k X_l$ and the lower bound on $S_n$ irreps could be modified. However, this requires a more sophisticated treatment in order to compute $S_n$ characters and estimate the hook-length formula, both of which may require heavy combinatorics and representation theory \cite{Ingram1950,Rasala1977,Roichman1996,Lassalle2008,Giambruno2015,Pak2020} and are beyond the intended scope of this paper. More importantly, we prove in the Appendix \ref{sec:Locality} that any ensemble with  bounded locality would eventually cease to be an exact $k$-design for some $k$. In other words, contrary to the conventional case without symmetry, it is impossible to achieve $k$-designs of arbitrarily large $k$ with any $O(1)$-local ensembles.


\subsection{Circuit locality for universality and unitary $k$-designs with $\text{SU}(d)$ symmetry}\label{sec:Locality}

Here we present detailed results on how locality limits the ability of achieving universality as well as $k$-designs, which is of fundamental interest from both physical and practical perspectives as motivated in the main text. It is well-known that 2-local unitaries are universal, meaning that they are able to generate any unitary to arbitrary precision for any local dimension \cite{Brylinski2001,Vlasov2001,Sawicki2016}; furthermore, they are powerful enough for the generation of $k$-designs for any $k$ \cite{Brown_2010,harrow2016local,harrow2023approximate,Haferkamp2021,Oszmaniec2022}. Under $\text{SU}(d)$ symmetry, we have showed in this work that the group CQA consisting of 4-local generators is universal when ignoring relative phase factors \cite{Zheng2021SpeedingUL} and forms an exact $\text{SU}(d)$-symmetric $k$-design for any $k<n(n-3)/2$. 
 We now explore the opposite direction and consider the degree of locality needed to achieve universality or designs of arbitrary $k$. We demonstrate that in order to achieve an arbitrarily large $k$ we have to replenish more phase factors and, in the end, operators with any bounded locality can never fulfill this task.

To begin with, we discuss the locality needed to achieve universality with all phase factors taken into account, i.e., generating $\mathcal{U}_\times$ rather than $S\mathcal{U}_\times$. This question has been addressed in Refs.~\cite{MarvianNature,MarvianSU2} for qubits as a no-go theorem which states that ensembles of local gates can never generate all relative phases. We prove the case for general qudits by $S_n$ representation theory as a preparation for describing the key results for $k$-designs. Recall that the number of $S_n$ irreps corresponding to Young diagrams with at most $d$ rows is given by the partition function $p(n,d)$ in Definition \ref{def:PartitionFunction}.

\begin{definition}
	Any {finite or infinite} set $S$ of $\text{SU}(d)$-symmetric unitaries (gates) on an $n$-qudit system is said to be \emph{exactly universal} under $\text{SU}(d)$ symmetry if $\langle S \rangle$, the Lie group generated by $S$, is precisely equal to $\mathcal{U}_\times$. It is \emph{approximately universal} under $\text{SU}(d)$ symmetry if $\langle S \rangle$ is dense in $\mathcal{U}_\times$, or the closure $\overline{\langle S \rangle} = \mathcal{U}_\times$.
\end{definition}

\begin{theorem}\label{Thm:UniversalityLocality}
	Given a finite or infinite set $S$ of $\text{SU}(d)$-symmetric unitary gates, in order to achieve either exact or approximate universality, $S$ must contain unitaries with locality $\gamma$ being \emph{at least} $\Omega(\log p(n,d))$, where $p(n,d)$ is the number of $S_n$ irreps corresponding to Young diagrams with at most $d$ rows.
\end{theorem}
\begin{proof}
	We first consider exact universality. Let $\mathcal{V}_\gamma$ be the group generated by all $\gamma \geq 4$ $\text{SU}(d)$-symmetric unitaries. It is shown in Refs.~\cite{MarvianSU2,MarvianNature,MarvianSUd} that the Lie algebra of this group is spanned by arbitrary complex linear combinations $H = \sum_i c_i \sigma_i$ of permutations $\sigma_i$ supported on $N$ qudits as long as the combination is anti-Hermitian. Then $H = (H - c_H) + c_H$ where
	\begin{align}
		c_H = \bigoplus_\lambda \frac{\operatorname{tr}_\lambda H}{\dim S^\lambda} I_\lambda
	\end{align}
	 is defined by the trace of $H$ in each irrep block $S^\lambda$ ($S_n$ characters) with multiplicities (cf. \eqref{eq:PhaseFactor}). Since $H$ is anti-Hermitian, $\operatorname{tr}_\lambda H$ must be a real linear combination of $\operatorname{tr}_\lambda \sigma$ times the imaginary unit.

	The group $\mathcal{V}_\gamma$ contains CQA by definition and, therefore, its Lie algebra can be written as \cite{Zheng2021SpeedingUL}:
	\begin{align}
		\mathfrak{L}( S\mathcal{U}_\times) \oplus \mathfrak{z}_N,
	\end{align}
	where $\mathfrak{z}_N$ is spanned by $i c_\sigma$ for all $\gamma$-local $\sigma$. As introduced in Section \ref{sec:SnTheory}, $c_\sigma = c_{\sigma'}$ for $\sigma$ and $\sigma'$ of the same cycle type (see Definition \ref{def:CycleType}) and they act as a scalar matrix in any equivalent $S_n$ irrep. Let us record these scalars in the following matrix:
	\begin{align}\label{eq:CharacterMatrix}
		C_n = \begin{pmatrix}
			c_{11} & \cdots & c_{1, p(n)} \\ \vdots & \ddots & \vdots \\ c_{p(n), 1} & \cdots & c_{p(n), p(n)} 
		\end{pmatrix},
	\end{align}   
	where the column indices label inequivalent irreps and the row indices represent different cycle types because the number of inequivalent irreps equals $p(n)$, the number  of partitions of $m$, which is also the number of different cycle types. The key point in proving our statement is to realize that the above matrix is invertible or, equivalently, the column vectors form a basis for phase factors from inequivalent $S_n$ irreps. Recall that in Theorem \ref{thm:CenterBases}, we presented three different center bases for phase factors and the one used here is just a rescale of $c_\lambda$ by the number of all permutations $\sigma$ of the given cycle type $\lambda$. When we only have $\gamma$-local permutations, $\mathfrak{z}_N$ defined above is spanned by the first $p(N)$ column vectors of $C_n$.
	
	On the other hand, Schur--Weyl duality says that only $S_n$ irreps corresponding to Young diagrams with no more than $d$ rows can be found in the qudit system. Therefore, the Lie algebra of $\mathcal{U}_\times$ is spanned by $\mathfrak{L}(S\mathcal{U}_\times)$ with column vectors from a certain invertible $p(n,d) \times p(n,d)$ submatrix from $C_n$. Unfortunately, there is in general no way to locate this submatrix, as computing the entries $c_{ij}$ of $C_n$ requires evaluating $S_n$ characters and computing $p(n,d)$ precisely is infeasible. However, by using the exponential lower bound  given below Definition \ref{def:PartitionFunction}, we can establish the following simple necessary condition for $\mathcal{V}_\gamma$ to be universal:
	\begin{align}
		p(\gamma) \geq p(n,d) \quad\implies\quad \gamma = \Omega(\log p(n,d)).
	\end{align}
	When $d = 2$, the required locality is at least $\log n$ (Theorem \ref{thm: YJM-center-basis} in Appendix \ref{sec:CQAk} gives a precise bound  $2\lfloor n/2 \rfloor = \Theta(n)$ for qubits). However, in the extreme case when $d \geq n$ reconstructing $\mathcal{U}_\times$ demands $\sigma$ of all cycle types, including gates permuting all qudits.

	To deal with approximate universality, assume the locality $\gamma$ of a (finite or infinite) generating set $S$ does not satisfy $\gamma = \Omega(\log p(n,d))$. Since $\mathcal{V}_\gamma$ is a compact subgroup and since $S \subset \mathcal{V}_\gamma$, 
	\begin{align}
		\overline{\langle S \rangle} \subset \mathcal{V}_\gamma \subsetneqq \mathcal{U}_\times,
	\end{align}
	which leads to a contradiction.
\end{proof}


We now state the result for $k$-designs.

\begin{theorem}\label{Thm:DesignLocality} 
	Given a finite or infinite ensemble $\mathcal{E}$ of $\text{SU}(d)$-symmetric unitaries (gates), in order to generate a $k$-design under $\text{SU}(d)$ symmetry for $k \to \infty$ in both exact and approximate senses, $\mathcal{E}$ must contain unitaries with locality $\gamma$ being \emph{at least} $\Omega(\log p(n,d))$, where $p(n,d)$ is the number of $S_n$ irreps corresponding to Young diagrams with at most $d$ rows. 
\end{theorem}
\begin{proof}
	Similar to the  proof for universality, we first assume that $\mathcal{E} = \mathcal{V}_\gamma$ with $\gamma \geq 4$ and then  decompose its Lie algebra into $\mathfrak{L}( S\mathcal{U}_\times) \oplus \mathfrak{z}_N$. By Lemma \ref{lemma:VariableChange}, we compute $T_k^{\mathcal{V}_\gamma}$ by
	\begin{align}
		\begin{aligned}
			& \int_{\gamma_j} \int_{S\mathcal{U}_\times} (e^{-i \sum_j \gamma_j c_j} V)^{\otimes k} \otimes (e^{i \sum_j \gamma_j c_j} \bar{V})^{\otimes k} dV d\gamma \\
			= & \int_{\gamma_j} (e^{-i \sum_j \gamma_j c_j})^{\otimes k} \otimes (e^{i \sum_j \gamma_j c_j} )^{\otimes k} d\gamma \int_{S\mathcal{U}_\times} V^{\otimes k} \otimes \bar{V}^{\otimes k} dV.
		\end{aligned}
	\end{align}
    To analyze the above integral, we first expand $V^{\otimes k} \otimes \bar{V}^{\otimes k}$ with respect to $S_n$ irrep blocks and then apply the Littlewood--Richardson rule and Schur orthogonality as we did in Section \ref{sec:LR}. Let 
    \begin{align}
    	(r_1,\ldots,r_{p(n,d)}, s_1,\ldots,s_{p(n,d)})
    \end{align}
    denote the orders of tensor product of unitaries $V_i$ and $\overline{V}_i$ on the block $S^{\lambda_i}$ expanded from $V^{\otimes k} \otimes \overline{V}^{\otimes k}$, so $\sum_i r_i = \sum_i s_i = k$ and $r_i, s_i$ are non-negative. Then the Lemma \ref{lemma:LR} demands the following conditions:
	\begin{align}\label{eq:LRCondition}
		d_i \vert r_i - s_i \text{ for all } i, \text{ and } r_{i_0} \neq s_{i_0} 
	\end{align}
	for at least one $i_0$. As discussed in Section \ref{sec:LR}, the integral is nonzero over $S\mathcal{U}_\times$ but vanishes over $\mathcal{U}_\times$. 
	
	Then, we move on to show that the phase integral involving $e^{-i \sum_j \gamma_j c_j}$ is also nonzero for large $k$, which implies that $T_k^{\mathcal{V}_\gamma} \neq T_k^{\mathcal{U}_{\times}}$. By definition, multiplying with the orders $r_i, s_i$ of tensor products, these phases are combined as the following matrix product
	\begin{align}\label{eq:LinearSystem}
		\begin{pmatrix} r_1 - s_1, & r_2 - s_2, & \cdots, & r_{p(n,d)} - s_{p(n,d)} \end{pmatrix} \begin{pmatrix}
			c_{11} & \cdots & c_{1, p(\gamma)} \\ c_{2 1} & \cdots & c_{2, p(\gamma)} \\ \vdots & & \vdots \\ c_{p(n,d), 1} & \cdots & c_{p(n,d), p(\gamma)} 
		\end{pmatrix},
	\end{align} 
    where the $c_i$ are written as column vectors $(c_{ij})$ from \eqref{eq:CharacterMatrix}. If each combination coefficient is zero, taking the exponential only yields the identity matrix and the phase integral ends up being nonzero. This situation happens if we can find a {nontrivial integral solution} $x_i = r_i - s_i$ such that \eqref{eq:LinearSystem} equals $0$. As a reminder, the condition $\sum_i r_i - \sum_i s_i = 0$ is included as the first equation since $c_1 = I$.  

	By Proposition \ref{prop:Integer}, the $c_{ij}$ are rational numbers, and then, by a Gaussian elimination procedure, such a solution always exists when $p(\gamma) < p(n,d)$. Let
	\begin{align}
		r_i = \begin{cases} d_1 \cdots d_{p(n,d)} \cdot x_i, & x_i > 0 \\ 0, & x_i \leq 0\end{cases},   \quad  s_i = \begin{cases} 0, & x_i > 0 \\ -d_1 \cdots d_{p(n,d)} \cdot x_i, & x_i \leq 0		\end{cases}.
	\end{align}
	This satisfies all the above conditions including \eqref{eq:LRCondition} with $k \geq d_1 \cdots d_{p(n,d)}$ and $T_k^{\mathcal{V}_\gamma} \neq T_k^{\mathcal{U}_{\times}}$.
	
	Given a generic $\text{SU}(d)$-symmetric ensemble $\mathcal{E}$ of  $\gamma$-local unitaries, it must be contained in $\mathcal{V}_\gamma$ by definition. Similarly to the proof for Proposition \ref{prop: FramePotential-Commutant}, any $M \in \operatorname{Comm}_k(\mathcal{V}_\gamma)$ is a unit eigenvector of $T_k^{\mathcal{E} \dag} T_k^{\mathcal{E}}$, regardless of whether or not $T_k^{\mathcal{E}}$ is Hermitian. Therefore, if $\mathcal{E}$ forms an $\epsilon$-approximate $k$-design for arbitrary $k$ in the sense of \eqref{def:AppDesign}, $\mathcal{V}_\gamma$ will also do by the following inclusion relation:
    \begin{align}
       \operatorname{Comm}_k(\mathcal{U}_\times) \subset \operatorname{Comm}_k(\mathcal{V}_\gamma) \subset W_{k,\mathcal{E}}^{\lambda = 1} = 	\operatorname{Comm}_k(\mathcal{U}_\times). 
    \end{align}
    In this case, according to our earlier proof, the locality of $\mathcal{V}_\gamma$, and hence of $\mathcal{E}$, is at least $\Omega(\log p(n,d))$.
\end{proof}
As a result,  we establish the following result.
\begin{corollary}
	For an $n$-qudit system, the generation of unitary $k$-designs  under $\text{SU}(d)$ symmetry for arbitrarily large $k$ requires quantum circuits of unbounded locality that grows with the total number $p(n,d)$ of $S_n$ irreps. In other words, it is impossible for circuits with any finite locality independent of the system size $n$ and the local dimension $d$ to generate unitary $k$-designs  under $\text{SU}(d)$ for  sufficiently large $k$. 
\end{corollary}


\subsection{Unitary designs with $\text{SU}(d)$ symmetry of arbitrary order from general-order CQA group}\label{sec:CQAk}

Given the no-go result for generating $k$-designs of arbitrary order with local ensembles under $\text{SU}(d)$ symmetry, the remaining question is whether they can be constructed with certain ensembles at the inevitable price of losing locality to some degree. In this subsection, we answer this by providing detailed descriptions of a recipe for achieving $k$-designs for arbitrary $k$, based on generalizing the idea of CQA. 

As clarified previously, $\mathrm{CQA} \supset S\mathcal{U}_\times$. There are still independent relative phase factors that cannot be manipulated through $\mathrm{CQA}$ or $S\mathcal{U}_\times$, which prevents these two subgroups from being universal or $k$-designs for arbitrary $k$. 
To replenish the phase factors, a direct implementation of $\Pi_\mu \vcentcolon = \frac{1}{n!} \sum_{\sigma \in S_n} \chi_\mu(\sigma) \sigma$ introduced in Theorem \ref{thm:CenterBases} demands $n!$ permutations from $S_n$ and is therefore infeasible. Based on our previous discussion in Appendix \ref{sec:CQA2}, adding permutations $\sigma \in S_n$ of different cycle types into the generating set of CQA produces more phase factors. However, as explained in Theorems \ref{Thm:CQA2} and \ref{Thm:UniversalityLocality}, there remains a significant obstacle in determining whether ${ c_\sigma }$ spans the necessary independent phases from the system. We find that the problem can be solved by considering general YJM elements. 

Let us first consider the case of qubits. When $d = 2$, the question of how to bridge the locality and the missing relative phase factors is comprehensively addressed in Ref.~\cite{MarvianSU2} and we provide an alternative approach here using YJM elements. After that, we show the formation of $\text{SU}(d)$ symmetric $k$-designs, for arbitrary large $k$, using $k$-th order YJM elements.  

\begin{theorem} \label{thm: YJM-center-basis}
	Consider $P_l = (\sum^{n}_{i = 1} X_i)^l$. The set $\{P_l\}$ with $l = 0,\ldots, \lfloor \frac{n}{2} \rfloor$ constitutes a basis of the center of $\mathbb{C}[S_n]$ restricted to the permutation representation on an $n$-qubit system with two-row Young diagrams. 
\end{theorem}
\begin{proof}
	Note that we set $l = 0,\ldots, \lfloor \frac{n}{2} \rfloor$ because there are $p(n,2) = \lfloor \frac{n}{2} \rfloor + 1$ different two-row Young diagrams or inequivalent irreps arising from the decomposition of the Hilbert space of an $n$-qubit system (see Definition \ref{def:PartitionFunction}). Accordingly, a center basis consists of $p(n,2)$ basis elements as mentioned in Theorem \ref{Thm:UniversalityLocality}. The most natural basis elements for the center are the orthonormal projections 
	\begin{align}
		\Pi_\lambda = \frac{\dim S^\lambda}{n!} \sum_{g \in S_n} \bar{\chi}_\lambda(g) \rho_\lambda(\sigma)
	\end{align}
	defined in Theorem \ref{thm:CenterBases} by $S_n$ characters. However, summing over all $n!$ elements from the symmetric group $S_n$ is undesirable.
	
	By Eq.~\eqref{eq:P_l} and \eqref{eq:P_l2}, we have
	\begin{align}\label{eq:P_lExpansion}
		P_l = \sum^{\lfloor n/2 \rfloor}_{i = 0} (\alpha_\lambda)^l \Pi_{\lambda_i},
	\end{align}
    where $\alpha_\lambda$ is the sum of all components of content vectors of $\lambda$. Then we consider the following \emph{Vandermonde matrix}: 
	\begin{align}
		V(\alpha_{\lambda_i}) = \left(\begin{array}{cccc}
			1 & 1 & \cdots & 1 \\
			\alpha_{\lambda_0} & \alpha_{\lambda_1}  & \cdots & \alpha_{\lambda_{\lfloor n/2 \rfloor}} \\
			\vdots & \vdots & \ddots & \vdots \\
			\alpha^{\lfloor n/2 \rfloor}_{\lambda_0} & \alpha^{\lfloor n/2 \rfloor}_{\lambda_1}  & \cdots & \alpha^{\lfloor n/2 \rfloor}_{\lambda_{\lfloor n/2 \rfloor}}
		\end{array}\right).
	\end{align}
	It can be easily seen that the row entries of $V(\alpha_{\lambda_i})$ are just coefficients of the linear expansion in \eqref{eq:P_lExpansion}. In other words, the matrix $V(\alpha_{\lambda_i})$ transforms $\{\Pi_{\lambda_i}\}$ to $\{P_l\}$.
	Clearly, $\{\Pi_{\lambda_i}\}$ is a basis. If the transformation $V(\alpha_{\lambda_i})$ is invertible, $\{P_l\}$ is also a valid basis. The invertibility of $V(\alpha_{\lambda_i})$ is revealed by its determinant, which equals
	\begin{align}
		V\left(\alpha_{\lambda_i}\right) = (-1)^{((\lfloor n/2 \rfloor + 1))((\lfloor n/2 \rfloor))/2} \Pi_{s < t} (\alpha_{\lambda_s} - \alpha_{\lambda_t}).
	\end{align}
Since we are considering two-row Young diagrams here, Lemma \ref{lemma: total-ordering2} asserts that the dominance relation in Definition \ref{def:Dominance} is totally ordered. Hence, $\alpha_{\lambda_s} - \alpha_{\lambda_t} \neq 0$ as long as $\lambda_s \neq \lambda_t$. Therefore, the Vandermonde matrix is invertible, confirming the statement. 
\end{proof}

The trick of using the Vandermonde matrix is also used as a crucial step in Ref.~\cite{Zheng2021SpeedingUL} to identify that the Lie algebra of CQA contains that of $S\mathcal{U}_\times$. When proving Theorem \ref{Thm:UniversalityLocality}, we mentioned that it is generally unclear how to determine a $p(n,d) \times p(n,d)$ invertible submatrix from \eqref{eq:CharacterMatrix} constituting a center basis on a generic qudit system. When $d = 2$, with the above technique involving properties of YJM elements and content vectors, we know that all cycles generated by YJM elements up to the $\lfloor n/2 \rfloor$-th order are indispensable. Consequently, the largest required locality of generators is $2\lfloor n/2 \rfloor$, leading to the following corollary.

\begin{corollary}\label{cor: universal-condition-qubits}
	On an $n$-qubit system $(d = 2)$, any $\text{SU}(2)$-symmetric generating set $S$ that achieves universality needs must contain unitaries acting on all qubits when $n$ is even, or on $n-1$ qubits when $n$ is odd.
\end{corollary}

As mentioned in Section \ref{sec:ExactDesign} in the main text, there are equivalent bases of the center leading to the same result as the above established in Ref.~\cite{MarvianSU2}.


We now address the main problem of generation of $\text{SU}(d)$-symmetric unitary $k$-designs. To this end, let us consider the following definition:

\begin{definition}
	The $k$-th order CQA group $\mathrm{CQA}^{(k)}$ is generated by incorporating YJM elements up to $k$-th order products:
	\begin{align}\label{eq:CQA-kth-order}
		\mathrm{CQA}^{(k)} = \Big\langle \exp(-i \sum_{i_1,\ldots, i_k} \beta_{i_1 \cdots i_k} X_{i_1} \cdots X_{i_k}), \quad \exp(-i\gamma H_S) \Big\rangle,
	\end{align}
	where $H_S$ is the summation of all adjacent transpositions (also see Definition \ref{def:CQA}).
\end{definition}

We now consider $k$-th order YJM elements and prove that $T_k^{\mathrm{CQA}^{(k)}} = T_k^{\mathcal{U}_\times}$. Recall that, in Appendix \ref{sec:CQA2}, we have verified that 
\begin{align}
	\begin{aligned}
		T_k^{\mathrm{CQA}} & = \int_{\gamma_j} \int_{S\mathcal{U}_\times} (e^{-i \sum_j \gamma_j c_j} V)^{\otimes k} \otimes (e^{i \sum_j \gamma_j c_j} \overline{V})^{\otimes k} dV d\gamma \\
		& = \int_{\gamma_j} (e^{-i \sum_j \gamma_j c_j})^{\otimes k} \otimes (e^{i \sum_j \gamma_j c_j} )^{\otimes k} d\gamma \int_{S\mathcal{U}_\times} V^{\otimes k} \otimes \overline{V}^{\otimes k} dV. 
	\end{aligned}
\end{align}
Rewriting the phase integral under basis $\{ P_l \}$, we obtain
\begin{align}
	\bigoplus_{\lambda_1,\ldots,\lambda_k, \mu_1,\ldots, \mu_k}\int^{2\pi}_{0} d\gamma_0 \cdots d\gamma_k \bigotimes_{i = 1,\ldots,k} e^{-i (\sum_l \gamma_l \alpha_{\lambda_i}^l)} I_{\lambda_i} \bigotimes_{j=1,\ldots,k} e^{i (\sum_l \gamma_l \alpha_{\mu_j}^l)} I_{\mu_j}.
\end{align}

We have illustrated in previous subsections that the integrals expanded from $T_k^{\mathcal{U}_\times}$ vanish unless $\{\lambda_1,\ldots, \lambda_k\} = \{\mu_1,\ldots, \mu_k\}$. Note that this encompasses the situations with equivalent copies of any $\lambda_i,\mu_j$ and any reordering on these irreps when taking tensor products. To show $T_k^{\operatorname{CQA}^{(k)}} = T_k^{\mathcal{U}_\times}$, we simply check the phase integral as we did in Appendix \ref{sec:CQA2}. The above expansion on each variable $\gamma_l$ reads $(\alpha_{\mu_0}^l + \cdots + \alpha_{\mu_k}^l) - (\alpha_{\lambda_0}^l + \cdots + \alpha_{\lambda_k}^l)$. The total integral will vanish if there exists some $l > 0$ such that the difference is nonzero (it equals zero for $\gamma_0$ since $l = 0$). To inspect the situation, let us consider the following linear equations for any given integers $p_1,\ldots,p_k$:  
\begin{align} \label{eq: sum-of-power-poly}
	\begin{aligned}
		\alpha_0 + \alpha_1 + \cdots + \alpha_k & = p_1, \\
		\alpha_0^2 + \alpha_1^2 + \cdots + \alpha_k^2 & = p_2, \\
		\vdots & \\
		\alpha_0^k + \alpha_1^k + \cdots + \alpha_k^k & = p_k.
	\end{aligned}
\end{align}
As encountered in Appendix \ref{sec:SnCommutant}, this is known as the \emph{moment problem}, and using basic algebraic geometry \cite{schmudgen2020ten}, we know that Eq.~\eqref{eq: sum-of-power-poly} admits, up to permutation, a unique solution $\{\alpha_l\}$. Since $\alpha_\lambda \neq \alpha_\mu$ for distinct two-row Young diagrams $\lambda$ and $\mu$ by Lemma \ref{lemma: total-ordering2}, this unique solution must correspond to a particular set of Young diagrams and we can conclude with the following theorem:

\begin{theorem}\label{Thm:CQAkQubit}
	For any $k$, the $k$-th order CQA group, $\mathrm{CQA}^{(k)}$, forms an exact SU(2)-symmetric unitary $k$-design. Since CQA becomes universal with $\lfloor n/2 \rfloor$-th order YJM elements, $\mathrm{CQA}^{(\lfloor n/2 \rfloor)}$ forms an exact SU(2)-symmetric $k$-design for arbitrary $k$.  
\end{theorem}


Note that the above analysis only holds for qubits. Even in the case of qutrits $d = 3$, the dominance relation ceases to be a total ordering and there are lots of examples violating the central condition $\alpha_\lambda \neq \alpha_\mu$ for $\lambda \neq \mu$ used in the preceding proofs. Even so, we are able to extend the conclusion to qudits with arbitrary $d$ by making use of YJM elements and Okounkov--Vershik approach \cite{Okounkov1996} to compare $\operatorname{Comm}_k(\mathrm{CQA})$ and $\operatorname{Comm}_k(\mathcal{U}_\times)$ as introduced in Appendices \ref{sec:SnTheory} and \ref{sec:CommutantTheory}. For intuition, we first consider the following example.

\begin{example}
	Let us understand that CQA forms a 1-design in a different way from Theorem \ref{Thm:CQA2}. By definition, CQA contains the YJM elements $X_i$. As the main result in Ref.~\cite{Okounkov1996}, Okounkov and Vershik have successfully verified that YJM elements $X_i \in \mathbb{C}[S_n]$ generate the \emph{Gelfand--Tsetlin subalgebra} $\operatorname{GZ}_n \subset \mathbb{C}[S_n]$. To put it more simply, let us consider the direct sum of all inequivalent $S_n$ irreps. Then finite linear combinations and products of the matrix representations of $X_i$, e.g.,~\eqref{eq:YJMExample}, yield arbitrary diagonal matrices on the representation space. This claim is stronger than Theorem \ref{thm:CenterBases} where phase matrices are constant scalars in each $S_n$ irrep block.
	
	Then our proof follows: Any $M \in \operatorname{Comm}_1(\mathrm{CQA})$ commutes with all group elements from $S\mathcal{U}_\times$ and, more importantly, commutes with all YJM elements, which enforces it to commute with arbitrary diagonal matrices including the phase matrices. Therefore, $\operatorname{Comm}_1(\mathrm{CQA}) = \operatorname{Comm}_1(\mathcal{U}_\times)$. 
	
	With the same assumption, let us check the situation for $k = 2$. We have 
	\begin{align}
		M \cdot (\exp(-i t X_j ) )^{\otimes 2} = (\exp(-i t X_j ) )^{\otimes 2} \cdot M
	\end{align}
	for arbitrary $j = 1,\ldots,n$ and $t \in \mathbb{R}$. By taking derivatives, we obtain
	\begin{align}
		[M, X_j \otimes I + I \otimes X_j ] =0,
	\end{align}
	where $[ \cdot\ , \cdot ]$ is the Lie bracket. The necessary condition presented above for $M \in \operatorname{Comm}_1(\mathcal{U}_\times)$ now converts to
	\begin{align}
		[M, D \otimes I + I \otimes D ] = 0,
	\end{align}
    where $D$ is an arbitrary diagonal matrix. Unfortunately, there is no guarantee that finite linear combinations and products of $\{X_j \otimes I + I \otimes X_j\}$ produces $D \otimes I + I \otimes D$. For instance,
    \begin{align}
    	(X_i \otimes I + I \otimes X_i) \cdot (X_j \otimes I + I \otimes X_j) = (X_i X_j) \otimes I + I \otimes (X_i X_j) + X_i \otimes X_j + X_j \otimes X_i
    \end{align}
    contains undesired terms like $X_i \otimes X_j + X_j \otimes X_i$. Lemma \ref{lemma:CQAk} proved below ensures that adding second-order YJM elements fixes this problem and that this generalizes to arbitrary $k > 1$.
\end{example}

\begin{lemma}\label{lemma:CQAk}
	Consider the tensor product $\mathcal{H}^{\otimes k}$ of an $n$-qudit system $\mathcal{H} = V^{\otimes n}$. Taking all matrix representations of $X_{i_1} \cdots X_{i_s}$ for  $1 \leq i_1 \leq \cdots \leq i_s \leq n$ (YJM elements are commutative with each other) and $s \leq k$ of the form
	\begin{align}
		(X_{i_1} \cdots X_{i_s}) \otimes I \otimes \cdots \otimes I + I \otimes (X_{i_1} \cdots X_{i_s}) \otimes \cdots \otimes I + I \otimes I \otimes \cdots \otimes (X_{i_1} \cdots X_{i_s}) 
	\end{align}
    is sufficient to generate $D \otimes I \otimes \cdots \otimes I + I \otimes D \otimes \cdots \otimes I + I \otimes I \otimes \cdots \otimes D$ for an arbitrary diagonal matrix $D$ on the representation space.
\end{lemma}
\begin{proof}
	Using notation from representation theory, define
	\begin{align}
		\rho(X_{i_1} \cdots X_{i_r}) \vcentcolon = (X_{i_1} \cdots X_{i_r}) \otimes I \otimes \cdots \otimes I + I \otimes (X_{i_1} \cdots X_{i_r}) \otimes \cdots \otimes I + I \otimes I \otimes \cdots \otimes (X_{i_1} \cdots X_{i_r}), 
	\end{align}
	where $\rho$ is actually the $k$-fold tensor product representation of the Lie algebra $\mathfrak{gl}(H)$. We also denote by $(X_{i_1} \cdots X_{i_r})^\alpha, \alpha = 1,2,\ldots,k$ the single term
	\begin{align}
		I \otimes \cdots \otimes I \otimes (X_{i_1} \cdots X_{i_r}) \otimes I \cdots \otimes I 
	\end{align}
	with $X_{i_1} \cdots X_{i_r}$ located at the $\alpha$-th tensor product. 
	
	When $k = 1$, obviously,
	\begin{align}\label{eq:ExampleK1}
		\rho(X_{i_1} \cdots X_{i_r}) =  \rho( X_{i_1} \cdots X_{i_{r-2}} X_{i_{r-1}} ) \rho(X_{i_r}).
	\end{align}
	When $k = 2$, it is straightforward to check that
	\begin{align}\label{eq:ExampleK2}
		\begin{aligned}
			2 \rho(X_{i_1} \cdots X_{i_r}) = &~ \rho( X_{i_1} \cdots X_{i_{r-2}} X_{i_{r-1}} ) \rho(X_{i_r}) + \rho( X_{i_1} \cdots X_{i_{r-2}} X_{i_{r}} ) \rho(X_{i_{r-1}}) \\
			& + \rho( X_{i_1} \cdots X_{i_{r-2}}) \rho(X_{i_{r-1}} X_{i_r}) - \rho( X_{i_1} \cdots X_{i_{r-2}} ) \rho(X_{i_{r-1}} ) \rho(X_{i_r}).
		\end{aligned}
	\end{align}
    So we can argue by induction that, as long as we employ all first and second-order YJM elements, $\rho(X_{i_1} \cdots X_{i_r})$ for arbitrary $r > 2$ can be generated.
    
    When $k = 3$, we first rewrite Eq.~\eqref{eq:ExampleK2} on $H^{\otimes 3}$ for $r - 1$:
    \begin{align}
    	\begin{aligned}
    		2 \rho(X_{i_1} \cdots X_{i_{r-1}})  =~& \rho( X_{i_1} \cdots X_{i_{r-3}} X_{i_{r-2}} ) \rho(X_{i_{r-1}}) + \rho( X_{i_1} \cdots X_{i_{r-3}} X_{i_{r-1}} ) \rho(X_{i_{r-2}}) \\
    		& + \rho( X_{i_1} \cdots X_{i_{r-3}}) \rho(X_{i_{r-2}} X_{i_{r-1}}) - \rho( X_{i_1} \cdots X_{i_{r-3}} ) \rho(X_{i_{r-2}} ) \rho(X_{i_{r-1}}) \\
    		& + \sum_{\alpha \neq \beta \neq \gamma} ( X_{i_1} \cdots X_{i_{r-3}} )^\alpha  (X_{i_{r-2}} )^\beta (X_{i_{r-1}} )^\gamma. 
    	\end{aligned}
    \end{align} 
    The last term emerges because tensor products with three nontrivial components are allowed in $H^{\otimes 3}$. Multiplying both sides with $\rho(X_{i_r})$ and noting that
    \begin{align}
    	\begin{aligned}
    		& \sum_{\alpha \neq \beta \neq \gamma} ( X_{i_1} \cdots X_{i_{r-3}} )^\alpha  (X_{i_{r-2}} )^\beta (X_{i_{r-1}} )^\gamma \rho(X_{i_r}) \\
    		= & \sum_{\alpha \neq \beta \neq \gamma} \Big( ( X_{i_1} \cdots X_{i_{r-3}} X_{i_r} )^\alpha  (X_{i_{r-2}} )^\beta (X_{i_{r-1}} )^\gamma + ( X_{i_1} \cdots X_{i_{r-3}} )^\alpha  (X_{i_{r-2}} X_{i_r} )^\beta (X_{i_{r-1}} )^\gamma \\
    		& + ( X_{i_1} \cdots X_{i_{r-3}} )^\alpha  (X_{i_{r-2}} )^\beta (X_{i_{r-1}} X_{i_r} )^\gamma \Big),
    	\end{aligned}
    \end{align}
    we have
    \begin{align}\label{eq:ExampleK3}
    	\begin{aligned}
    		& \rho( X_{i_1} \cdots X_{i_{r-3}} ) \rho(X_{i_{r-2}} ) \rho(X_{i_{r-1}} ) \rho(X_{i_r}) \\
    		= & \sum_{\alpha \neq \beta \neq \gamma} \Big( ( X_{i_1} \cdots X_{i_{r-3}} X_{i_r} )^\alpha  (X_{i_{r-2}} )^\beta (X_{i_{r-1}} )^\gamma + ( X_{i_1} \cdots X_{i_{r-3}} )^\alpha  (X_{i_{r-2}} X_{i_r} )^\beta (X_{i_{r-1}} )^\gamma \\
    		& + ( X_{i_1} \cdots X_{i_{r-3}} )^\alpha  (X_{i_{r-2}} )^\beta (X_{i_{r-1}} X_{i_r} )^\gamma \Big) \\
    		& + \rho( X_{i_1} \cdots X_{i_{r-3}} X_{i_{r-2}} ) \rho(X_{i_{r-1}}) \rho(X_{i_r}) + \rho( X_{i_1} \cdots X_{i_{r-3}} X_{i_{r-1}} ) \rho(X_{i_{r-2}}) \rho(X_{i_r}) \\
    		& + \rho( X_{i_1} \cdots X_{i_{r-3}}) \rho(X_{i_{r-2}} X_{i_{r-1}}) \rho(X_{i_r}) \\
    		& - 2 \rho(X_{i_1} \cdots X_{i_{r-1}}) \rho(X_{i_r}). 
    	\end{aligned}
    \end{align}
    
    Again, we rewrite Eq.~\eqref{eq:ExampleK2} on $H^{\otimes 3}$, but for $\rho( X_{i_1} \cdots X_{i_{r-3}} X_{i_r} ) \rho(X_{i_{r-2}} ) \rho(X_{i_{r-1}})$,
    \begin{align}
    	\begin{aligned}
    		2 \rho(X_{i_1} \cdots X_{i_r}) =~& \rho( X_{i_1} \cdots X_{i_{r-3}} X_{i_{r-2}} X_{i_r} ) \rho( X_{i_{r-1}}) + \rho( X_{i_1} \cdots X_{i_{r-3}} X_{i_{r-1}} X_{i_r} ) \rho(X_{i_{r-2}} )  \\
    		& + \rho( X_{i_1} \cdots X_{i_{r-3}} X_{i_r} ) \rho(X_{i_{r-2}} X_{i_{r-1}}) - \rho( X_{i_1} \cdots X_{i_{r-3}} X_{i_r} ) \rho(X_{i_{r-2}} ) \rho(X_{i_{r-1}}) \\
    		& + \sum_{\alpha \neq \beta \neq \gamma} ( X_{i_1} \cdots X_{i_{r-3}} X_{i_r} )^\alpha  (X_{i_{r-2}} )^\beta (X_{i_{r-1}} )^\gamma.
    	\end{aligned}
    \end{align}
    Recall that YJM elements commute with each other, so there is no distinction between the left-hand side term from the above equation and that from Eq.~\eqref{eq:ExampleK2}. Similarly, 
    \begin{align}
    	\begin{aligned}
    		2 \rho(X_{i_1} \cdots X_{i_r}) = &~ \rho( X_{i_1} \cdots X_{i_{r-3}} X_{i_{r-2}} X_{i_r} ) \rho( X_{i_{r-1}}) + \rho( X_{i_1} \cdots X_{i_{r-3}} X_{i_{r-1}} ) \rho( X_{i_{r-2}} X_{i_r} )  \\
    		& + \rho( X_{i_1} \cdots X_{i_{r-3}} ) \rho(X_{i_{r-2}} X_{i_{r-1}} X_{i_r} ) - \rho( X_{i_1} \cdots X_{i_{r-3}} ) \rho(X_{i_{r-2}} X_{i_r} ) \rho(X_{i_{r-1}}) \\
    		& + \sum_{\alpha \neq \beta \neq \gamma} ( X_{i_1} \cdots X_{i_{r-3}} )^\alpha  (X_{i_{r-2}} X_{i_r} )^\beta (X_{i_{r-1}} )^\gamma, \\
    		2 \rho(X_{i_1} \cdots X_{i_r}) = &~ \rho( X_{i_1} \cdots X_{i_{r-3}} X_{i_{r-1}} X_{i_r} ) \rho( X_{i_{r-2}}) + \rho( X_{i_1} \cdots X_{i_{r-3}} X_{i_{r-2}} ) \rho( X_{i_{r-1}} X_{i_r} )  \\
    		& + \rho( X_{i_1} \cdots X_{i_{r-3}} ) \rho(X_{i_{r-2}} X_{i_{r-1}} X_{i_r} ) - \rho( X_{i_1} \cdots X_{i_{r-3}} ) \rho(X_{i_{r-1}} X_{i_r} ) \rho(X_{i_{r-2}}) \\
    		& + \sum_{\alpha \neq \beta \neq \gamma} ( X_{i_1} \cdots X_{i_{r-3}} )^\alpha  (X_{i_{r-1}} X_{i_r} )^\beta (X_{i_{r-2}} )^\gamma,
    	\end{aligned}
    \end{align}
    where third-order YJM elements $\rho(X_{i_{r-2}} X_{i_{r-1}} X_{i_r} )$ and $\rho(X_{i_{r-2}} X_{i_{r-1}} X_{i_r} )$ are used for $k = 3$. Substituting these identities into Eq.~\eqref{eq:ExampleK3}, we find that it equals $\rho(X_{i_1} \cdots X_{i_r})$ when $k = 3$. Since we now have access to the tensor product representations of arbitrarily higher order products of YJM elements, an immediate application of Okounkov--Vershik theorem \cite{Okounkov1996} indicates we can generate tensor product representations of arbitrary diagonal matrices.  By induction, this statement holds for larger $k$. 
\end{proof}

Finally, we arrive at the following conclusion:

\begin{theorem}\label{Thm:CQAk}
	For a general $n$-qudit system, the $k$-th order CQA group $\mathrm{CQA}^{(k)}$ forms an exact $k$-design with $\text{SU}(d)$ symmetry, i.e., $T_k^{\mathrm{CQA}^{(k)}} = T_k^{\mathcal{U}_\times}$. 
\end{theorem}


\section{Generating unitary $k$-designs with $\text{SU}(d)$ symmetry by CQA ensembles}\label{sec:AppDesign}

Having discussed the group $k$-design properties through rigorous group representation theory, we now proceed to study the practical generation of $k$-designs under $\text{SU}(d)$ symmetry, where we desire circuits generated by local ensembles to converge to $k$-designs in an approximate sense. In the following, we define a series of ensembles $\mathcal{E}$ sampling from one-parameter subgroups or the time evolution of YJM elements (Definition \ref{def:A-CQAEnsemble}) and prove that they converge to $\text{SU}(d)$-symmetric $k$-design in the sense of Theorem \ref{thm:Approximate}. In particular, we analyze the efficiency of convergence, i.e., the scaling behavior of time (the circuit depth) needed for the convergence with respect to the system size. Mathematically, we can do so by bounding the spectral gap of $T_k^{\mathcal{E}}$ or analyzing its frame potential. However, as introduced in the main text and Appendix \ref{sec:SnCommutant}, most previously known theories and methods related to this problem (see, e.g.~Refs.~\cite{Knabe1988,Nachtergaele1996,Cirac2006,Oliveira2design2007a,Oliveira2design2007b,Znidaric2008,Harrow2design2009,Brown_2010,Gosset2016,harrow2016local,hunter2019unitary,Haferkamp2021,brian2022linear}) are invalid under $\text{SU}(d)$ symmetry. 
In the following, we present detailed discussions on this, along with several new theoretical and numerical approaches built on $S_n$ representation theory \cite{Sagan01,Goodman2009,Tolli2009}.


\subsection{CQA ensemble and its variants}\label{sec:CQAEnsemble}

We first introduce various sampling strategies that define ensembles, or circuits, of $\text{SU}(d)$-symmetric unitaries, based on the CQA model. The most basic method is to sample from the \emph{one-parameter subgroups} of transpositions, i.e.,
\begin{align}\label{eq:Circle}
	\{\exp(-it (i,j)); t \in \mathbb{R}\} \subset \mathcal{U}_\times \equiv H_\times.
\end{align}
Since the eigenvalues of any transposition $(i,j)$ are just $\pm 1$, the one-parameter subgroup is compact and isomorphic with the circle $S^1 \cong \operatorname{U}(1)$. Then, it suffices to restrict $t \in [0,2\pi]$. As a reminder, there is no guarantee that an arbitrary one-parameter subgroup is compact, e.g., 
\begin{align}
	\Big\{ \begin{pmatrix} e^{-it} & 0 \\ 0 & e^{-iat} \end{pmatrix}, t \in \mathbb{R} \Big\} \subset \operatorname{SU}(2),
\end{align}
where $a$ is an irrational number, is non-compact, and in order to sample each element from it, $t$ has to be taken from the whole $\mathbb{R}$. Analogously, we may sample from the following subgroup defined by second-order YJM elements:
\begin{align}\label{eq:Torus}
	\{\exp(-i \sum_{k,l} \beta_{kl} X_k X_l); \beta_{kl} \in \mathbb{R}\} \subset \mathcal{U}_\times  \equiv H_\times.
\end{align} 
Again, since YJM elements always have integer eigenvalues (as discussed in Appendix \ref{sec:SnTheory}) and commute with each other, the subgroup is compact and isomorphic to a certain torus as a product of $S^1 \cong \operatorname{U}(1)$. Then $\beta_{kl} \in [0,2\pi]$.

On the other hand, we may sample local $\text{SU}(d)$-symmetric unitaries supported on arbitrary $\gamma$ qudits in an abstract sense. By Schur--Weyl duality, when $\gamma = 2$, this sampling method is identical to that in \eqref{eq:Circle}. Later, we illustrate how to sample some 4-local unitaries explicitly, as the CQA model is generated by 4-local unitaries. 

\begin{definition}\label{def:A-CQAEnsemble}
	The \emph{CQA random walk ensemble} is defined as follows in each step of the random walk on the quantum circuit (see the Remark in Appendix \ref{sec:RandomWalk}):
	\begin{enumerate}
		\item We first sample an element parametrized as $\exp(-i\sum_{k,l} \beta_{kl} X_k X_l)$ by second-order YJM elements with $\beta_{kl} \in [0,1]$.
		
		\item Then, we randomly select an integer $j \in \{1,\ldots,n-1\}$ and then sample an element from the one-parameter subgroup determined by $\exp(-it(j,j+1))$ with $t \in [0, 2\pi]$. 
		
		\item We sample again from the subgroup defined by $\exp(-i\sum_{k,l} \beta_{kl} X_k X_l)$.
	\end{enumerate}
\end{definition}

\begin{definition}\label{def:A-V4CQAEnsemble}
	In each step of the random walk, we arbitrarily select four integers $i_1,i_2,i_3$, and $i_4$ from $[n]$ and then uniformly sample a 4-local $\text{SU}(d)$-symmetric unitary supported on qudits labeled by these integers. This assembles the \emph{CQA local random unitary ensemble $\mathcal{E}_{\mathcal{V}_4}$}.
\end{definition}

Based on discussions from Appendix \ref{sec:A-ExactDesign}, it is natural to anticipate that these two ensembles converge to unitary $k$-designs under $\text{SU}(d)$ symmetry for all $k < n(n-3)/2$. We are going to present the proof details. Moreover, in Section \ref{sec: cqa-convergence}, we exemplify our numerical methods on the convergence time of a 2-local ensemble on qubits ($d = 2$), we also provide a comprehensive definition here before formally elucidating our results. 

\begin{definition}\label{def:SWAPEnsemble}
	Forgetting the YJM elements, the \emph{SWAP ensemble} $\mathcal{E}_{\operatorname{eSWAP}}$ in each step of the random walk only samples $\exp(-it(j,j+1))$ as a \emph{1D chain} in the language of quantum many body theory or samples $\exp(-it(i,j))$ with arbitrary $1 \leq i,j \leq n$, known as \emph{all-to-all interaction random circuits}. 
\end{definition}  

\begin{lemma}\label{lemma:Ensemble1}
	The $k$-fold channel $T_k^{\mathcal{E}_{\mathrm{CQA}}}$ defined in each step of the CQA ensemble satisfies
	\begin{align}
		T_k^{\mathcal{E}_{\mathrm{CQA}}} = T_k^{\mathrm{YJM}} \frac{1}{n-1} (T_k^{(1,2)} + \cdots + T_k^{(n-1,n)}) T_k^{\mathrm{YJM}},
	\end{align}
    where $T_k^{\mathrm{YJM}}$ and $T_k^{(j,j+1)}$ are the $k$-fold channels twirled over tori in \eqref{eq:Torus} and circles in \eqref{eq:Circle} respectively. The operator  $T_k^{\mathcal{E}_{\mathrm{CQA}} }$ is positive semidefinite with eigenvalues bounded in the interval $[0,1]$. Similar results hold for $T_k^{\mathcal{E}_{\mathcal{V}_4}}$ and $T_k^{\mathcal{E}_{\operatorname{eSWAP}}}$.
\end{lemma}
\begin{proof}
	By definition,
	\begin{align}
		T_k^{\mathcal{E}_{\mathrm{CQA}}}(M) & = \int_{\mathcal{E}_{\mathrm{CQA}}} V^{\otimes k} M V^{\dagger \otimes k} dV \notag \\
		& = \sum_{j = 1}^{n-1} \frac{1}{n-1} \int \Big(V_{\mathrm{YJM}} V_{(j,j+1)} V_{\mathrm{YJM}} \Big)^{\otimes k} M \Big(V_{\mathrm{YJM}} V_{(j,j+1)} V_{\mathrm{YJM}} \Big)^{\dagger \otimes k} dV_{\mathrm{YJM}} dV_{(j,j+1)} \\
		& = (T_k^{\mathrm{YJM}} \frac{1}{n-1} (T_k^1 + \cdots + T_k^{n-1}) T_k^{\mathrm{YJM}}) (M) \notag.
	\end{align}
	Since $T_k^{(i,j)}, T_k^{\mathrm{YJM}}$ are twirled over compact subgroups, they are Hermitian projectors (see Appendix \ref{sec:k-channel}). Then by basic linear algebra, $T_k^{\mathcal{E}_{\mathrm{CQA}}}$ is positive semidefinite and its largest eigenvalue is  $\leq 1$.
\end{proof}

\begin{lemma}\label{lemma:Ensemble2}
	Let $\operatorname{Comm}_k(\mathrm{YJM})$ denote the commutant of the torus defined in \eqref{eq:Torus} with $\operatorname{Comm}_k(i,j)$ defined analogously. The following identity of the commutant holds for any $k$:
	\begin{align}
		& \operatorname{Comm}_k(\mathrm{YJM}) \cap \bigcap_{1 \leq j \leq n-1} \operatorname{Comm}_k(j,j+1) = \operatorname{Comm}_k(\mathrm{CQA}), \\
		& \bigcap_{1 \leq j \leq n-1} \operatorname{Comm}_k(j,j+1) = \bigcap_{1 \leq i < j \leq n} \operatorname{Comm}_k(i,j) = \operatorname{Comm}(\operatorname{eSWAP}). 
	\end{align}
\end{lemma}
\begin{proof}
	The statement holds trivially for $k = 1$ by the definitions of CQA and eSWAP (see the Remark in Appendix \ref{sec:CQA2}). For $k = 2$, since any unitary $U \in \mathrm{CQA}$ is a finite product with the form of 
	\begin{align} 
		\exp(-i \sum_{k,l} \beta_{kl} X_k X_l ) \exp(-i\gamma \sum_j (j,j+1)) \cdots
		\exp(-i \sum_{k,l} \beta'_{kl} X_k X_l) \exp(-i\gamma' \sum_j (j,j+1) ),
	\end{align}
	we have
	\begin{align}
		\begin{aligned}
			U \otimes U = (\exp(-i \sum_{k,l} \beta_{kl} X_k X_l ) )^{\otimes 2} & (\exp(-i\gamma \sum_j (j,j+1)) )^{\otimes 2} \cdots \\
			\cdots & (\exp(-i \sum_{k,l} \beta'_{kl} X_k X_l) )^{\otimes 2} (\exp(-i\gamma' \sum_j (j,j+1) ))^{\otimes 2}.
		\end{aligned}
	\end{align}
    By the identity 
    \begin{align}
    	(\exp A) \otimes (\exp B) = \exp( A\otimes I + I \otimes B),
    \end{align}
    for any $M \in \operatorname{Comm}_2(1,2) \cap \cdots \cap \operatorname{Comm}_2(n-1,n) \cap \operatorname{Comm}_2(\mathrm{YJM})$, proving that $M$ commutes with $U \otimes U$ for any $U \in \mathrm{CQA}$ is equivalent to showing that $M$ commutes with $A \otimes I + I \otimes A$ with $A$ taken from the Lie algebra of $\mathrm{CQA}$ (a similar trick is also used in Theorem \ref{Thm:CQAk}).
	
	By assumption,
	\begin{align}
		[M, (j,j+1) \otimes I + I \otimes (j,j+1)] = 0, \quad  [M, X_k X_l \otimes I + I \otimes X_k X_l] = 0.
	\end{align}
	On the other hand, due to the definition of the tensor product representation, the following identity holds for any matrices $A,B \in \operatorname{End}(V)$: 
	\begin{align}
		[A \otimes I + I \otimes A, B \otimes I + I \otimes B] = [A,B] \otimes I + I \otimes [A,B].
	\end{align}
	Therefore, $M$ commutes with the tensor product representation of transpositions and YJM elements, which span the Lie algebra of $\mathrm{CQA}$, and hence the proof follows. The converse direction is obvious.
	
	For eSWAP, we have the following argumentation:
	\begin{align}
		\begin{aligned}
			& [M, (i,i+1) \otimes I + I \otimes (i,i+1)] = 0  \\
			\implies\quad & [M, \big( (i,i+1) \otimes I + I \otimes (i,i+1) \big)^2] = 2[M, (i,i+1) \otimes (i,i+1) + I \otimes I] = 0 \\
			\implies\quad & [M, (j,j+1) \otimes (j,j+1)] = 0  \\
			\implies\quad & [M, (j,j+2) \otimes I + I \otimes (j,j+2)]  \\
			&=  [M, (j+1,j+2) \otimes (j,j+1) \big( (j,j+1) \otimes I + I \otimes (j,j+1) \big) (j,j+1) \otimes (j+1,j+2)] = 0.
		\end{aligned}
	\end{align}
	Inductively, we derive that $M$ commutes with $(i,j) \otimes I + I \otimes (i,j)$ for any $i \neq j$. The case for general $k$ can be demonstrated similarly.
\end{proof}

\begin{theorem}\label{thm:Approximate}
	For all $k$, the ensemble $\mathcal{E}_{\mathrm{CQA}}$ forms approximate $k$-designs with respect to the group CQA. Analogously, for all $k$, $\mathcal{E}_{\mathcal{V}_4}$ and $\mathcal{E}_{\operatorname{eSWAP}}$ forms approximate $k$-designs with respect to the group $\mathcal{V}_4$ and eSWAP respectively.
\end{theorem}
\begin{proof}
	As we carefully introduce in the in Section \ref{sec: k-designs-intro}  and Appendix \ref{sec:RandomWalk}, it is only when $T_k^{\mathcal{E}_{\mathrm{CQA}}}$ and $T_k^{\mathrm{CQA}}$ share the same unit eigenspace that $\mathcal{E}_{\mathrm{CQA}}$ can converge to unitary $k$-designs with respect to CQA under multiple actions of  the ensemble. For compact groups such as CQA, the unit eigenspace of $T_k^{\mathrm{CQA}}$ just the commutant algebra $\operatorname{Comm}_k(\mathrm{CQA})$ by the invariance of Haar measure, but it has no specific name for a general ensemble like $\mathcal{E}_{\mathrm{CQA}}$. We simply denote that unit eigenspace by $W_{k,\mathcal{E}_{\mathrm{CQA}}}^{\lambda = 1}$ and, by definition $W_{k,\mathcal{E}_{\mathrm{CQA}}}^{\lambda = 1}$ contains $\operatorname{Comm}_k(\mathcal{E}_{\mathrm{CQA}})$. Then, we need to demonstrate the converse inclusion:
	\begin{align}
		\operatorname{Comm}_k(\mathrm{CQA}) \supset W_{k,\mathrm{CQA}}^{\lambda = 1}.
	\end{align}
	To this end, let $M$ be any (super-)operator such that $T_k^{\mathcal{E}_{\mathrm{CQA}}} (M) = M$. That is, $M \in W_{k,\mathcal{E}_{\mathrm{CQA}}}^{\lambda = 1}$. By Lemma \ref{lemma:Ensemble1}, 
	\begin{align}
		T_k^{\mathrm{YJM}} (M) = T_k^{\mathrm{YJM}} T_k^{\mathcal{E}_{\mathrm{CQA}}} (M) = (T_k^{\mathrm{YJM}})^2 \frac{1}{n-1} (T_k^{(1,2)} + \cdots + T_k^{(n-1,n)}) T_k^{\mathrm{YJM}} (M) = T_k^{\mathcal{E}_{\mathrm{CQA}}} (M) = M, 
	\end{align}
	which means that $M \in \operatorname{Comm}_k(\mathrm{YJM})$. Then,
	\begin{align}
		T_k^{\mathcal{E}_{\mathrm{CQA}}} (M) = T_k^{\mathrm{YJM}} \frac{1}{n-1} \frac{1}{n-1} (T_k^{(1,2)} + \cdots + T_k^{(n-1,n)}) T_k^{\mathrm{YJM}} (M) = T_k^{\mathrm{YJM}} \frac{1}{n-1} (T_k^{(1,2)} + \cdots + T_k^{(n-1,n)}) (M).  
	\end{align}
	Assume that there is a certain $i_0 \in \{1,\ldots,n-1\}$ such that $M \notin \operatorname{Comm}_k(i_0,i_0+1)$. Since $T_k^{(i_0,i_0+1)}$ is a Hermitian projector, its operator norm is exactly 1 and then
	\begin{align}
		\frac{1}{n-1} \Big\Vert (T_k^{(1,2)} + \cdots + T_k^{(n-1,n)} ) (M) \Big\Vert \leq \frac{1}{n-1} \sum_{j = 1}^{n-1} \Vert T_k^{j,j+1} (M) \Vert < \Vert M \Vert,
	\end{align}
	which contradicts the fact that $T_k^{\mathcal{E}_{\mathrm{CQA}}} (M) = M$. Therefore, 
	\begin{align}
		M \in \operatorname{Comm}_k(1,2) \cap \cdots \cap \operatorname{Comm}_k(n-1,1) \cap \operatorname{Comm}_k(\mathrm{YJM}) = \operatorname{Comm}_k(\mathrm{CQA})
	\end{align}
    by Lemma \ref{lemma:Ensemble2}. Since $T_k^{\mathcal{E}_{\mathrm{CQA}}}$ is positive semidefinite, any other non-unit eigenvalues are strictly bounded within $[0,1)$. As we have instantiated in \eqref{eq:EigenvaluesExample}, $\mathcal{E}_{\mathrm{CQA}}$ is able to converge to unitary $k$-designs with respect to CQA. 
\end{proof}

As a reminder, the above inequality may not hold for non-Hermitian projectors like
	\begin{align}
		Pv = \begin{pmatrix} 1 & 0 \\ 1 & 0 	\end{pmatrix} \begin{pmatrix} 1 \\ 0 \end{pmatrix} = \begin{pmatrix} 1 \\ 1 \end{pmatrix}.
	\end{align}
The norm of the image can even increase. On the other hand, the property 
\begin{align}
    M \in W_{k,\mathrm{CQA}}^{\lambda = 1} \Leftrightarrow M \in \operatorname{Comm}_k(1,2) \cap \cdots \cap \operatorname{Comm}_k(n-1,1) \cap \operatorname{Comm}_k(\mathrm{YJM})
\end{align}
can be understood as a counterpart to the notion of \emph{frustration-freeness} in quantum many-body physics \cite{Tasaki2020}, which has also been widely used in the previous study of random circuits without symmetry \cite{Znidaric2008,Brown_2010,harrow2016local,Haferkamp2021,harrow2023approximate}. Together with Theorem \ref{Thm:CQA2} and \ref{Thm:CQAk}, we conclude the following.

\begin{corollary}
	For an $n$-qudit system with $n \geq 9$ and $d < n$, the ensemble $\mathcal{E}_{\mathrm{CQA}}$, as well as $\mathcal{E}_{\mathcal{V}_4}$, can converge to $k$-design to with respect to $\mathcal{U}_\times$ on general qudits for all $k < n(n-3)/2$. In the case of qubits ($d = 2$), the ensemble $\mathcal{E}_{\operatorname{eSWAP}}$ can converge to $k$-designs for $k < n(n-3)/2$.
\end{corollary} 

\begin{remark}
	It is always desirable to use $\mathcal{E}_{\operatorname{eSWAP}}$ as it involves 2-local unitaries and most of our numerical computations are based on $\mathcal{E}_{\operatorname{eSWAP}}$. However,  $\mathcal{E}_{\operatorname{eSWAP}}$ cannot generate unitary $k$-designs for $d > 2$ and $k > 1$ \cite{MarvianSUd}. We now introduce how to identify counterexamples explicitly by $S_n$ representation theory, which may provide a deep understanding on this issue.
	
	In Refs.~\cite{Marin1,Marin2}, it is indicated that eSWAPs cannot generate $\text{SU}(S^\lambda)$ if $\lambda = (\lambda_i)$ is not a hook, i.e., $\lambda_2 > 1$ and equals its conjugate (see Appendix \ref{sec:SnTheory}, such cases only happens when $d > 2$). For instance, we take 
	\begin{align}
		\lambda_0 = \ytableausetup{boxsize=1.25em} \ydiagram{3,2,1} = \lambda_0',
	\end{align}
	which uniquely corresponds to an $S_6$ irrep (with copies) from the six-qudit system. Recall that, in Appendix \ref{sec:CQA2}, we compare $T_k^{\mathcal{U}_\times}$ and $T_k^{\mathrm{CQA}}$ by expanding those tensor products $V^{\otimes k} \otimes \bar{V}^{\otimes k}$ with respect to $S_n$ irreps and then taking integrals. 
	
	Similarly, we now expand $T_k^{\mathcal{E}_{\operatorname{eSWAP}}}$ by different irrep blocks. In particular, we can count the number of unit eigenvalues of $T_k^{\mathcal{E}_{\operatorname{eSWAP}}}$ when restricted to $S^{\lambda_0 }$ through the following integral:
	\begin{align}\label{eq:321couterexample}
		\frac{1}{2\pi} \int_0^{2\pi} (e^{-it(1,2)} \big\vert_{S^{\lambda_0}})^{\otimes k} \otimes (e^{it(1,2)} \big\vert_{S^{\lambda_0}})^{\otimes k} dt.
	\end{align}
	For instance, when $k = 2$, we compute by using Young orthogonal form that the above matrix has {three} unit eigenvalues. On the other hand, it is straightforward to see, by Theorem \ref{Thm:Commutant}, that the corresponding expanded term from $T_k^{\mathcal{U}_\times}$,
	\begin{align}
		\int_{\operatorname{U}(S^{\lambda_0})} (U_{S^{\lambda_0}})^{\otimes 2} \otimes (\overline{U}_{S^{\lambda_0}})^{\otimes 2} d(U_{S^{\lambda_0}}),
	\end{align}
	only has {two} unit eigenvalues (see Eq.~\eqref{eq:GeneralPermutation}), resembling the commutant of a 2-design without symmetry. Therefore, $(T_2^{\mathcal{E}_{\operatorname{eSWAP}}})^p$ can never converge to $T_k^{\mathcal{U}_\times}$ because it has strictly more unit eigenvalues. Taking $T_2^{\mathrm{YJM}} T_2^{\mathcal{E}_{\operatorname{eSWAP}}} T_2^{\mathrm{YJM}}$ by $T_2^{\mathrm{YJM}}$ resolves this problem and makes the eigenvectors coincide. Other counterexamples can be found in a similar fashion on Young diagrams like $\lambda_0$, and a comprehensive explanation for the numerical computation of the eigensystem of \eqref{eq:321couterexample} is given in Appendix \ref{sec:Approximate}.
\end{remark}


\subsection{Several attempts to evaluate the convergence speed of $\mathcal{E}_{\mathrm{CQA}}$ }\label{sec:Approximate}

Eventually, we would like to understand how quickly the aforementioned ensembles converge to an $\epsilon$-approximate SU($d$)-symmetric $k$-design. This is closely related to the phenomenon of \emph{scrambling} that has been widely studied in quantum many-body dynamics \cite{Znidaric07:2,Znidaric2008,Liu_2018,junyu2017chaos,junyu2020chargescrambler}. We will now consider the spectral gap method for $k = 1$ and $k = 2$, and then discuss the fundamental difficulties associated with the general case.

Since any transposition or SWAP $\tau = (i,j) \in S_n$ satisfies $\tau^2 = I$, its time evolution can be expanded by the Euler identity
\begin{align}
	e^{-i\theta \tau} = \cos\theta I - i\sin\theta \tau. 
\end{align} 
Then
\begin{align}\label{eq:EulerIntegral1}
	T_{k = 1}^\tau & = \frac{1}{2\pi} \int_0^{2\pi} e^{-i\theta \tau} \otimes e^{i\theta \tau} d\theta = \frac{1}{2\pi}  \int_0^{2\pi} (\cos\theta I - i\sin\theta \tau) \otimes (\cos\theta I + i\sin\theta \tau) d\theta = \frac{1}{2} (I \otimes I + \tau \otimes \tau),
\end{align} 
and
\begin{align}\label{eq:EulerIntegral2}
	\begin{aligned}
		T_{k = 2}^\tau & = \frac{1}{2\pi} \int_0^{2\pi} (e^{-i\theta \tau})^{\otimes 2} \otimes (e^{i\theta \tau})^{\otimes 2} d\theta = \frac{1}{2\pi} \int_0^{2\pi} (\cos\theta I + i\sin\theta \tau)^{\otimes 2} \otimes (\cos\theta I - i\sin\theta \tau)^{\otimes 2} d\theta  \\
		= & ~\frac{1}{2\pi} \int_0^{2\pi} \Big(\cos^4\theta I \otimes I \otimes I \otimes I + \sin^4\theta \tau \otimes \tau \otimes \tau \otimes \tau + \cos^2\theta \sin^2\theta \big(I \otimes \tau \otimes I \otimes \tau + I \otimes \tau \otimes \tau \otimes I \\
		& + \tau \otimes I \otimes I \otimes \tau + \tau \otimes I \otimes \tau \otimes I - I \otimes I \otimes \tau \otimes \tau - \tau \otimes \tau \otimes I \otimes I \big) \Big) d\theta \\
		= & ~\frac{1}{8} (3 IIII + 3\tau\tau\tau\tau + I\tau I\tau + I\tau\tau I + \tau I I\tau + \tau I\tau I - I I \tau\tau - \tau\tau I I ),
	\end{aligned}
\end{align} 
where we have omitted the tensor product notation in the final line for conciseness. The expansions of $T_k^\tau$ for general $k$ can be derived using $\sin^2\theta + \cos^2\theta = 1$ and
\begin{align}
	& \frac{1}{2\pi} \int_0^{2\pi} \cos^{2k}\theta d\theta = \frac{1}{2\pi} \int_0^{2\pi} \sin^{2k}\theta d\theta = \frac{1}{2^{2k}} \binom{2k}{k}, \\
	& \frac{1}{2\pi} \int_0^{2\pi} \cos^{2k-1}\theta \sin\theta d\theta = \frac{1}{2\pi} \int_0^{2\pi} \cos\theta \sin^{2k-1}\theta d\theta = 0.
\end{align} 

We now explain various methods that we have considered. To begin with, we introduce our numerical computations for the magnitude of the second largest eigenvalue of $T_{k = 1}^{\mathcal{E}_{\mathrm{CQA}}}$ to understand the convergence to 1-design, and then elucidate the main mathematical obstacles when tackling general unitary $k$-designs under $\text{SU}(d)$ symmetry.


\subsubsection*{Methods from quantum many-body theory}
As in the proof of Lemma \ref{lemma:Ensemble2}, Eq.~\eqref{eq:EulerIntegral1} indicates that
\begin{align}
	M \in \operatorname{Comm}_1(\mathcal{E}_{\operatorname{eSWAP}}) \subset \operatorname{End}(V^{\otimes n})
\end{align}
commutes with all adjacent transpositions $\tau = (j,j+1)$ and hence commutes with the group $S_n$. Therefore, it is sufficient to employ 
\begin{align}
	T_k^{\mathcal{E}_{\operatorname{eSWAP}}} = \frac{1}{n} (T_k^{(1,2)} + \cdots + T_k^{(n-1,n)} + T_k^{(1,n)}) 
\end{align}
to generate 1-designs for either qubits or general qudits. Obviously, determining the second largest eigenvalue of the positive semidefinite operator $T_k^{\mathcal{E}_{\operatorname{eSWAP}}}$ is equivalent to determining the \emph{spectral gap} of 
\begin{align}\label{eq:A-1DHamiltonian}
	H \vcentcolon = n(I - T_k^{\mathcal{E}_{\operatorname{eSWAP}}} ) = \sum_i (I -  T_k^{(i,i+1)} ) = \sum_i P_i,
\end{align}
which is further tantamount to finding some $\gamma = \Delta(H) > 0$ such that
\begin{align}
	H^2 \geq \gamma H.
\end{align} 
As introduced in Section IV.B of the main text, this observation leads us to apply a classical method due to Knabe \cite{Knabe1988}, which was originally devised to estimate the spectral gap of 1D quantum spin chains with {periodic boundary conditions}.  As a recap, let us define the \emph{bulk Hamiltonian} consisting of all $P_j,\ldots,P_{m+j-1}$ terms from Eq.~\eqref{eq:A-1DHamiltonian}:
\begin{align}\label{eq:bulk}
	h_{m,j} = \sum_{i = j}^{m + j - 1} P_i.
\end{align}
The improved Knabe local gap bound \cite{Gosset2016} indicates that 
\begin{align}\label{eq:A-KnabeBound}
	 \Delta(H) \geq \frac{5(m^2 + 3m + 2)}{6(m^2 + 2m -3)} \Big( \Delta(h_{m,j}) - \frac{6}{(m+1)(m+2)} \Big).
\end{align}
To obtain a valid lower bound on the gap, we need to find a certain $m = 2,3,\ldots$ such that $\Delta(h_{m,j}) > 6/{(m+1)(m+2)}$. 

Suppose that $m = 2$. It then suffices to compute the gap of 
\begin{align}
	P_1 + P_2 = 2I - T_{k = 1}^{(1,2)} - T_{k = 1}^{(2,3)} = I - \frac{1}{2} \Big( (1,2)^{\otimes 2} + (2,3)^{\otimes 2} \Big).
\end{align}
by the permutation invariance of $h_{2,j}$. We present an observation that is simple but crucial for constructing our computational method:

\begin{observation}
	The symmetric group $S_3$ only admits the 1-dimensional trivial and sign irreps and the 2-dimensional standard irrep labeled by $(3), (1^3)$, and $(2,1)$, respectively. The Young orthogonal forms (see \eqref{eq:YoungOrthogonal}) of $\tau = (1,2)$ and $(2,3)$ on the direct sum of all these irreps are as follows:
	\begin{align}\label{eq:YoungOrthogonalExample}
		\renewcommand\arraystretch{1.25}
		(1,2) = \begin{pmatrix} 1 & 0 & 0 & 0 \\ 0 & 1 & 0 & 0 \\ 0 & 0 & -1 & 0 \\ 0 & 0 & 0 & -1 \end{pmatrix}, \quad  (2,3) = \begin{pmatrix} 1 & 0 & 0 & 0 \\ 0 & -\frac{1}{2} & \frac{\sqrt{3}}{2} & 0 \\ 0 & \frac{\sqrt{3}}{2} & \frac{1}{2} & 0 \\ 0 & 0 & 0 & -1 \end{pmatrix}.
	\end{align}
    We argue in Section IV.B of the main text that according to the $S_n$ branching rule \cite{Fulton1997,Sagan01,Goodman2009,Tolli2009}, for arbitrary $n \geq 3$ and $j = 1,\ldots,n$, $\Delta(h_{2,j})$ is equal to the gap of
    \begin{align}
    	\renewcommand\arraystretch{1.25}
    	 I - \frac{1}{2} \left( \begin{pmatrix} 1 & 0 & 0 & 0 \\ 0 & 1 & 0 & 0 \\ 0 & 0 & -1 & 0 \\ 0 & 0 & 0 & -1 \end{pmatrix}^{\otimes 2} + \begin{pmatrix} 1 & 0 & 0 & 0 \\ 0 & -\frac{1}{2} & \frac{\sqrt{3}}{2} & 0 \\ 0 & \frac{\sqrt{3}}{2} & \frac{1}{2} & 0 \\ 0 & 0 & 0 & -1 \end{pmatrix}^{\otimes 2} \right).
    \end{align}  
\end{observation}

Exact diagonalization yields $\Delta(h_{2,j}) = {1}/{2}$, in which case the Knabe bound \eqref{eq:A-KnabeBound} is not applicable. Therefore, we proceed to test the case for $m = 3$ with the following Young orthogonal forms of direct sums of $S_4$ irreps:
\begin{align}
	\renewcommand\arraystretch{1.3}
	\newcommand\scalemath[2]{\scalebox{#1}{\mbox{\ensuremath{\displaystyle #2}}}} \scalemath{0.7}{
		(1,2) = \begin{pmatrix} 1 & 0 & 0 & 0 & 0 & 0 & 0 & 0 & 0 & 0 \\
			0 & 1 & 0 & 0 & 0 & 0 & 0 & 0 & 0 & 0 \\
			0 & 0 & -1 & 0 & 0 & 0 & 0 & 0 & 0 & 0 \\
			0 & 0 & 0 & 1 & 0 & 0 & 0 & 0 & 0 & 0 \\
			0 & 0 & 0 & 0 & -1 & 0 & 0 & 0 & 0 & 0 \\
			0 & 0 & 0 & 0 & 0 & 1 & 0 & 0 & 0 & 0 \\
			0 & 0 & 0 & 0 & 0 & 0 & 1 & 0 & 0 & 0 \\
			0 & 0 & 0 & 0 & 0 & 0 & 0 & -1 & 0 & 0 \\
			0 & 0 & 0 & 0 & 0 & 0 & 0 & 0 & -1 & 0 \\
			0 & 0 & 0 & 0 & 0 & 0 & 0 & 0 & 0 & -1  \end{pmatrix},
		(2,3) = \begin{pmatrix} 1 & 0 & 0 & 0 & 0 & 0 & 0 & 0 & 0 & 0 \\
			0 & -\frac{1}{2} & \frac{\sqrt{3}}{2} & 0 & 0 & 0 & 0 & 0 & 0 & 0 \\
			0 & \frac{\sqrt{3}}{2} & \frac{1}{2} & 0 & 0 & 0 & 0 & 0 & 0 & 0 \\
			0 & 0 & 0 & 1 & 0 & 0 & 0 & 0 & 0 & 0 \\
			0 & 0 & 0 & 0 & \frac{1}{2} & \frac{\sqrt{3}}{2} & 0 & 0 & 0 & 0 \\
			0 & 0 & 0 & 0 & \frac{\sqrt{3}}{2} & -\frac{1}{2} & 0 & 0 & 0 & 0 \\
			0 & 0 & 0 & 0 & 0 & 0 & -\frac{1}{2} & \frac{\sqrt{3}}{2} & 0 & 0 \\
			0 & 0 & 0 & 0 & 0 & 0 & \frac{\sqrt{3}}{2} & \frac{1}{2} & 0 & 0 \\
			0 & 0 & 0 & 0 & 0 & 0 & 0 & 0 & -1 & 0 \\
			0 & 0 & 0 & 0 & 0 & 0 & 0 & 0 & 0 & -1 	\end{pmatrix},  
		(3,4) = \begin{pmatrix}
			1 & 0 & 0 & 0 & 0 & 0 & 0 & 0 & 0 & 0 \\
			0 & 1 & 0 & 0 & 0 & 0 & 0 & 0 & 0 & 0 \\
			0 & 0 & -1 & 0 & 0 & 0 & 0 & 0 & 0 & 0 \\
			0 & 0 & 0 & -\frac{1}{3} & 0 & \frac{2 \sqrt{2}}{3} & 0 & 0 & 0 & 0 \\
			0 & 0 & 0 & 0 & 1 & 0 & 0 & 0 & 0 & 0 \\
			0 & 0 & 0 & \frac{2 \sqrt{2}}{3} & 0 & \frac{1}{3} & 0 & 0 & 0 & 0 \\
			0 & 0 & 0 & 0 & 0 & 0 & -1 & 0 & 0 & 0 \\
			0 & 0 & 0 & 0 & 0 & 0 & 0 & -\frac{1}{3} & \frac{2 \sqrt{2}}{3} & 0 \\
			0 & 0 & 0 & 0 & 0 & 0 & 0 & \frac{2 \sqrt{2}}{3} & \frac{1}{3} & 0 \\
			0 & 0 & 0 & 0 & 0 & 0 & 0 & 0 & 0 & -1 
	\end{pmatrix} }
\end{align}
Note that these matrices are made up from submatrices defined on inequivalent $S_m$ irreps. Therefore, we can do the computation efficiently  using tensor products of submatrices.

For larger $m$, although we can expand the tensor product by $S_m$ irreps, it immediately becomes hard to diagonalize $h_{m,j}$ because the dimension of a single irrep may scale exponentially (see \eqref{eq:dim-Comparision} and \eqref{eq:dim-Comparision2}), let alone its 4-fold tensor product that appears in the 2-design analysis. Instead of finding the complete spectra, we examine Eq.~\eqref{eq:EulerIntegral2}, where $\tau$ are restricted to {part of} the $S_n$ irreps with moderate dimensions such as those assigned on $S^{(m)} \otimes S^{(m-1,1)} \otimes S^{(m-2, 2)} \otimes S^{(m-1,1)}$. Inevitably, as reported in the numerical results in Section \ref{sec: cqa-convergence}, there are eigenvalues smaller than ${6}/{(m+1)(m+2)}$, indicating that  $\Delta(h_{m,j})$ could not surpass the local gap threshold for larger $m$. The decay of $\Delta(h_{m,j})$ also provides evidence against the feasibility of using \emph{Nachtergaele's martingale method} \cite{Nachtergaele1996} which, detects constant-gapped (an $\Omega(n^{-1})$ gap after normalization) 1D quantum spin chains.

\begin{remark}
	One may still ask whether Knabe's local gap bound applies to the CQA ensemble on general qudits with
	\begin{align}
		T_k^{\mathcal{E}_{\mathrm{CQA}}} = T_k^{\mathrm{YJM}} \Big( \frac{1}{n-1} \sum_{1 \leq j \leq n-1} T_k^{\tau_j} \Big) T_k^{\mathrm{YJM}}.
	\end{align}
	The answer is the negative because we cannot define a bulk Hamiltonian $h_{m,j}$, such as Eq.~\eqref{eq:bulk}, that is exactly supported on $m$ sites of the system. Given $\tau_1 = (i,i+1), \tau_2 = (j,j+1)$ with $i+1 < j$, due to the intertwining with $T_k^{\mathrm{YJM}}$, $T_k^{\mathrm{YJM}} T_k^{\tau_1} T_k^{\mathrm{YJM}}$ and $T_k^{\mathrm{YJM}} T_k^{\tau_2} T_k^{\mathrm{YJM}}$ no longer commute.
	
	To clarify this, let us restrict ourselves to the case when $k = 2$ on a simple tensor product like $S^{\lambda \otimes 4}$ with Young basis element $\ket{\alpha_T}$ (we temporarily omit the multiplicity index). Let $\ket{\alpha_{T_r}}, r = 1,2,3,4$ be basis vectors where $\tau_1$ and $\tau_2$ act according to the following orthogonal form:
	\begin{align}
		& (i,i+1) \ket{\alpha_{T_1}} = \frac{1}{r_1} \ket{\alpha_{T_1}} + \sqrt{1 - \frac{1}{r_1^2}} \ket{\alpha_{T_2}}, \quad (i,i+1) \ket{\alpha_{T_2}} = \frac{1}{r_1} \ket{\alpha_{T_2}} + \sqrt{1 - \frac{1}{r_1^2}} \ket{\alpha_{T_3}}, \\
		& (j,j+1) \ket{\alpha_{T_1}} = \frac{1}{r_2} \ket{\alpha_{T_1}} + \sqrt{1 - \frac{1}{r_2^2}} \ket{\alpha_{T_3}}, \quad (j,j+1) \ket{\alpha_{T_2}} = \frac{1}{r_2} \ket{\alpha_{T_2}} + \sqrt{1 - \frac{1}{r_2^2}} \ket{\alpha_{T_4}}.
	\end{align}
	For instance, such a basis vector can be found using the following Young tableau:
	\begin{align}
		\newcommand\scalemath[2]{\scalebox{#1}{\mbox{\ensuremath{\displaystyle #2}}}} \scalemath{0.5}{
			\ytableausetup{boxsize=2.5em}
			\begin{ytableau}
				1 & 3 & 5 & \none[\dots] & \none[\dots]
				& i+1 & \none[\dots] & \none[\dots] & j+1 & \none[\dots] & \none[\dots] & n \\
				2 & 4 & \none[\dots] & \none[\dots] & i & \none[\dots] & \none[\dots] & j & \none[\dots] & \none[\dots] & n-1
		\end{ytableau}  }
	\end{align}
	Then one can check by definition that  
	\begin{align}\label{eq:non-commute1}
		\begin{aligned}
			& \Big\langle E_{\ket{\alpha_{T_1}}, \ket{\alpha_{T_3}}} \otimes E_{\ket{\alpha_{T_3}}, \ket{\alpha_{T_1}}}, ( T_k^{\mathrm{YJM}} T_k^{\tau_2} T_k^{\mathrm{YJM}} ) ( T_k^{\mathrm{YJM}} T_k^{\tau_1} T_k^{\mathrm{YJM}}) E_{\ket{\alpha_{T_1}}, \ket{\alpha_{T_1}}} \otimes E_{\ket{\alpha_{T_2}}, \ket{\alpha_{T_2}}} \Big\rangle \\
			\neq & \Big\langle E_{\ket{\alpha_{T_1}}, \ket{\alpha_{T_3}}} \otimes E_{\ket{\alpha_{T_3}}, \ket{\alpha_{T_1}}}, ( T_k^{\mathrm{YJM}} T_k^{\tau_1} T_k^{\mathrm{YJM}} ) ( T_k^{\mathrm{YJM}} T_k^{\tau_2} T_k^{\mathrm{YJM}}) E_{\ket{\alpha_{T_1}}, \ket{\alpha_{T_1}}} \otimes E_{\ket{\alpha_{T_2}}, \ket{\alpha_{T_2}}} \Big\rangle,
		\end{aligned}
	\end{align}
	which proves the non-commutativity. 
\end{remark}

\begin{remark}
	For similar reasons, we find that the martingale method \cite{Nachtergaele1996} is also invalid for studying the spectral gap of $T_k^{\mathcal{E}_{\mathrm{CQA}}}$ for general qudits. Adjusting notations in Ref.~\cite{Nachtergaele1996} to our case, we set 
	\begin{align}
		H_{\Lambda_m} = \sum_{i = 1}^m I - T_k^{\mathrm{YJM}} T_k^{\tau_i} T_k^{\mathrm{YJM}}, \quad
		H_{\Lambda_m \setminus \Lambda_{m - l}} = \sum_{i = m -l+1}^m I - T_k^{\mathrm{YJM}} T_k^{\tau_i} T_k^{\mathrm{YJM}},
	\end{align}
	where $l \leq m \leq n$. Let $G_m, G_{\Lambda_m \setminus \Lambda_{m - l}}$ denote the operators that project vectors into the ground state spaces of $H_m, H_{\Lambda_m \setminus \Lambda_{m - l}}$ respectively. Assume that $m_1 \leq m_2 - l$. Then $G_{m_1}$ and $G_{\Lambda_{m_2} \setminus \Lambda_{m_2 - l}}$ do not commute, which violates the requirement in Ref.~\cite{Nachtergaele1996}: 
	\begin{align}\label{eq:non-commute2}
		\begin{aligned}
			& \Big\langle E_{\ket{\alpha_{T_1}}, \ket{\alpha_{T_3}}} \otimes E_{\ket{\alpha_{T_3}}, \ket{\alpha_{T_1}}}, ( G_{m_1} G_{\Lambda_{m_2} \setminus \Lambda_{m_2 - l}} ) E_{\ket{\alpha_{T_1}}, \ket{\alpha_{T_1}}} \otimes E_{\ket{\alpha_{T_2}}, \ket{\alpha_{T_2}}} \Big\rangle \\
			\neq & \Big\langle E_{\ket{\alpha_{T_1}}, \ket{\alpha_{T_3}}} \otimes E_{\ket{\alpha_{T_3}}, \ket{\alpha_{T_1}}}, (  G_{\Lambda_{m_2} \setminus \Lambda_{m_2 - l}} G_{m_1} ) E_{\ket{\alpha_{T_1}}, \ket{\alpha_{T_1}}} \otimes E_{\ket{\alpha_{T_2}}, \ket{\alpha_{T_2}}} \Big\rangle,
		\end{aligned}
	\end{align}
	where the Young basis vectors are selected, for example, as
	\begin{align}
		\newcommand\scalemath[2]{\scalebox{#1}{\mbox{\ensuremath{\displaystyle #2}}}} \scalemath{0.5}{
			\ytableausetup{boxsize=2.5em}
			\begin{ytableau}
				1 & 3 & 5 & \none[\dots] & \none[\dots]
				& i+1 & \none[\dots] & \none[\dots] & j+1 & j+2 & j+3 & \none[\dots] & n \\
				2 & 4 & \none[\dots] & \none[\dots] & i & \none[\dots] & \none[\dots] & j
		\end{ytableau}  }.
	\end{align}   
\end{remark}


\subsubsection*{Properties of YJM elements}

There is a potential approach based on $S_n$ representation theory \cite{Okounkov1996,Tolli2009} that does not rely on the numerical calculation of local gaps. Let us replace $\mathcal{E}_{\operatorname{eSWAP}}$ defined by a 1-dimensional adjacent transposition chain by the ensemble consisting of time evolutions generated by $(1,n),(2,n),\ldots,(n-1,n)$ (which corresponds to a star graphically). By Eq.~\eqref{eq:EulerIntegral1}, 
\begin{align}\label{eq:Star}
	\begin{aligned}
		T_{k=1}^{X_n} & = \frac{1}{n - 1} (T_{k=1}^{(1,n)} + T_{k=1}^{(2,n)} + \cdots + T_{k=1}^{(n-1,n)} + T_k^{(1,n)}) \\
		& = \frac{1}{2}I + \frac{1}{2(n-1)} ( (1,n)^{\otimes 2} + (2,n)^{\otimes 2} + \cdots + (n-1,n)^{\otimes 2}).
	\end{aligned}
\end{align}
Since each $(i,j)^{\otimes 2}$ is just given by the tensor product representation of $S_n$ on the $n$-qudit system, $T_{k=1}^{X_n}$ is interpreted as the matrix representation of elements from the group algebra $\mathbb{C}[S_n]$. 
Upon careful examination of the definition, we observe that $T_{k=1}^{X_n}$ is just the representation of the $n$-th YJM element $X_n$ (with the normalization factor ${1}/({n-1})$), hence the notation. As a remarkable property of YJM elements mentioned in Appendix \ref{sec:SnTheory}, their eigenvalues can be directly read off as components of content vectors of standard Young tableaux. For $X_n$, the largest possible eigenvalue is $n-1$ and the second largest one is $n-2$ corresponding to eigenstates (Young basis vectors) with the following standard tableaux: 
\begin{align}
	\newcommand\scalemath[2]{\scalebox{#1}{\mbox{\ensuremath{\displaystyle #2}}}} \scalemath{0.5}{
	\ytableausetup{boxsize=2.5em}
	\begin{ytableau}
		1 & 2 & 3 & \none[\dots]
		& n - 1 & n \\
	\end{ytableau}  \qquad \qquad
    \begin{ytableau}
    	1 & 2 & 3 & \none[\dots] & n - 1  \\ n
    \end{ytableau} }
\end{align}
Taking the normalization factor into account, we have
\begin{align}
	\lambda_2(T_{k=1}^{X_n}) \leq 1 - \frac{1}{n - 1} \ \text{ and } \ \Delta(T_{k=1}^{X_n}) \geq  \frac{1}{n - 1}.
\end{align}  
Using \eqref{eq:InequalitySecondEigenvalue} from the Remark of Appendix \ref{sec:FramePotential}, we conclude that when 
\begin{align}
	p \geq (n-1) \Big(  2n \log d + \log \frac{1}{\epsilon} \Big),
\end{align}
the aforementioned ensemble forms an $\epsilon$-approximate 1-design. For $k > 1$ however, Eq.~\eqref{eq:EulerIntegral2} generally does not induce a well-defined tensor product representation of YJM elements, and  more sophisticated treatments of $S_n$ representation theory are expected for addressing this problem.

Instead, let $\mathcal{E}_{S_n}$ denote the ensemble consisting of merely 2-local SWAPs. In each step of the random walk, we select $i$ randomly and apply $(i,n)$, as well as the identity channel, with probability 1/2. Then, we have
\begin{align}\label{eq:SnDesign}
	T_k^{\mathcal{E}_{S_n}} = \frac{1}{n-1} \sum_{i=1}^{n-1} \frac{1}{2}( I^{\otimes k} + (i,n)^{\otimes k} ),
\end{align}
the spectral gap of which is exactly the same as $T_{k=1}^{X_n}$ discussed above, which is independent of $k$ due to the properties of YJM elements. Since SWAPs generate $S_n$, using similar arguments from Theorem \ref{thm:Approximate}, we see that the ensemble $\mathcal{E}_{S_n}$ efficiently converges to unitary $k$-designs with respect to $S_n \subsetneqq \mathcal{U}_\times$. This is not surprising because a more general result in probability theory \cite{Diaconis1988,Varju2015,Meckes2019}says that the distribution induced via random walks using SWAPs efficiently converges in measure to the uniform distribution of $S_n$.


\subsubsection*{Random circuits with all-to-all interaction}

Besides the 1D chain ($\mathcal{E}_{\operatorname{eSWAP}}$) or  star ($\mathcal{E}_{X_n}$) architectures discussed above, it is also natural to consider employing all possible transpositions $(i,j)$ in defining the ensemble, corresponding to the so-called \emph{ all-to-all} circuit architecture, the associated interaction or adjacency graph of which is the \emph{complete graph}. As mentioned in Section \ref{sec:CQAEnsemble}, by Schur--Weyl duality, this is simply the $\mathcal{V}_2$ ensemble (cf.~Definition \ref{def:A-V4CQAEnsemble}) with
\begin{align}
	T_k^{\mathcal{E}_{\mathcal{V}_2}} = \frac{2}{(n - 1)n} \sum_{1 \leq i < j \leq n} T_k^{(i,j)}.
\end{align}

Analogous to Knabe's original derivation of 1-dimensional local gap threshold \cite{Knabe1988}, we have the following lemma for the complete graph case.
\begin{lemma}
	Let
	\begin{align}\label{eq:A-Complete}
		H \vcentcolon = \frac{2}{n(n-1)}(I - T_k^{\mathcal{E}_{\mathcal{V}_2}} ) = \sum_{1 \leq i < j \leq n} (I - T_k^{(i,j)}) = \sum_{1 \leq i < j \leq n}  P_k^{(i,j)}
	\end{align}
	and let $h_r = \sum_{i,j \neq r} P_k^{(i,j)}$ be a local bulk Hamiltonian supported on the $n-1$ vertices except for $r$. Suppose that $\Delta(h_r) = \gamma_{n-1}$. Then the gap of $H$ admits the lower bound
 \begin{align}
     \Delta(H)\geq\frac{1}{n-3}( (n-2)\gamma_{n-1} - 1).
 \end{align} More generally, given any collection of $m$ qubits, and letting $h_{i_1,\ldots,i_m} = \sum_{i < j \in \{i_1,\ldots,i_m\}} P_k^{(i,j)}$, then
	\begin{align}\label{eq:A-LocalGap}
		\gamma_n \geq 1 + \frac{n-2}{m-2}(\gamma_m - 1).
	\end{align}
\end{lemma}
\begin{proof}
	It is known that the inequality $H^2 \geq \gamma_n H$ implies that $\gamma_n$ is a lower bound of the gap of $H$. Expanding $H^2$ directly, we obtain
	\begin{align}
		H^2 - \frac{1}{n-3}\sum_{1 \leq r \leq n} h_r^2 + \frac{1}{n-3} H \geq 0.
	\end{align}
	With the assumption on the local gap of $h_r$,
	\begin{align}
		H^2 & \geq \frac{1}{n-3}\sum_{1 \leq r \leq n} h_r^2 - \frac{1}{n-3} H \\
		& \geq \frac{1}{n-3} \sum_{1 \leq r \leq n} \gamma_{n-1} h_r - \frac{1}{n-3} H = \frac{1}{n-3} (n-2) \gamma_{n-1}H - H = \frac{1}{n-3} \big( (n-2)\gamma_{n-1} - 1 \big) H.
	\end{align}
	As the inequality 
	\begin{align}
		\gamma_n \geq \frac{1}{n-3}\big( (n-2)\gamma_{n-1} - 1 \big) & \Leftrightarrow (n-3)\gamma_n \geq (n-2)\gamma_{n-1} - (n-2) + (n-3) \Leftrightarrow \frac{\gamma_n - 1 }{\gamma_{n-1} - 1} \geq \frac{n-2}{n-3} 
	\end{align}
	holds generally for all $n$,  by induction, we conclude that $\gamma_n \geq 1 + (n-2)\gamma_3$ for the lowest possible $m = 3$.
\end{proof}

\begin{remark}
	According to this lemma, if $\gamma_m \geq 1$, then $\gamma_n \geq 1$ for all $n \geq m$. However, if $\gamma_m < 1$, the bound of $\gamma_n$ would eventually decrease to a non-positive number and become invalid. Our numerical computation using Young orthogonal form shows that for $m=3,\ldots,10$, the gap of the bulk Hamiltonian locally supported on $m$-qubits is always ${1}/{2}$, which does not qualify for the above lemma. This naturally suggests that the gap remains to be $1/2$ for an arbitrary number of qubits. 
    As a reminder, the martingale method \cite{Nachtergaele1996} can be applied to lattices on complete graphs like the present case, so we should examine its validity again.  Let
	\begin{align}
		H_{\Lambda_m} = \sum_{1 \leq i < j \leq m} I - T_k^{(i,j)}, \quad 	H_{\Lambda_m \setminus \Lambda_{m-l}} = \Big( \sum_{j=1}^{m-l} I - T_k^{(j,m-l+1)} + \cdots + \sum_{j=1}^{m-1} I - T_k^{(j,m)} \Big),
	\end{align}  
	with $G_{\Lambda_m}, G_{\Lambda_m \setminus \Lambda_{m-l}}$ being projections onto the ground state subspace. One can also see by definition that $G_{\Lambda_m} = G_{\Lambda_m \setminus \Lambda_{m-l}}$.
	
	Let $m_1 \leq m_2 - l$ as in our previous remark. We check the commutativity of these projections on $S_{m_2}$ irrep blocks such as $S^{\lambda \otimes 2} \otimes S^{\mu \otimes 2}$ with $\lambda \neq \mu$. Given the Young basis elements $\ket{\alpha_{T_1}}, \ket{\alpha_{T_2}}$ of $S^\lambda, S^\mu$ respectively, according to our illustration in Appendix \ref{sec:SnCommutant}, we have
	\begin{align}\label{eq:non-commute3}
		\begin{aligned}
			& \Big\langle E_{\ket{\alpha_{T_1}}, \ket{\alpha_{T_2'}}} \otimes E_{\ket{\alpha_{T_2}}, \ket{\alpha_{T_1'}}},   G_{\Lambda_{m_2} \setminus \Lambda_{m_2 - l}} G_{\Lambda_{m_1}} E_{\ket{\alpha_{T_1}}, \ket{\alpha_{T_1}}} \otimes E_{\ket{\alpha_{T_2}}, \ket{\alpha_{T_2}}} \Big\rangle \\
			= & \Big\langle E_{\ket{\alpha_{T_1}}, \ket{\alpha_{T_2'}}} \otimes E_{\ket{\alpha_{T_2}}, \ket{\alpha_{T_1'}}},  G_{\Lambda_{m_2}} G_{\Lambda_{m_1}}  E_{\ket{\alpha_{T_1}}, \ket{\alpha_{T_1}}} \otimes E_{\ket{\alpha_{T_2}}, \ket{\alpha_{T_2}}} \Big\rangle = 0 \\
			\neq & \Big\langle E_{\ket{\alpha_{T_1}}, \ket{\alpha_{T_2'}}} \otimes E_{\ket{\alpha_{T_2}}, \ket{\alpha_{T_1'}}}, G_{\Lambda_{m_1}} G_{\Lambda_{m_2} \setminus \Lambda_{m_2 - l}}  E_{\ket{\alpha_{T_1}}, \ket{\alpha_{T_1}}} \otimes E_{\ket{\alpha_{T_2}}, \ket{\alpha_{T_2}}} \Big\rangle,
		\end{aligned}
	\end{align}
	where the standard tableaux corresponding to the above Young basis elements can be taken as (for $m_1 = 3$ and $m_2 = 6$)
	\begin{align}
		\newcommand\scalemath[2]{\scalebox{#1}{\mbox{\ensuremath{\displaystyle #2}}}} \scalemath{0.5}{
		\ytableausetup{boxsize=2.5em}
		T_1 = \begin{ytableau}
				1 & 3 & 5 & 6 \\
				2 & 4 
		\end{ytableau}, \qquad
	    T_2 = \begin{ytableau}
		1 & 2 & 4 & 5 & 6 \\
		3  
		\end{ytableau} , \qquad
	    T_2' = \begin{ytableau}
	    	1 & 2 & 5 & 6 \\
	    	3 & 4 
	    \end{ytableau}, \qquad
	    T_1' = \begin{ytableau}
	    	1 & 3 & 4 & 5 & 6 \\
	    	2  
	    \end{ytableau} }.
	\end{align}   
	
	There is also a recursion relation developed in Ref.~\cite{Haferkamp2021} to bound the second largest eigenvalue of all-to-all random circuits without symmetry. As mentioned in Section \ref{sec: cqa-convergence}, since it is generally impossible to rewrite a generalized permutation (see also Appendix \ref{sec:SnCommutant}) into an $n$-fold tensor product, this strategy using induction on the number of qudits is inapplicable in our case.  The reason is that for systems with $1,2,...,n-1,n$ qudits, the underlying permutation actions are given by 
    \begin{align}
        S_1 \subset S_2 \subset \cdots \subset S_{n-1} \subset S_n,
    \end{align}
    which yields distinct irreps governed by these groups and makes it difficult to apply any induction hypothesis from the case of $n-1$ directly to the case of $n$,
\end{remark}


\subsection{Numerical methods for qudits}\label{sec:qudits}

We now describe how to compute the (partial) spectra of $T_k^{\mathcal{E}_{\mathrm{CQA}}}$ or $T_k^{\mathcal{E}_{\mathcal{V}_4}}$ with higher order YJM elements or cycles from $S_n$, which is indispensable for qudits with $d > 2$. 

For YJM elements, since they are diagonal matrices with explicit diagonal entries (see Appendix \ref{sec:SnTheory}), 
\begin{align}
	T_k^{\mathrm{YJM}} = \int_0^{2\pi} (e^{-i \sum_{i,j} \beta_{ij} X_i X_j} )^{\otimes k} \otimes (e^{i \sum_{i,j} \beta_{ij} X_i X_j} )^{\otimes k} d\beta_{ij} = \prod_{i,j} \int_0^{2\pi} (e^{-i \beta_{ij} X_i X_j} )^{\otimes k} \otimes (e^{i \beta_{ij} X_i X_j} )^{\otimes k} d\beta_{ij}, 
\end{align} 
where $\prod_{i,j}$ denotes a matrix product, can be straightforwardly calculated. This is how we compare the spectra of
\begin{align}
	T_2^{\mathrm{YJM}} \frac{1}{5} (T_2^{1,2} + \cdots T_2^{5,6}) T_2^{\mathrm{YJM}} \ \text{ and } \ T_2^{1,2} + \cdots T_2^{5,6}
\end{align}
restricted on the $2$-fold tensor product of the irrep $S^{(3,2,1)}$ in the end of Appendix \ref{sec:CQAEnsemble}, which demonstrates the necessity of using second-order YJM elements to generate unitary $k$-designs on general qudits.

However, conjugating with YJM elements ($T_k^{\mathrm{YJM}}$) always breaks the permutation invariance of the $k$-fold channel of the  ensemble concerned. For instance, now there is no reason to consider
\begin{align}
	T_k^{\mathrm{YJM}} (T_k^{i,i+1} + T_k^{i+1,i+2}) T_k^{\mathrm{YJM}}  \ \text{ and } \ T_k^{\mathrm{YJM}} (T_k^{j,j+1} + T_k^{j+1,j+2}) T_k^{\mathrm{YJM}} 
\end{align} 
similar matrices, so computing local gaps for the first few terms from $T_k^{\mathcal{E}_{\mathrm{CQA}}}$ is not even sufficient for applying Kanbe's theory, as well as any other theory relying on the permutation invariance. 

However, there is still a major problem associated with $\mathcal{E}_{\mathcal{V}_4}$. By Schur--Weyl duality, any 4-local $\text{SU}(d)$-symmetric unitaries can, in principle, be produced by $S_4$ group elements, but there is no constructive strategy, contrary to the 2-local case, where we can simply take $\exp(-it(i,j))$. One may parametrize a general 4-local unitary by generalized Euler angles, but it is also not applicable here taking $\text{SU}(d)$ symmetry into account. To remedy the problem, we reformulate the definition of $\mathcal{E}_{\mathcal{V}_4}$ as follows.

\begin{definition}
	Define $\mathcal{E}_4$ to be a 4-local unitary circuit ensemble where in each step of the random walk, we randomly select four integers $i_1,i_2,i_3$, and $i_4$ from $1,2,\ldots,n$. Then we uniformly sample unitaries determined by 2-cycles, 3-cycles, and $(2,2)$-cycles permuting these integers:
	\begin{align}
		\exp\Big(-it(i_1,i_2)\Big), \quad \exp\Big(-it \big( (i_2 i_3 i_4) + (i_4 i_3 i_2) \big) \Big), \quad \exp\Big(-it(i_1,i_3)(i_2,i_4)\Big).
	\end{align}
\end{definition}

Note that a 3-cycle such as $(i_2, i_3, i_4)$ is not Hermitian in general, but $(i_2, i_3, i_4) + (i_4, i_3, i_2)$ is a well-defined 3-local $\text{SU}(d)$-symmetric Hamiltonian. More importantly, since second-order products of YJM elements only produce cycles defined as above (see Appendix \ref{sec:CQA2}), following a similar argumentation from Lemma \ref{lemma:Ensemble2} and Theorem \ref{thm:Approximate}, we confirm that this modified ensemble converges to a $k$-design if $k < n$. To compute
\begin{align}
	& T_k^{(i_1,i_3)(i_2,i_4)} = \int (e^{-i t(i_1,i_3)(i_2,i_4)} )^{\otimes k} \otimes (e^{i t(i_1,i_3)(i_2,i_4) } )^{\otimes k} dt \\
	& T_k^{(i_1, i_3, i_4)} = \int (e^{-it\big( (i_2, i_3, i_4) + (i_4, i_3, i_2) \big) } )^{\otimes k} \otimes (e^{i t\big( (i_2, i_3, i_4) + (i_4, i_3, i_2) \big) } )^{\otimes k} dt
\end{align}
constituting $T_k^{\mathcal{E}_4}$, we note that $( (i_1,i_3)(i_2,i_4))^2 = I$ and $\big( (i_2, i_3, i_4) + (i_4, i_3, i_2) \big)^2 = (i_2, i_3, i_4) + (i_4, i_3, i_2) + 2I$. Therefore, these Hermitian matrices only admit integer eigenvalues and hence the parameter $t$ is restricted to $[0, 2\pi]$. Besides, $T_k^{(i_1,i_3)(i_2,i_4)}$ is expanded in the same way as 2-cycle transpositions, e.g., Eq.~\eqref{eq:EulerIntegral2}.

To compute the second integral, simply note that
\begin{align}
	\begin{aligned}
		& \big( 2I - (i_2, i_3, i_4) - (i_4, i_3, i_2) \big)^2 = 3\big( 2I - (i_2, i_3, i_4) - (i_4, i_3, i_2) \big) \Big) \\
		\implies\quad & \big( 2I - (i_2, i_3, i_4) - (i_4, i_3, i_2) \big)^m = 3^{m-1} \big( 2I - (i_2, i_3, i_4) - (i_4, i_3, i_2) \big) \\
		\implies\quad & e^{i t\big( 2I - (i_2, i_3, i_4) - (i_4, i_3, i_2) \big) } = \frac{1}{3} (e^{3it} - 1) \big( 2I - (i_2, i_3, i_4) - (i_4, i_3, i_2) \big) + I.
	\end{aligned}
\end{align}
Therefore, by substituting
\begin{align}
    \begin{aligned}
        e^{-it\big( (i_2, i_3, i_4) + (i_4, i_3, i_2) \big) } & = e^{-2it} e^{-i t\big( (i_2, i_3, i_4) - (i_4, i_3, i_2) - 2I\big)} \\
        & = \frac{1}{3} (2e^{it} + e^{-2it} )I - \frac{1}{3} (e^{it} - e^{-2it} ) \big( (i_2, i_3, i_4) + (i_4, i_3, i_2) \big)
    \end{aligned}
\end{align}
into the integral, we can easily obtain the answer.


\end{document}